\documentclass[twocolumn]{aastex62}

\received{2019 December 16}
\accepted{2020 March 25}
\submitjournal{ApJ}
\usepackage{amsmath}
\usepackage{amssymb}
\usepackage{graphicx,graphics}
\usepackage{rotating}
\usepackage{natbib}
\usepackage{color}
\usepackage{comment}
\usepackage{apjfonts}
\usepackage{listings}
\usepackage{booktabs}
\usepackage{hyperref}
\hypersetup
{
    pdfauthor={Yuki Kimura},
    pdftitle={Kimura et al. 2020},
}
\shorttitle{HSC SSP AGN Optical Variability}
\shortauthors{Y. Kimura et al}
\begin{document}
\title{Properties of AGN Multiband Optical Variability in the HSC SSP Transient Survey}
\correspondingauthor{Yuki Kimura}
\email{yuki.kimura@astr.tohoku.ac.jp}
\author[0000-0002-6740-1242]{Yuki Kimura}
\affiliation{Astronomical Institute, Tohoku University, 6-3 Aramaki, Aoba-ku, Sendai, Miyagi 980-8578, Japan}
\author{Toru Yamada}
\affiliation{Astronomical Institute, Tohoku University, 6-3 Aramaki, Aoba-ku, Sendai, Miyagi 980-8578, Japan}
\affiliation{Institute of Space Astronautical Science, Japan Aerospace Exploration Agency, Sagamihara, Kanagawa 252-5210, Japan}
\author{Mitsuru Kokubo}
\affiliation{Astronomical Institute, Tohoku University, 6-3 Aramaki, Aoba-ku, Sendai, Miyagi 980-8578, Japan}
\author{Naoki Yasuda}
\affiliation{Kavli Institute for the Physics and Mathematics of the Universe (WPI), Institutes for Advanced Study, University of Tokyo, Kashiwa, Chiba 277-8583, Japan}
\author{Tomoki Morokuma}
\affiliation{Institute of Astronomy, Graduate School of Science, The University of Tokyo, 2-21-1 Osawa, Mitaka, Tokyo 181-0015, Japan}
\author{Tohru Nagao}
\affiliation{Research Center for Space and Cosmic Evolution, Ehime University, 2-5 Bunkyo-cho, Matsuyama, Ehime 790-8577, Japan}
\author{Yoshiki Matsuoka}
\affiliation{Research Center for Space and Cosmic Evolution, Ehime University, 2-5 Bunkyo-cho, Matsuyama, Ehime 790-8577, Japan}
\begin{abstract}
We study variability of active galactic nuclei (AGNs) by using the deep optical multiband photometry data obtained from the Hyper Suprime-Cam Subaru Strategic Program (HSC SSP) survey in the COSMOS field.
The images analyzed here were taken with 8, 10, 13, and 15 epochs over three years in the $g$, $r$, $i$, and $z$ bands, respectively. 
We identified 491 robust variable AGN candidates, down to $i=25$ mag and with redshift up to $4.26$. 
Ninety percent of the variability-selected AGNs are individually identified with the X-ray sources detected in the Chandra COSMOS Legacy survey.
We investigate their properties in variability by using structure function analysis and find that the structure function for low-luminosity AGNs ($L_{\mathrm{bol}}\lesssim10^{45}$~erg~s$^{-1}$) shows a positive correlation with luminosity, which is the opposite trend for the luminous quasars.
This trend is likely to be caused by larger contribution of the host galaxy light for lower-luminosity AGNs.
Using the model templates of galaxy spectra, we evaluate the amount of host galaxy contribution to the structure function analysis and find that dominance of the young stellar population is needed to explain the observed luminosity dependence.
This suggests that low-luminosity AGNs at $0.8\lesssim z\lesssim1.8$ are predominantly hosted in star-forming galaxies.
The X-ray stacking analysis reveals the significant emission from the individually X-ray undetected AGNs in our variability-selected sample. 
The stacked samples show very large hardness ratios in their stacked X-ray spectrum, which suggests that these optically variable sources have large soft X-ray absorption by dust-free gas.
\end{abstract}
\section{Introduction} \label{sec1}
Active galactic nuclei (AGNs) show stochastic luminosity variation in all wavelength ranges on timescales from several hours to many years \citep{ulr97}.
Recent studies show that the variability amplitudes of luminous AGNs, i.e., quasars, is anticorrelated with their luminosities as well as the observed wavelengths in the UV/optical range \citep[e.g.,][]{van04}.
Since the optical radiation of a quasar is likely to be dominated by the accretion disk, such variability properties can be related to the disk dynamics.
Several models, like accretion disk instability \citep{ree84, kaw98}, X-ray reprocessing of disk thermal emission \citep{kro91}, and inhomogeneous disk model \citep{dex11}, have been suggested to explain the mechanism behind the AGN optical variability, but the primary  origin is still under debate \citep[e.g.,][]{kok14,kok15}.

While several methods to identify AGNs, such as X-ray detection, broad emission line detection, classification with narrow line ratios, and mid-infrared (MIR) color diagnostics, are prevalent, variability-based AGN selection method provides another way to detect type-I AGNs.
For low-luminosity AGNs, it is difficult to detect their faint X-ray radiation and also difficult to identify them by the line ratios or their colors due to the overwhelming host galaxy light \citep{hai16,mez18,bal18,bal19}.

If we consider the depth of the Chandra COSMOS Legacy survey \citep{civ16}, one of the currently deepest X-ray data sources with substantially large sky coverage ($>1$~deg$^2$ area), the limiting X-ray flux corresponds to an AGN luminosity of $L_{\mathrm{0.5-2\ keV}} \sim 10^{42.5}$~erg~s$^{-1}$ at redshift 1 \citep{mar16}.
This X-ray luminosity can be converted to a bolometric luminosity of $L_{\mathrm{bol}} \sim 10^{43.5}$~erg~s$^{-1}$ assuming a bolometric correction factor of $\sim10$ \citep{lus12}.
This bolometric luminosity of AGNs can be related with the Eddington luminosity $L_{\mathrm{Edd}}$ by introducing the Eddington ratio $\lambda_{\mathrm{Edd}}$, where $L_{\mathrm{bol}}=\lambda_{\mathrm{Edd}}L_{\mathrm{Edd}}$, and the Eddington luminosity can be converted to a black hole mass with $1.26\times10^{38}$~erg~s$^{-1}$ $(M_{\mathrm{BH}}/M_{\odot})$.
Assuming a typical Eddington ratio ($\lambda_{\mathrm{Edd}}\sim0.1$), the depth of the Chandra COSMOS Legacy survey can detect AGNs with a black hole mass larger than $10^{6.5}~M_{\odot}$ at $z\sim1$.
Additionally, such low black hole mass systems have emission lines with less than 2000~km~s$^{-1}$ width from the broad line region (BLR).
This means that such objects can be misclassified as type-II AGNs in the optical line diagnostics.

On the other hand, the variability method can be more efficient in identifying low-luminosity AGNs because lower-luminosity AGNs tend to show larger variability amplitudes \citep[e.g.,][]{van04}.
Recent studies show that the variability-based AGN selection method can detect low-mass black holes.
\citet{mor16} find a low-mass black hole ($M_{\mathrm{BH}}=2.7\times10^{6}~M_{\odot}$) whose line width of the broad H$\alpha$ emission is 1880 km~s$^{-1}$, from the high-cadence (1 hr) optical imaging data.
\citet{bal18,bal19} also find a few hundred low-mass AGNs with host galaxy stellar mass of $M_{\star}<10^{10}~M_{\odot}$ from the optical variability method.
Thus, the variability method, which enables us to search for low-mass black holes residing in low-luminosity type-I AGNs, is an essential tool to investigate the origin of the central black holes and their coevolution with the host galaxies.

For such low-luminosity AGNs, however, the effect of the host galaxies on the observed variability properties are not well investigated.
\citet{she11} show that the host galaxy light contributions appear in the UV-optical spectral energy distribution (SED) of the AGNs with a rest-frame $5100$ {\AA} luminosity less than about $10^{45}$~erg~s$^{-1}$.
How the host galaxy light affects the optical variability analysis should be further studied.

In order to detect faint AGNs with the variability method, deep multicolor imaging data with moderate cadence are essential. 
Since AGNs are relatively rare objects, wide sky coverage is also important to obtain a sufficient number of objects for statistical analysis.
Recently, Hyper Suprime-Cam Subaru Strategic Program \citep[HSC SSP;][]{aih18a,aih18b,miy18,kom18,kaw18,fur18,bos18,hau18,cou18}  time-domain observations were conducted as a part of the UltraDeep layer from May 2014 to April 2017 with five broadband filters ($g$, $r$, $i$, $z$, $y$) in the COSMOS field \citep{yas19}.
The main purpose of the UltraDeep layer in the HSC SSP survey is to probe high redshift galaxies and supernovae, and the limiting magnitude of each epoch in this field ($r\sim26$ mag) is much deeper than the previous variability surveys in the literature (e.g., $r_{\mathrm{P1}}\sim22$ for Pan-STARRS 1; \citealp{sim15}).
The observations were carried with fair cadence with more than eight epochs for each band.
Using the deep multi-epoch/band imaging data, we conduct an optical variability analysis to obtain a new sample of variability-selected AGNs especially for faint objects that have not been studied so far.
Our main purpose in this paper is to search for faint AGNs and study their optical variability properties.
We also discuss the effects of the host galaxy light contamination in the optical variability properties. 

This paper consists of the following sections.
First, we introduce the dataset of the HSC SSP UltraDeep COSMOS field and describe how we identify variable AGN candidates in Section~\ref{sec2}.
Then, we show the basic properties of these variability-selected AGNs in Section~\ref{sec3}.
The results of the X-ray stacking analysis for the variability-selected AGNs are also shown in this section.
The behavior of the optical variability amplitudes as a function of AGN physical parameters based on a structure function analysis are shown in Section~\ref{sec4}.
In Section~\ref{sec5}, we discuss the effects of host galaxy contamination on structure function analysis, and also examine the host galaxy properties of the low-luminosity variable AGNs.
We also discuss the interpretation of X-ray undetected variable AGNs.
We summarize our results in Section~\ref{sec6}.

Throughout this paper, we assume $\Lambda$CDM cosmological parameters of $\Omega_m=0.3$ and $\Omega_{\Lambda}=0.7$ and Hubble constant $H_0=70$~km~s$^{-1}$~Mpc$^{-1}$.
We use the AB magnitude system for all filters.

\section{Identification of Variable Objects} \label{sec2}
\subsection{Observed Data Set} \label{sec2_1}

We here briefly describe the HSC SSP survey and the data set used in our variability analysis.
HSC \citep{miy18, kom18, fur18} has 104 science CCDs covering a 1\fdg5 diameter field of view with pixel scale of 0.168 arcsecond pixel$^{-1}$.
The HSC SSP survey consists of three main layers, Wide, Deep, and UltraDeep.
These fields cover areas of 1400~deg$^2$ (spring and fall equatorial stripes, Hectomap; the coadded depth of $r$-band magnitude $\sim26$), 27~deg$^2$ (XMM-LSS, E-COSMOS, ELAIS-N1, DEEP2-F3; $r\sim27$), and 3.5~deg$^2$ (SXDS, COSMOS; $r\sim28$), respectively \citep{aih18b}.

In our variability analysis, we focus on the HSC UltraDeep Survey COSMOS field.
This layer is one of the deepest fields in the HSC SSP survey and is suitable for our variability study of less-luminous AGNs because the multiwavelength data, including the very deep X-ray observation \citep[the flux limits at 50\% completeness are $4.9\times10^{-16}$~erg~s$^{-1}$~cm$^{-2}$ in the soft band (0.5-2~keV) and $3.1\times10^{-15}$~erg~s$^{-1}$~cm$^{-2}$ in the hard band (2-10~keV) in the Chandra COSMOS Legacy survey;][]{civ16} are available.
The HSC SSP time-domain observations started in May 2014 and were completed in April 2017.
All of the single epoch imaging was reduced by using the HSC pipeline \citep{bos18} version 4.0.5 with the default configuration parameters.

Our variability analysis was conducted by using the HSC $g$, $r$, $i$, and $z$ band data.
We did not use the $y$ band data since the $y$ band data has a relatively shallow depth, and scattered light still remains in the coadd images \citep{yas19}.
The $r$ and $i$ band filters were replaced by new ones, referred to as the $r2$ band and $i2$ band with improved uniformity \citep{kaw18} on June 24, 2016 and February 2, 2016, respectively.
No notable systematic differences in the $r$ and $r2$ band nor the $i$ and $i2$ band photometries are found in our analysis.
For simplicity, we hereafter refer to both of them as the $r$ and $i$ band filters without distinction.
 
For the HSC data, the observed sky fields are specified by tracts and patches.
A tract represents a square field with 1\fdg5 on a side, and each tract is divided into $9\times9$ patches, each of which has $4200\times4200$ pixels (11\farcm76 on a side).
Each patch has an overlap of 200 pixels on each side with an adjacent patch.
In this paper, we refer to a patch as a $4000\times4000$ pixel field with no overlaps from contiguous patches.

\begin{figure}[t!]
\plotone{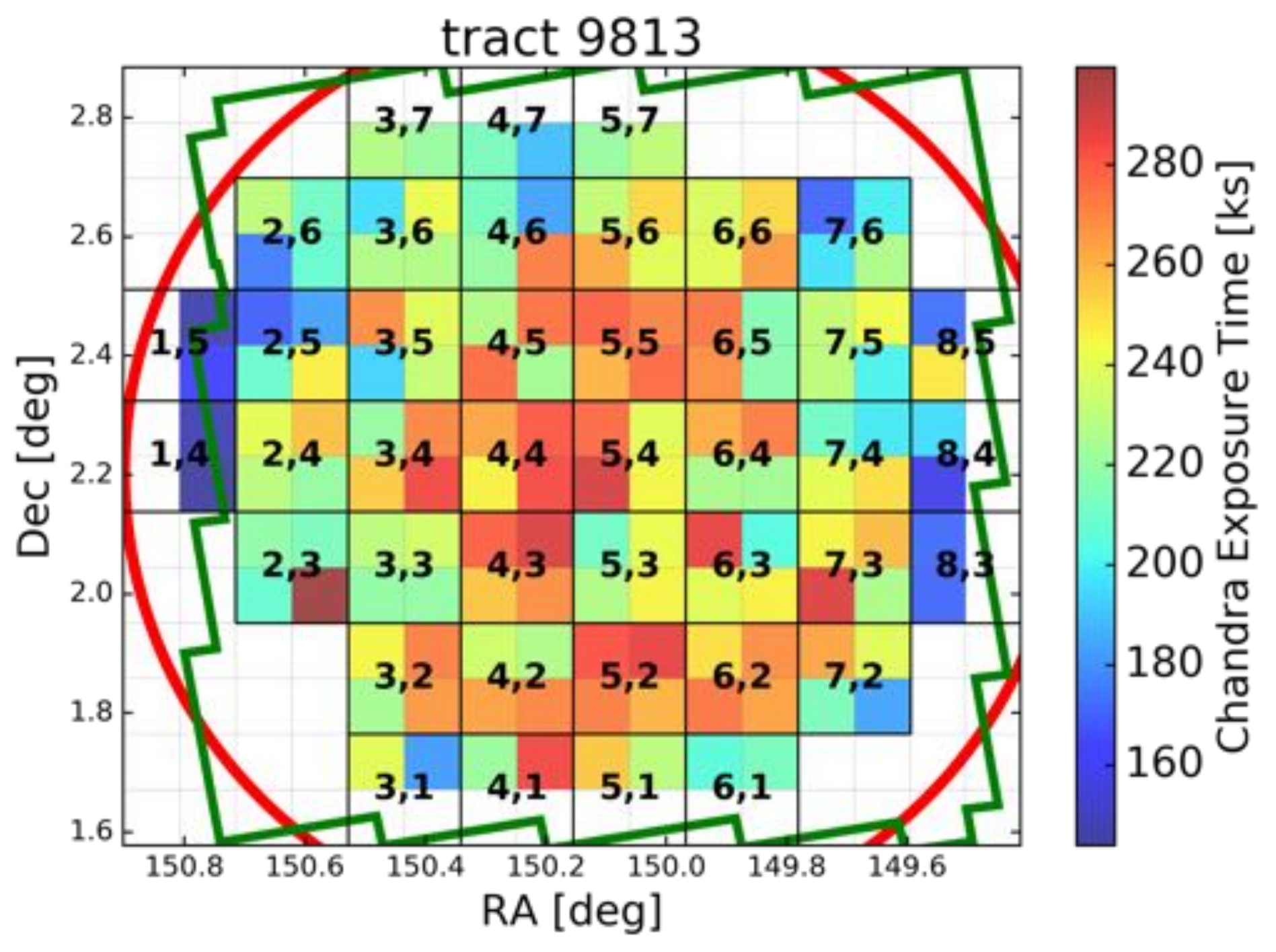}
\caption{
The HSC field of view in the COSMOS field is shown with the red circle.
The color bar shows the effective exposure time in the Chandra COSMOS Legacy survey.
In our variability analysis, we used color-mapped areas where the median values of the total Chandra X-ray exposure time in each sub-patch (i.e., a quarter of patch) field are larger than 150~ks.
The patch identification numbers are printed on the map.
The green solid line is the HST COSMOS survey area \citep{koe07}.
\label{fig1}
}
\end{figure}

We used the data in the `tract 9813' region, which contains a large part of the COSMOS field, and confined our analysis to the area of the 41 patch regions that overlaps with the deep Chandra X-ray observation where the total Chandra exposure time is larger than 150~ks \citep{civ16}.
The area where we performed the variability analysis is shown in Figure~\ref{fig1}.

We used only the data at the epoch with good seeing, where the FWHM of the point spread function (PSF) is less than 1\arcsec (see Section~\ref{sec2_2_2}).
The total number of epochs we used in the variability analysis consist of 8, 10, 13, and 15 for the $g$, $r$, $i$, and $z$ bands, respectively; thus, we have 28, 45, 78, and 105 pairs of epochs (we refer to these pairs as ``epoch-pairs'' in this paper) in each band filter, respectively. 
The data are summarized in Table~\ref{tbl1}.

\begin{deluxetable*}{cC|CCC|CCC|CCC|CCC}[!]
\tablecaption{Summary of the HSC SSP UltraDeep time-domain survey in the COSMOS field used in this paper
}
\tablehead{
 \colhead{} & \colhead{} \vline
 & \colhead{} & \colhead{$g$ band} & \colhead{} \vline
 & \colhead{} & \colhead{$r$ band} & \colhead{} \vline
 & \colhead{} & \colhead{$i$ band} & \colhead{} \vline
 & \colhead{} & \colhead{$z$ band} & \colhead{} \\
 \colhead{Date} & \colhead{MJD} \vline
 & \colhead{$\Delta t$} & \colhead{PSF} & \colhead{$m_{\rm{lim}}$} \vline
 & \colhead{$\Delta t$} & \colhead{PSF} & \colhead{$m_{\rm{lim}}$} \vline
 & \colhead{$\Delta t$} & \colhead{PSF} & \colhead{$m_{\rm{lim}}$} \vline
 & \colhead{$\Delta t$} & \colhead{PSF} & \colhead{$m_{\rm{lim}}$} \\
 (1) & (2) & (3) & (4) & (5) & (3) & (4) & (5) & (3) & (4) & (5) & (3) & (4) & (5)
}
\startdata
2014-03-28 & 56744 &  & - &  & 0 & 0.53 & 26.03 & 0 & 0.74 & 25.31 & 0 & 0.83 & 24.75 \\
2014-11-18 & 56979 & 0 & 0.72 & 26.24 &  & - &  &  & - &  &  & - &  \\
2015-01-16 & 57038 &  & - &  &  & - &  &  & - &  & 294 & 0.48 & 25.33 \\
2015-01-21 & 57043 &  & - &  &  & - &  & 299 & 0.53 & 25.66 &  & - &  \\
2015-03-18 & 57099 &  & - &  & 355 & 0.51 & 25.84 &  & - &  &  & - &  \\
2015-05-17 & 57159 & 180 & 0.90 & 25.89 &  & - &  &  & - &  &  & - &  \\
2015-05-21 & 57163 &  & - &  &  & - &  & 419 & 0.57 & 25.46 &  & - &  \\
2016-01-15 & 57402 &  & - &  &  & - &  &  & - &  & 658 & 0.82 & 24.67 \\
2016-03-07 & 57454 & 475 & 0.72 & 26.18 &  & - &  &  & - &  &  & - &  \\
2016-03-09 & 57456 &  & - &  & 712 & 0.94 & 25.51 &  & - &  &  & - &  \\
2016-03-12 & 57459 &  & - &  &  & - &  &  & - &  & 715 & 0.49 & 24.72 \\
2016-11-23 & 57715 &  & - &  &  & - &  &  & - &  & 971 & 0.69 & 24.50 \\
2016-11-25 & 57717 &  & - &  &  & - &  & 973 & 0.76 & 25.15 &  & - &  \\
2016-11-28 & 57720 &  & - &  & 976 & 0.72 & 25.60 &  & - &  &  & - &  \\
2016-11-29 & 57721 &  & - &  &  & - &  &  & - &  & 977 & 0.97 & 24.65 \\ 
2017-01-02 & 57755 & 776 & 0.65 & 25.96 &  & - &  & 1011 & 0.64 & 25.53 & 1011 & 0.72 & 24.58 \\
2017-01-21 & 57774 &  & - &  &  & - &  &  & - &  & 1030 & 0.49 & 25.20 \\
2017-01-23 & 57776 &  & - &  & 1032 & 0.77 & 25.60 & 1032 & 0.66 & 25.57 & & - &  \\
2017-01-30 & 57783 &  & - &  &  & - &  & 1039 & 0.71 & 25.14 & 1039 & 0.60 & 24.88 \\
2017-02-01 & 57785 & 806 & 0.61 & 25.58 &  & - &  &  & - &  &  & - &  \\
2017-02-02 & 57786 &  & - &  & 1042 & 0.61 & 25.59 & 1042 & 0.45 & 24.78 & & - &  \\ 
2017-02-21 & 57805 &  & - &  &  & - &  &  & - &  & 1061 & 0.61 & 24.68 \\
2017-02-23 & 57807 & & - & & 1063 & 0.87 & 25.59 &  & - &  &  & - &  \\
2017-02-25 & 57809 &  & - &  &  & - &  & 1065 & 0.65 & 25.02 &  & - &  \\
2017-03-04 & 57816 &  & - &  &  & - &  & 1072 & 0.59 & 25.48 & 1072 & 0.60 & 24.68 \\
2017-03-06 & 57818 &  & - &  & 1074 & 0.69 & 25.58 &  & - &  &  & - &  \\
2017-03-22 & 57834 & 855 & 0.79 & 25.97 &  & - &  &  & - & & 1090 & 0.54 & 24.71 \\
2017-03-23 & 57835 &  & - &  &  & - &  & 1091 & 0.62 & 24.98 &  & - &  \\
2017-03-25 & 57837 &  & - &  & 1093 & 0.90 & 25.37 &  & - &  &  & - &  \\
2017-03-29 & 57841 & 862 & 0.87 & 25.73 &  & - &  &  & - & & 1097 & 0.71 & 24.57 \\
2017-03-30 & 57842 &  & - &  &  & - &  & 1098 & 0.92 & 25.17 &  & - &  \\
2017-04-23 & 57866 &  & - &  & 1122 & 0.88 & 25.30 &  & - &  & 1122 & 0.76 & 24.37 \\
2017-04-26 & 57869 & 890 & 0.83 & 25.58 &  & - &  & & - & &  & - &  \\
2017-04-27 & 57870 &  & - &  &  & - &  & 1126 & 0.53 & 24.98 &  & - &  \\
2017-04-29 & 57872 &  & - &  &  & - &  &  & - &  & 1128 & 0.70 & 24.24
\enddata
\tablecomments{
Column (1): Observed date in the format of yyyy-mm-dd. 
Column (2): Modified Julian Date of the observed date.
Column (3): Time difference in days from the first observation epoch.
Column (4): Median FWHM value of PSF in arcsecond.
Column (5): Median limiting magnitude (S/N=5). The aperture radius is set to be 1.5 times the FWHM value of the matched PSF over the whole epoch in each filter.
\label{tbl1}
}
\end{deluxetable*}

\subsection{Data Analysis} \label{sec2_2}
\subsubsection{Targets for Photometry} \label{sec2_2_1}

\begin{deluxetable}{lc}[htbp!]
\tablecaption{HSC database selection flags and conditions
\label{tbl2}
}
\tablehead{
\colhead{Flag} &
\colhead{Boolean}
}
\startdata
{\rm detect\_is\_primary} & True\\
{\rm (g|r|i|z)flags\_pixel\_edge} & False\\
{\rm (g|r|i|z)flags\_pixel\_bad} & False\\
{\rm (g|r|i|z)flags\_pixel\_cr\_center} & False\\
{\rm (g|r|i|z)flags\_pixel\_saturated\_center} & False\\
{\rm (g|r|i|z)flags\_pixel\_interpolated\_center} & False\\
{\rm (g|r|i|z)flags\_pixel\_bright\_object\_any} & False\\
\enddata
\end{deluxetable}

We first collected object coordinates from the HSC SSP UltraDeep Survey multiband stacked catalog (hereafter HSC catalog) where we obtained aperture photometry in our variability analysis.
Using the HSC Catalog Archive Server (CAS), we selected isolated or deblended objects from the HSC catalog.
To construct a clean sample, we avoided objects with flags of bad pixels, cosmic-ray effects, and saturation.
In addition, we also avoided objects by using the bright-object flag, which indicates that the object is affected by the nearby bright stars.
The SQL selection conditions are summarized in Table~\ref{tbl2}.
We then imposed a magnitude threshold for the targets of photometry in the HSC catalog, $i=26$ mag, which is set to 0.3-1.2 mag deeper than the limiting magnitude of single epoch images (see Table~\ref{tbl1}).
In order to utilize the multiwavelength data in the COSMOS field, we also cross-matched the list with the COMOS2015 catalog \citep{lai16} by positions with the separation threshold of 0\farcs6.
Some objects listed in the COSMOS2015 catalog are matched with more than two objects listed in the HSC catalog, and we avoided such objects from the HSC catalog.
We finally selected a total of 271475 objects (parent sample) for the flux variation analysis that are in common between the HSC catalog (after flagging) and the COSMOS2015 catalog.

\subsubsection{PSF Matching} \label{sec2_2_2}

For the variability analysis, we corrected the PSF difference for each epoch-pair before performing aperture photometry.
As the PSF slightly varies over the wide HSC field of view, we measured the FWHM of the point sources in sub-patches with an area of $5\farcm6\times5\farcm6$ (i.e., a quarter of a patch field) to match the PSF of the images in the same filter in each epoch-pair.
Point sources are selected from the information of the second-order adaptive moment calculated by using the Hirata$-$Seljak$-$Mandelbaum (HSM) algorithm \citep{hir03, man05}.
A few tens of point sources are available in each sub-patch.
The median values of the FWHM over each image are listed in Table~\ref{tbl1}.
We matched the PSF sizes by using the IRAF \citep{iraf_86, iraf_93} $gauss$ task.

\subsubsection{Fixed Aperture Photometry} \label{sec2_2_3}

After PSF matching, we conducted aperture photometry using the IRAF $phot$ task.
In our variability analysis, we set the fixed aperture radius to be $1.5\times$ FWHM of the matched PSF.
We adopted the object positions listed in the HSC catalog (Section~\ref{sec2_2_1}) for the centers of the apertures.
The local average value of the sky background was evaluated in a circular annulus with the inner and outer radii of $2.0$ and $2.5$ times the FWHM, respectively.
The limiting magnitudes for each sub-patch field were also evaluated using multiple random apertures on the sky with the same radius.
If the objects have magnitudes below the limiting magnitudes, i.e., signal-to-noise ratio (S/N) $<5$, in both of the epochs, we flagged such targets as `$faint$' in the epoch-pair.

The aperture photometry may be affected by nearby sources, which may cause fake variation due to the slight change of seeing even after the PSF matching.
We flagged the objects when the aperture photometry is significantly affected by this contamination; if the surface brightness of adjacent sources at the aperture edge of the target source is larger than $2\sigma$ of the sky background, or the surface brightness of the target itself, we flagged such target as `$neighbor$' in the epoch-pair.

\subsubsection{Photometric Error of Flux Difference} \label{sec2_2_4}

\begin{figure}[t!]
\plotone{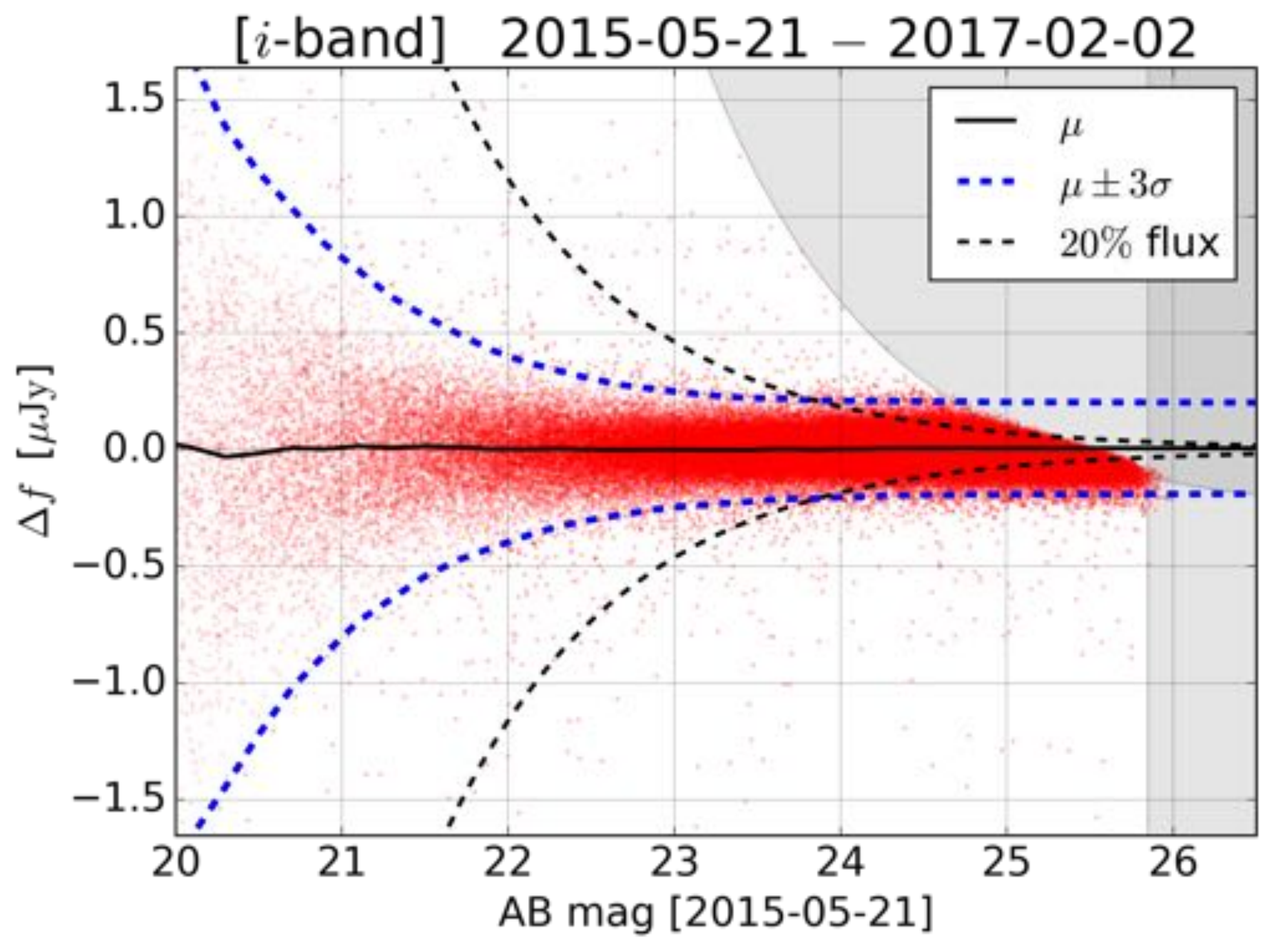}
\caption{
An example of flux differences ($\Delta f$) as a function of magnitude.
The black solid line is the mean $\mu$, and the blue dashed lines are $\pm3.0 \sigma$ deviations from the $\mu$ of the distribution of $\Delta f$ in each magnitude.
The black dashed lines are the $20 \%$ flux levels.
The shaded region is the lower S/N region ($\mathrm{S/N}<5$).
\label{fig2}
}
\end{figure}

We measured the photometric error of the flux difference in each epoch-pair.
First, we obtained $\Delta f$, defined as the flux difference for each object.
Then, we calculated the standard deviation of $\Delta f$ in each magnitude bin.
The standard deviation of $\Delta f$ was calculated by fitting a Gaussian function, $N_k\times\exp\left[-(\Delta f_k-\mu_k)^2/2\sigma_k^2\right]$, with three free parameters, normalization ($N$), mean value ($\mu$), and standard deviation ($\sigma$), of the distribution of $\Delta f$ in the $k$th magnitude bin.
Figure~\ref{fig2} shows an example of a $\Delta f$ distribution as a function of $i$ band magnitude in an epoch-pair of 2015-05-21 and 2017-02-02.

\subsection{Selection of Variable AGNs} \label{sec2_3}

In this subsection, we describe the method used to identify the candidates for variable AGNs.
Our selection is based on the probability of the target's flux differences for all of the epoch-pairs in each filter.
We introduce ensemble probability and how to evaluate the significance of variability in each filter (Section~\ref{sec2_3_1}).
Since AGNs have light curves that are correlated in wavelength, we then check the cross-correlation coefficients of light curves between different two variability-flagged filters (Section~\ref{sec2_3_2}).
After applying the criteria for the significance of the variability and cross-correlation coefficients, we conduct a visual inspection of the images and light curves for the variable candidates and remove spurious objects (Section~\ref{sec2_3_3}).
Finally, we check the variability detection rate in our method by using the variability-selected AGNs in the previous surveys (Section~\ref{sec2_3_4}).

\subsubsection{Ensemble Probability} \label{sec2_3_1}

To evaluate the significance of variability in each filter, we define `$ensemble$ $probability$' ($P_{\mathrm{band}}$), based on the observed flux differences, as
\begin{eqnarray}
P_{\mathrm{band}}(n)&=&\prod_{i}^{n} P_{i}\ (\Delta f_{i}^{\mathrm{(obj)}} \mid \mu_{i}, \sigma_{i} )\nonumber\\ 
&=&\prod_{i}^{n}\frac{1}{\sqrt{2\pi\sigma_{i}^2}}\exp\left[-\frac{(\Delta f_{i}^{\mathrm{(obj)}}-\mu_{i})^2}{2\sigma_{i}^2}\right],
\label{eq1}
\end{eqnarray}
where $\Delta f^{\mathrm{(obj)}}$ is the flux difference of a target, and $\mu$ and $\sigma$ are the mean and standard deviation, respectively, at the target magnitude in $i$th epoch-pair, which were calculated in Section~\ref{sec2_2_4}.
$n$ is the number of the epoch-pairs where the target is not flagged as $faint$ nor $neighbor$, described in Section~\ref{sec2_2_3}.
It is noted that the maximum values of $n$ are 28, 45, 78, and 105 for $g$, $r$, $i$, and $z$ bands, respectively.

The smaller ensemble probability means more significant flux variation.
We use a threshold value (in each filter separately) in our criteria for variable objects.
We set a threshold by considering the minimum ensemble probability of `$non$-$variable$' objects ($P_{\mathrm{min}}(n)$) that occupy a large fraction of the parent sample.
Non-variable objects are defined as follows: first we consider the following function to decide the critical value of the flux difference,
\begin{eqnarray}
F(n, x_{\mathrm{crit}})&\equiv&\prod_{i}^{n} P_{i}\left(\frac{|\Delta f_{i}^{\mathrm{(obj)}}-\mu_{i}|}{\sigma_i}\leq x_{\mathrm{crit}}\mid \mu_{i}, \sigma_{i}\right)\nonumber\\ 
&=&\left[\int_{-x_{\mathrm{crit}}}^{x_{\mathrm{crit}}} \frac{1}{\sqrt{2\pi}}\exp\left(-x^2/2\right)dx\right]^{n}.
\label{eq2}
\end{eqnarray}
This function gives the probability of objects that have never experienced an absolute flux difference ($|\Delta f_{i}^{\mathrm{(obj)}}-\mu_{i}|$) more than $x_{\mathrm{crit}}\sigma_i$ in all of the $n$ epoch-pairs.
In this paper, we set $F=0.95$, indicating that 95\% of the sources are assumed to be non-variable objects for the data in one band filter.
Once $F$ is set, the critical value $x_{\mathrm{crit}}$ only depends on the number of the epoch-pairs, e.g., $x_{\mathrm{crit}}\sim3.12$ for $n=28$, and $x_{\mathrm{crit}}\sim3.49$ for $n=105$ (the maximum number of epoch-pairs for the $g$ and $z$ bands, respectively).
Then, we find the objects that have never experienced an absolute flux difference more than $x_{\mathrm{crit}} \sigma_i$ in all of the $n$ epoch-pairs.
Hereafter, we refer to these objects as the `$non$-$Var$' sample.
Using the $non$-$Var$ sample, we calculate the minimum ensemble probability $P_{\mathrm{min}}(n)$ for each filter.
We then search for the objects that satisfy the following condition:
\begin{eqnarray}
P_{\mathrm{band}}(n)<P_{\mathrm{min}}(n),
\label{eq3}
\end{eqnarray}
and put a flag of `$variability$' in this filter to the objects.

\begin{figure}[t!]
\plotone{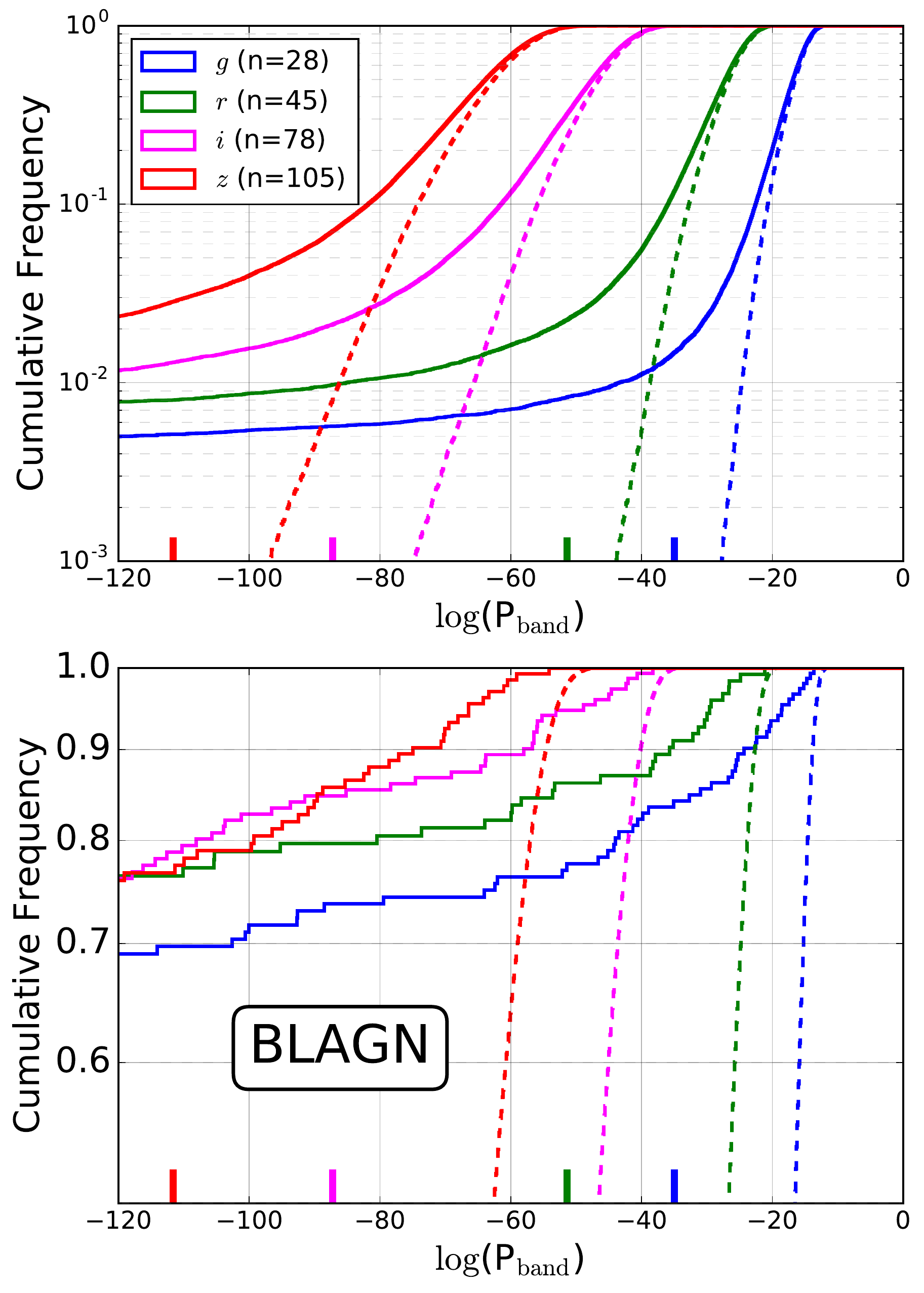}
\caption{
Cumulative distributions of the ensemble probabilities $P_{\mathrm{band}}(n)$ for the parent sample (top panel) and the known BLAGNs (bottom panel).
We show the objects that have the maximum number of epoch-pairs in each filter.
The solid lines and the dashed lines are calculated from the parent sample and $non$-$Var$ sample, respectively (blue: $g$ band, green: $r$ band, magenta: $i$ band, and red: $z$ band).
The short vertical lines are the minimum ensemble probabilities of the $non$-$Var$ sample (i.e., $P_{\mathrm{min}}(n)$) for each filter.
\label{fig3}
}
\end{figure}

The top panel of Figure~\ref{fig3} shows the cumulative distributions of ensemble probabilities in each filter ($n$ is the maximum epoch-pair in each filter).
The bottom panel of Figure~\ref{fig3} shows the same distribution as the top panel but only for the well-known (i.e., previously cataloged) broad line AGNs (BLAGNs) in the parent sample.
We use the X-ray catalog \citep[hereafter, Chandra catalog;][]{mar16}, the COSMOS2015 catalog, and the HSC catalog to select those objects from all of the following criteria: 
(i) $spec\_type=1$ from the Chandra catalog, which are BLAGNs (FWHM > 2000~km~s$^{-1}$) AGNs identified by spectroscopic information; (ii) $Qg\geq1.5$ from the Chandra catalog, which means clear spectroscopic redshift is available; (iii) $21\leq m_i \leq 24$ from the HSC catalog, where $m_i$ is the cmodel $i$ band magnitude; and (iv) not identified as stars.
The stars are selected from either of the following flags (hereafter, star-flags); $TYPE = 1$ from the COSMOS2015 catalog (identified from SED fitting), $star\_flag \geq 1$ from the Chandra catalog (spectroscopically, photometrically, and visually identified), or spectroscopic redshift $spec$-$z=0$ from the HSC catalog.
It is clearly shown in Figure \ref{fig3} that $\gtrsim75$\% of the BLAGNs are classified as significant variable objects in each filter.

We then apply the following conservative criterion:
\begin{eqnarray}
n_{\mathrm{band}} \geq 2,
\label{eq4}
\end{eqnarray}
where $n_{\mathrm{band}}$ is the number of $variability$-flagged filter bands.
From this criterion, we find 1744 variable candidates (0.64\% of the parent sample).
Although this criterion may remove real variable objects, it is found to be useful in removing the single-band fake variable sources that may be affected by spurious detections such as passing artificial satellites or bad pixels.
We check the fraction of the objects selected by this criterion from the known BLAGNs and find that 83\% of the BLAGNs satisfy the criterion in Equation~(\ref{eq4}) (other 4\% show $n_{\mathrm{band}}=1$).

\subsubsection{Cross-correlation of the Multiband Light Curves} \label{sec2_3_2}

\begin{figure*}[htbp!]
\gridline{
\fig{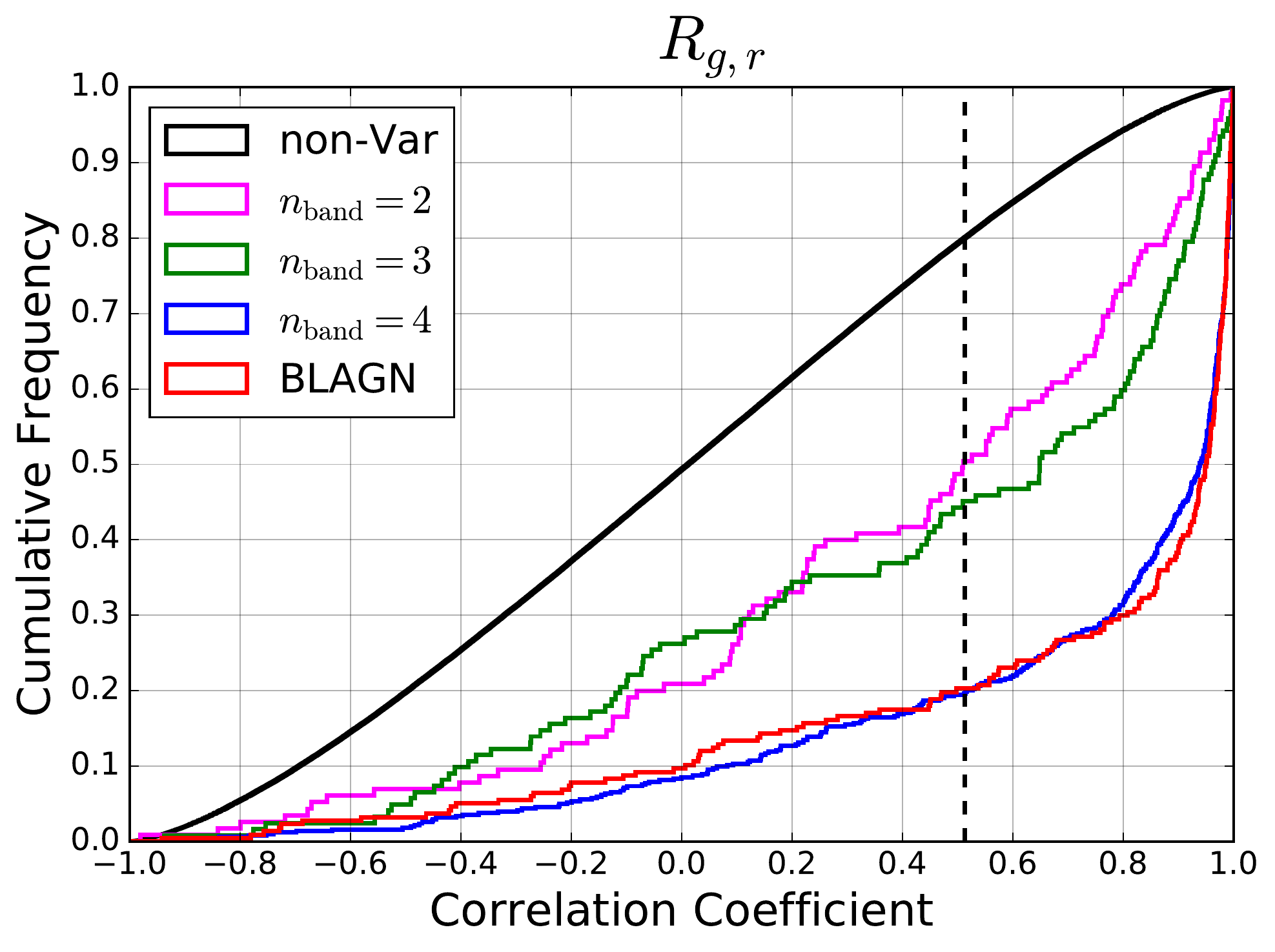}{0.33\textwidth}{(a) $g$ band \& $r$ band}
\fig{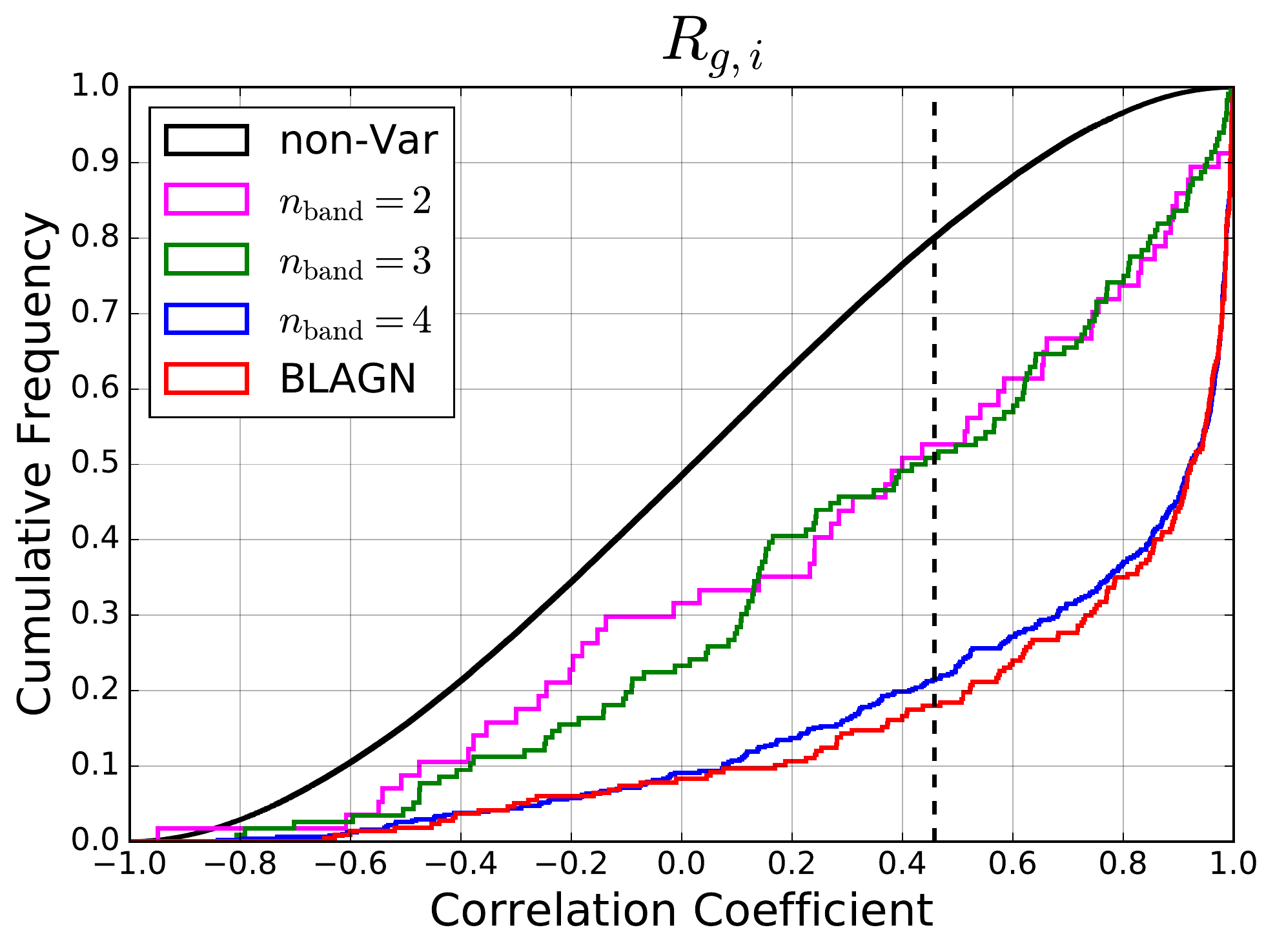}{0.33\textwidth}{(b) $g$ band \& $i$ band}
\fig{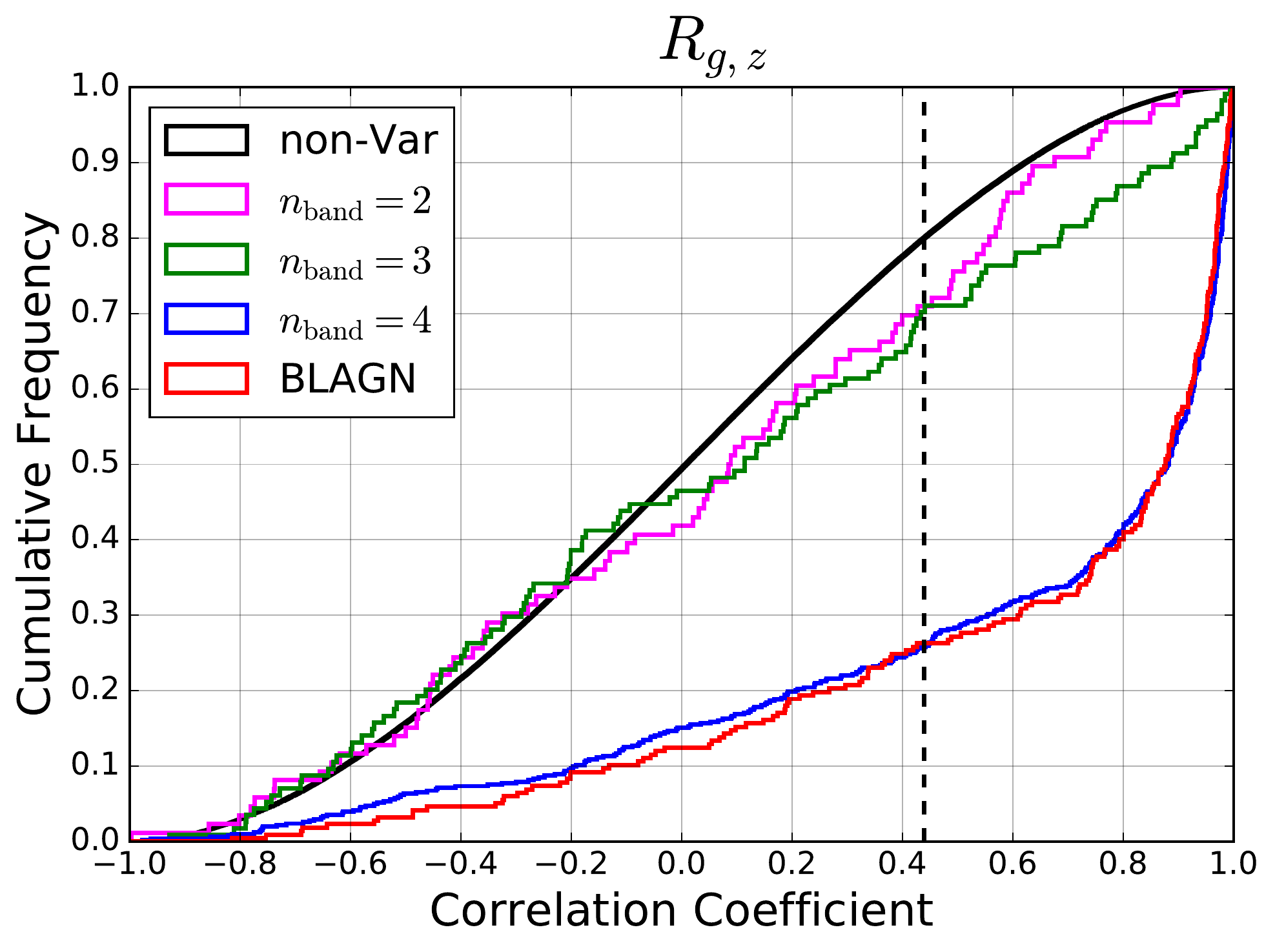}{0.33\textwidth}{(c) $g$ band \& $z$ band}
}
\gridline{
\fig{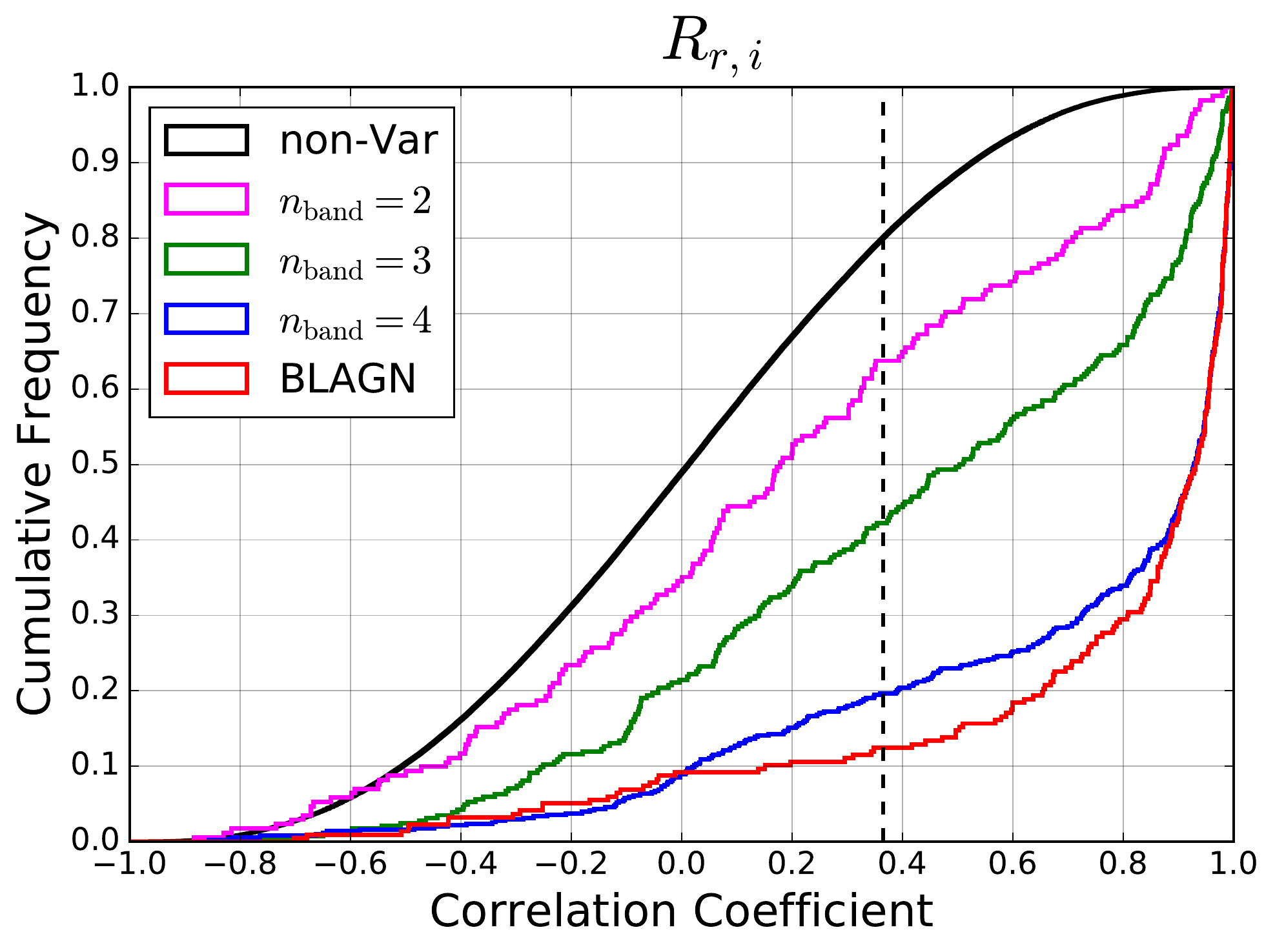}{0.33\textwidth}{(d) $r$ band \& $i$ band}
\fig{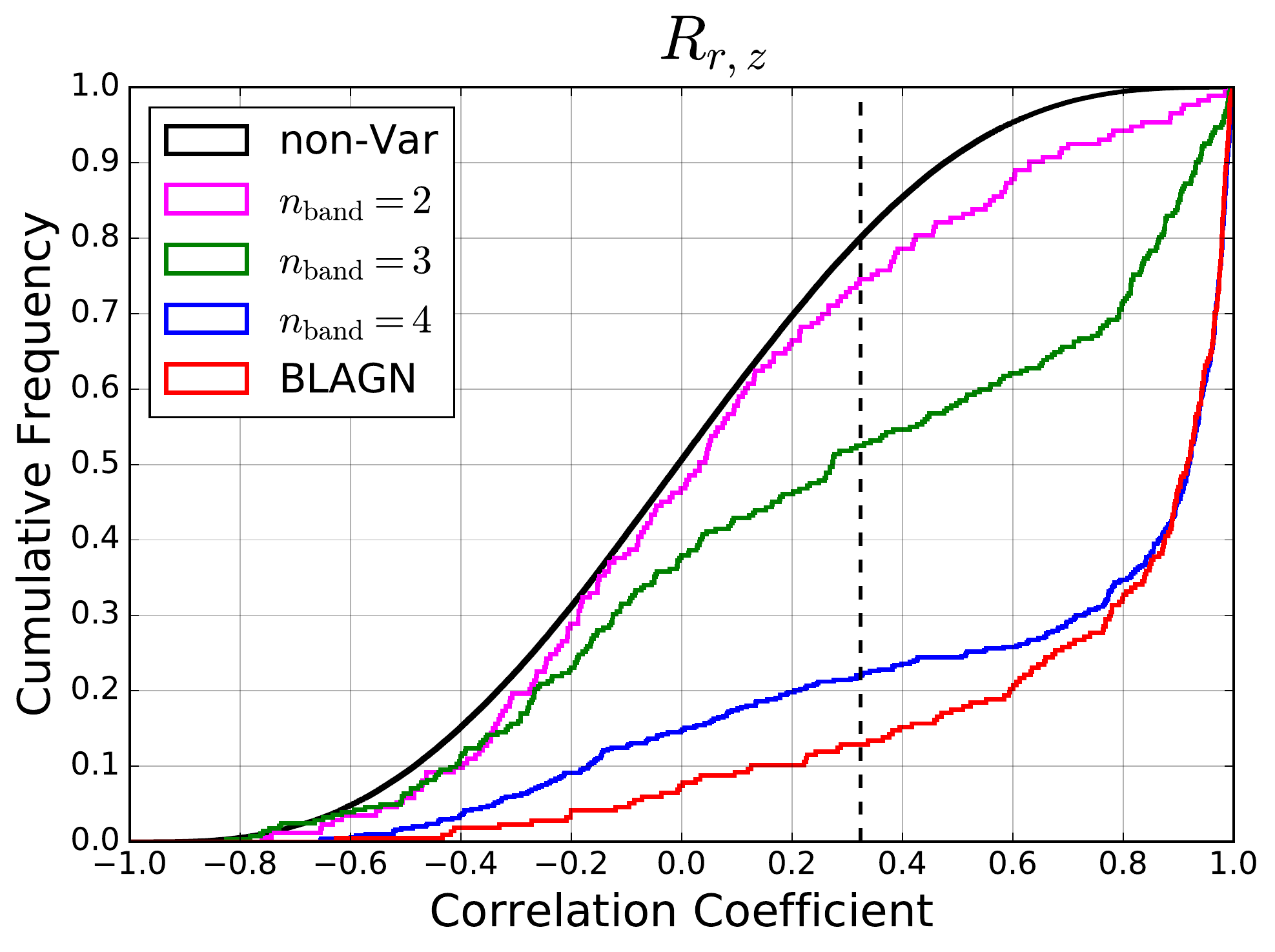}{0.33\textwidth}{(e) $r$ band \& $z$ band}
\fig{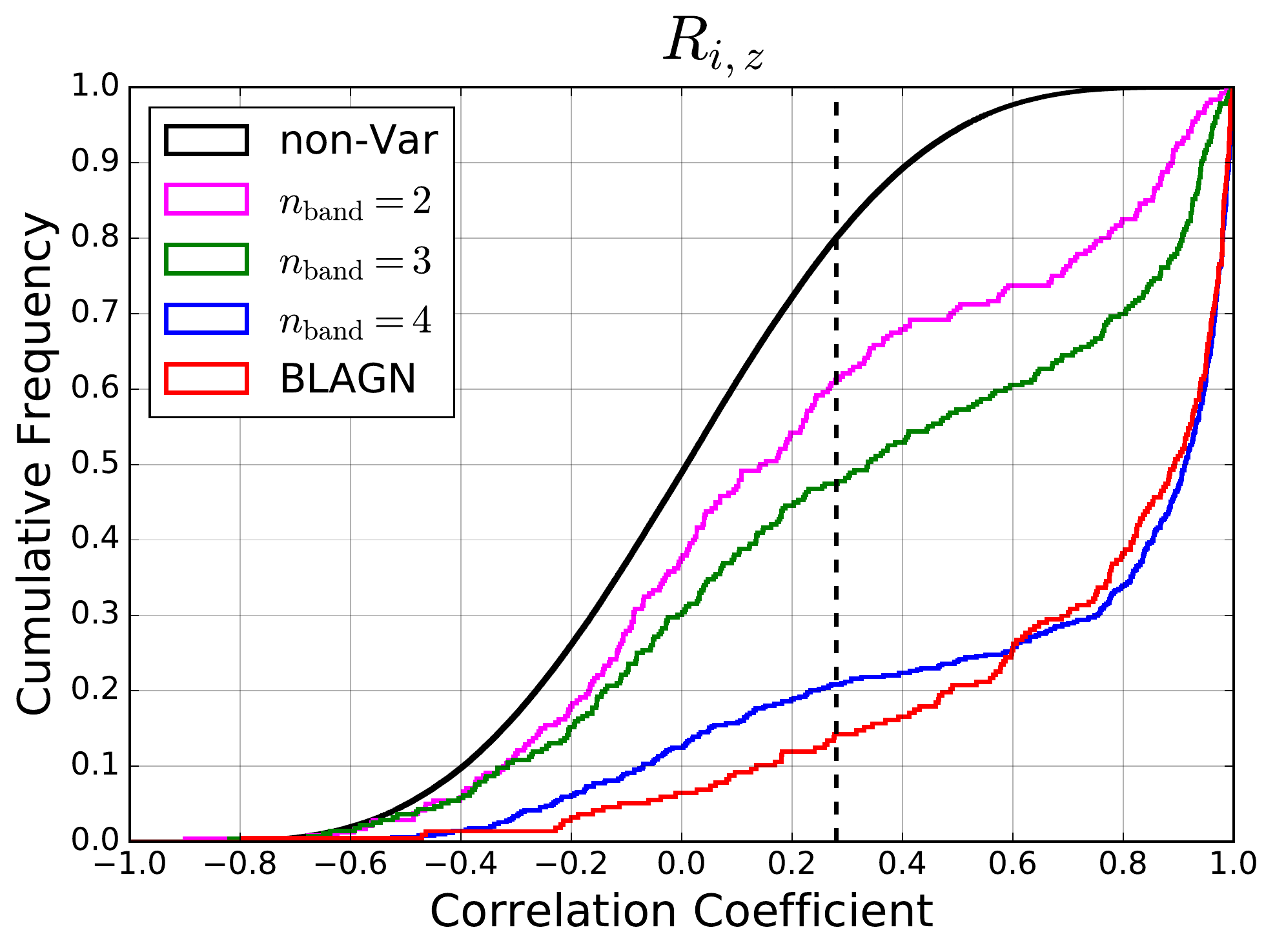}{0.33\textwidth}{(f) $i$ band \& $z$ band}
}
\caption{
Cumulative distributions of the cross-correlation coefficients between (a) the $g$ band and $r$ band, (b) the $g$ band and $i$ band, (c) the $g$ band and $z$ band, (d) the $r$ band and $i$ band, (e) the $r$ band and $z$ band, and (f) the $i$ band and $z$ band light curves, respectively.
The black line is the distribution for $non$-$Var$ objects, and the magenta, green, and blue lines are for variable objects with two, three, and four $variability$-flagged bands, respectively.
The red line is calculated from the sample of the known BLAGNs.
The vertical dashed lines indicate $R_{\mathrm{crit}}$ for each band pair, where $R > R_{\mathrm{crit}}$ is used as the criterion of the real cross-correlated variables.
\label{fig4}
}
\end{figure*}

To make the sample of variable AGNs more robust, we apply an additional selection criterion based on cross-correlation coefficients of the multiband light curves.

We calculate the cross-correlation coefficients of each pair of light curves in the $variability$-flagged filters for the variable candidates. 
It is noted that when $n_{\mathrm{band}}$ equals two, three, and four, we calculate cross-correlation coefficients of one, three, and six pairs of light curves, respectively.
Here, we consider that the two band photometries obtained within 5 days are quasi-simultaneous observations, which yields more than five data point pairs for calculating cross-correlations ($n_{\mathrm{pair}}$ in Table \ref{tbl3}).
Using these data point pairs, we calculate the cross-correlation coefficient $R_{\mathrm{A,B}}$ between band A and band B as
\begin{eqnarray}
R_{\mathrm{A,B}}=\frac{\sum_{i}^{n_{\mathrm{pair}}}\left(f_{\mathrm{A}, i}-\langle f_{\mathrm{A}}\rangle\right)\left(f_{ \mathrm{B}, i}-\langle f_{\mathrm{B}} \rangle\right)}{\sqrt{\sum_j^{n_{\mathrm{pair}}}\left(f_{\mathrm{A}, j}-\langle f_{\mathrm{A}}\rangle\right)^2}\sqrt{\sum_k^{n_{\mathrm{pair}}}\left(f_{\mathrm{B}, k}-\langle f_{\mathrm{B}}\rangle\right)^2} },
\label{eq5}
\end{eqnarray}
where $f$ is the observed flux and $\langle f\rangle$ is the mean flux over the data points.
In this calculation, aperture photometry was re-performed after the PSFs were matched to the largest one among all of the images over all epochs.

Figure~\ref{fig4} shows the cumulative distributions of the cross-correlation coefficients between two band pairs.
Each line represents the case for $non$-$Var$ sample (black), classified in both bands, two band $variability$-flagged objects (magenta), three band $variability$-flagged objects (green), and four band $variability$-flagged objects (blue).
The red line shows the case for the known BLAGNs, a large fraction of which show strong correlations in all of the band pairs.

\begin{deluxetable}{ccc}[t!]
\tablecaption{Cross-correlation coefficient criteria
\label{tbl3}
}
\tablehead{
\colhead{Band-pair} & \colhead{$n_{\mathrm{pair}}$} & \colhead{$R_{\mathrm{crit}}$} 
}
\startdata
($g$, $r$) & 5 & 0.513\\
($g$, $i$) & 6 & 0.457\\
($g$, $z$) & 6 & 0.440\\
($r$, $i$) & 8 & 0.365\\
($r$, $z$) & 9 & 0.324\\
($i$, $z$) & 12 & 0.280
\enddata
\end{deluxetable}

\begin{figure*}[t!]
\gridline{
\fig{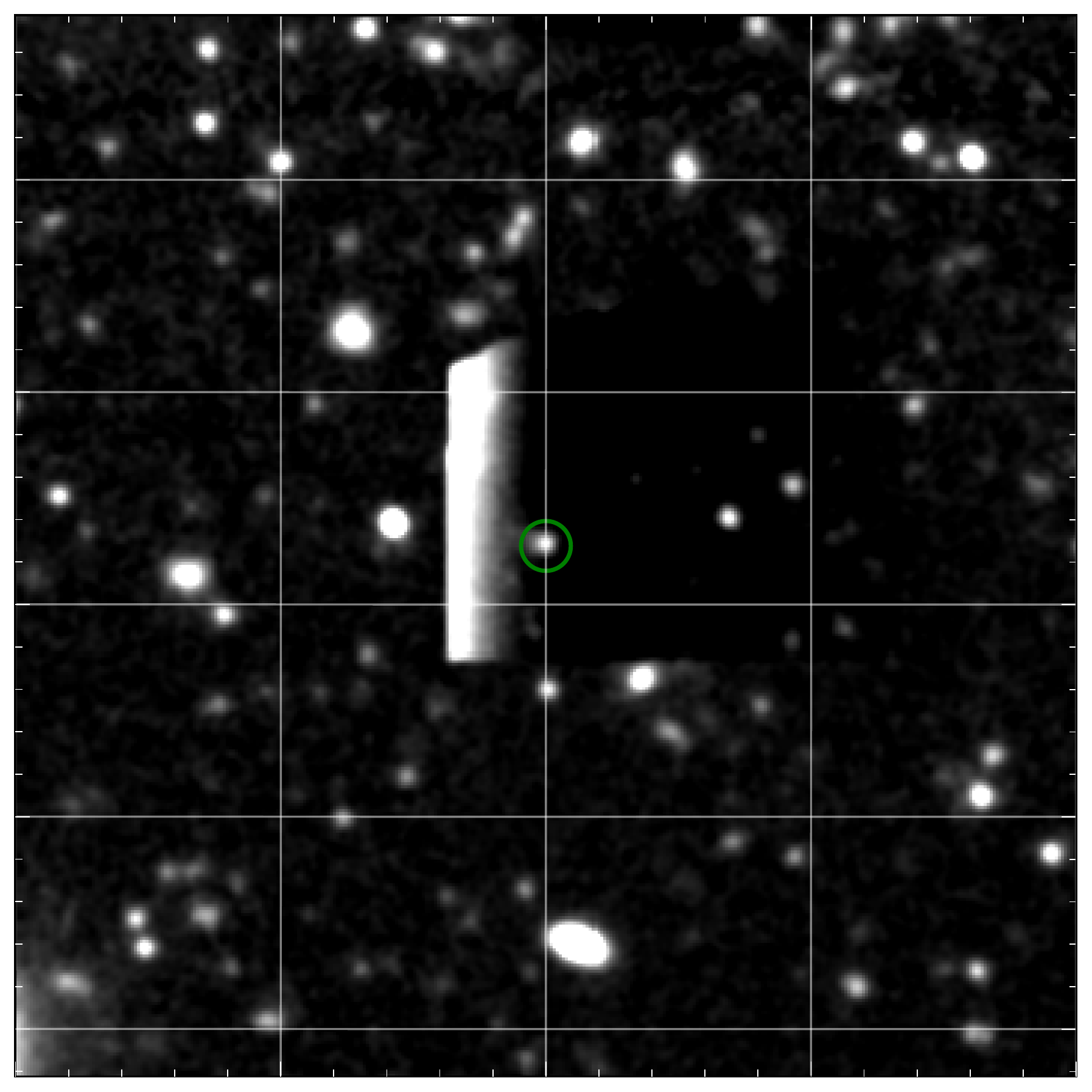}{0.235\textwidth}{bad pattern}
\fig{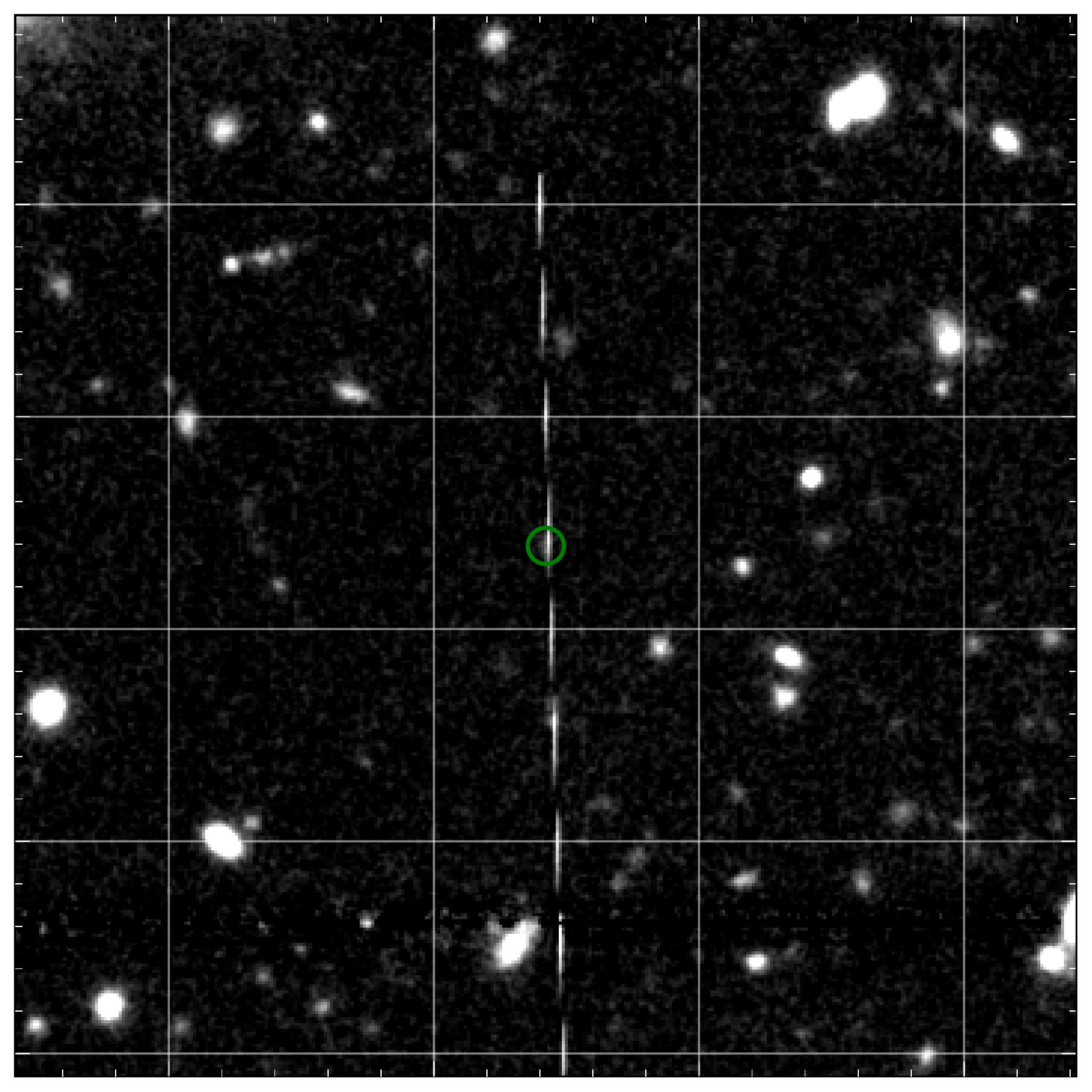}{0.235\textwidth}{satellite}
\fig{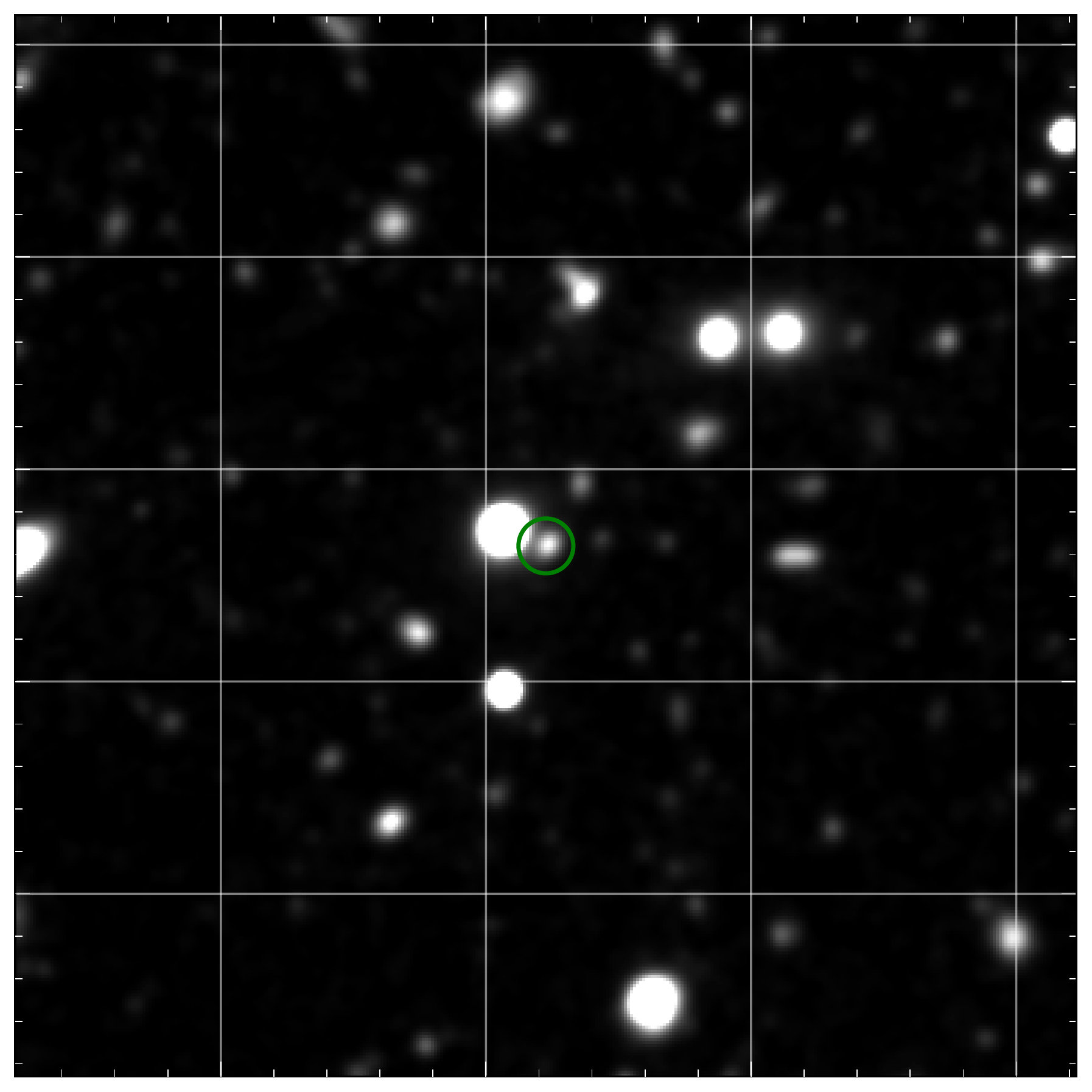}{0.235\textwidth}{neighbor object}
\fig{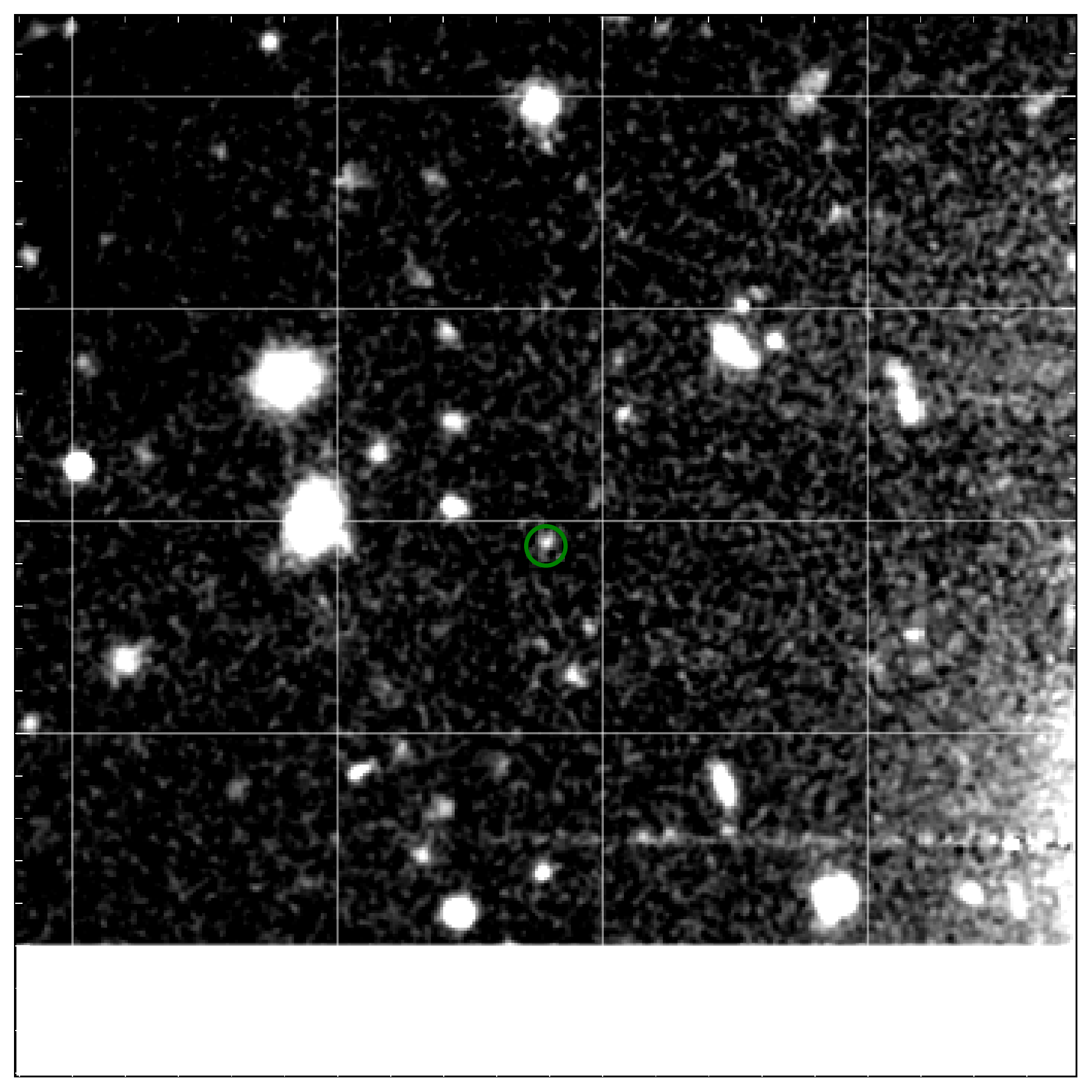}{0.235\textwidth}{bright object}
}
\caption{Postage stamp images of the examples of visually classified error-affected objects.
The image size is $1 \times 1$~arcmin$^{2}$.
The green circle plotted in the center of each image represents the aperture size used in photometry.
\label{fig5}
}
\end{figure*}

To select the variable objects that show strong correlation of multiband light curves, we set a criterion for each cross-correlation coefficient, in which only the top 20\% of the $non$-$Var$ sample shows the value.
Then, we set the flag of `$correlation$' in the band pair, whose correlation coefficient of the light curves is larger than the criteria.
These critical values ($R_{\mathrm{crit}}$) are shown by the black dashed lines in each panel of Figure~\ref{fig4} and listed in Table~\ref{tbl3}.
Approximately more than 75\% of the known BLAGNs satisfy this criterion in each cross-correlation coefficient as seen in Figure~\ref{fig4}.

We require for variable objects to satisfy this cross-correlation coefficient criterion in at least one band pair, namely, 
\begin{eqnarray}
n_{\mathrm{corr}} \geq 1,
\label{eq6}
\end{eqnarray}
where $n_{\mathrm{corr}}$ denotes the number of $correlation$-flagged band pairs.
After applying this criteria, we can recover $\sim82\%$ of all of the known BLAGNs and $\sim99\%$ of the BLAGNs that satisfy the criterion in Equation~(\ref{eq4}).
Finally, we obtain 1078 variable candidates in total (62\% of the objects that satisfy the criterion in Equation~(\ref{eq4})).


\begin{figure*}[t!]
\gridline{
\fig{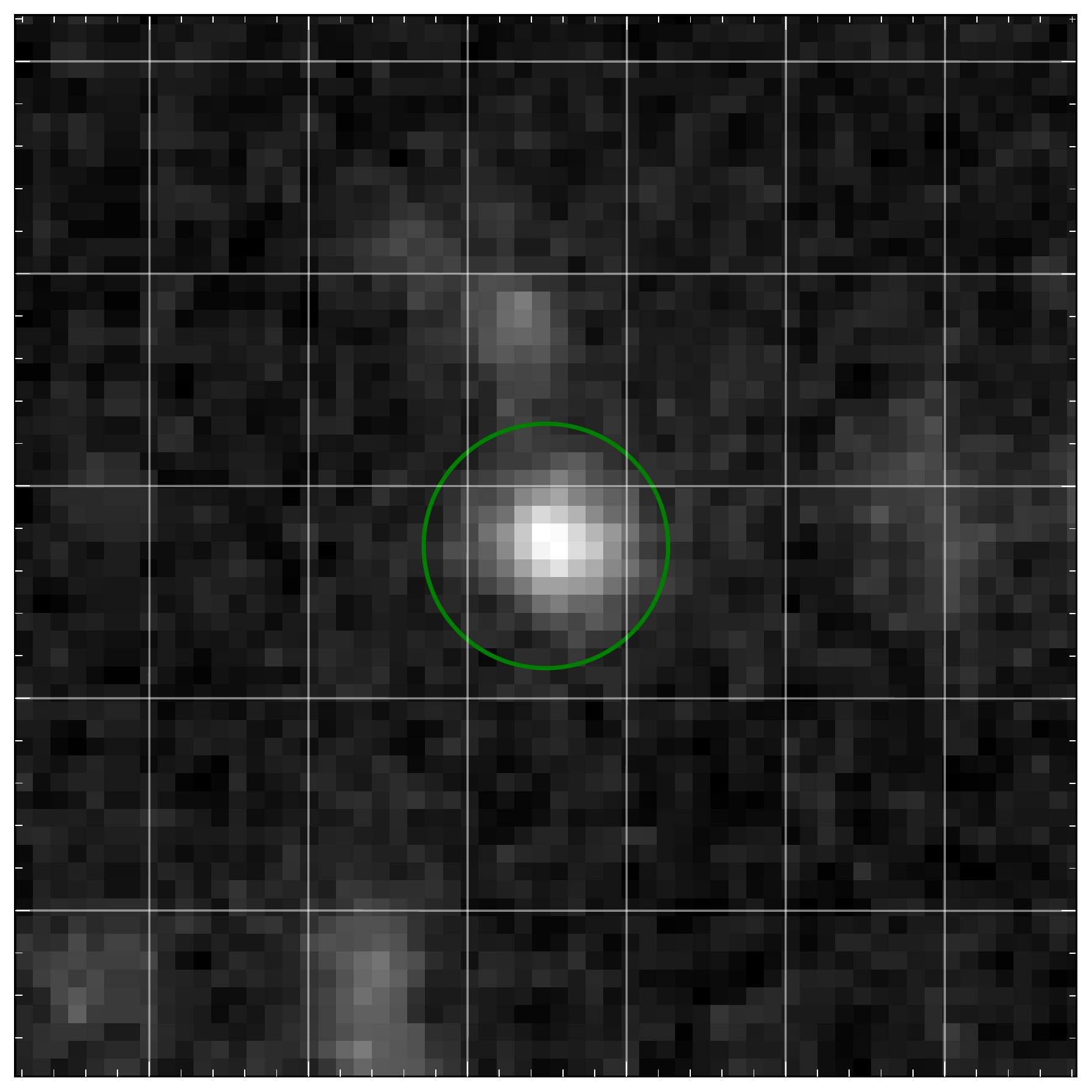}{0.23\textwidth}{$g$ band [2014-11-18]}
\fig{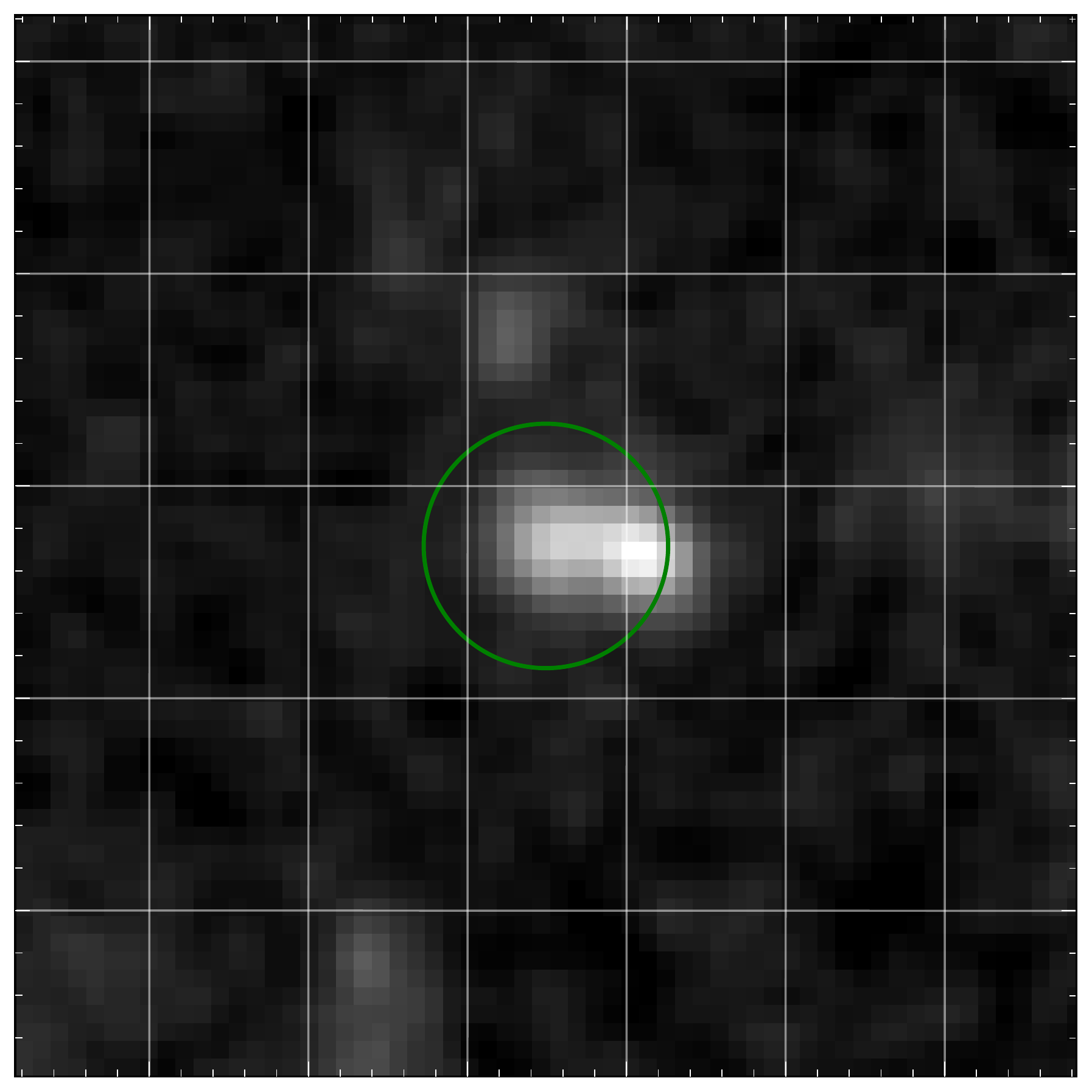}{0.23\textwidth}{$g$ band [2017-02-01]}
\fig{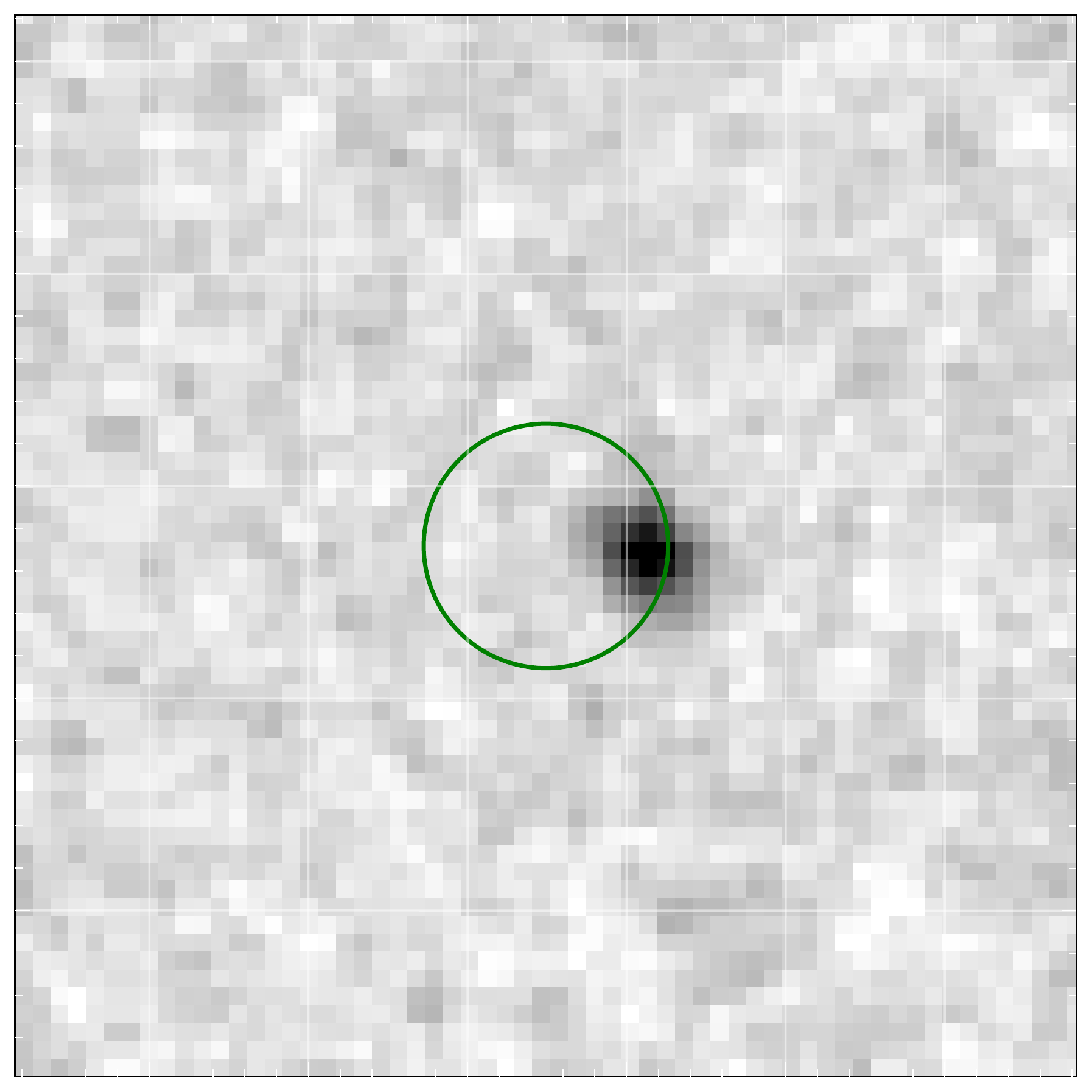}{0.23\textwidth}{subtracted}
\fig{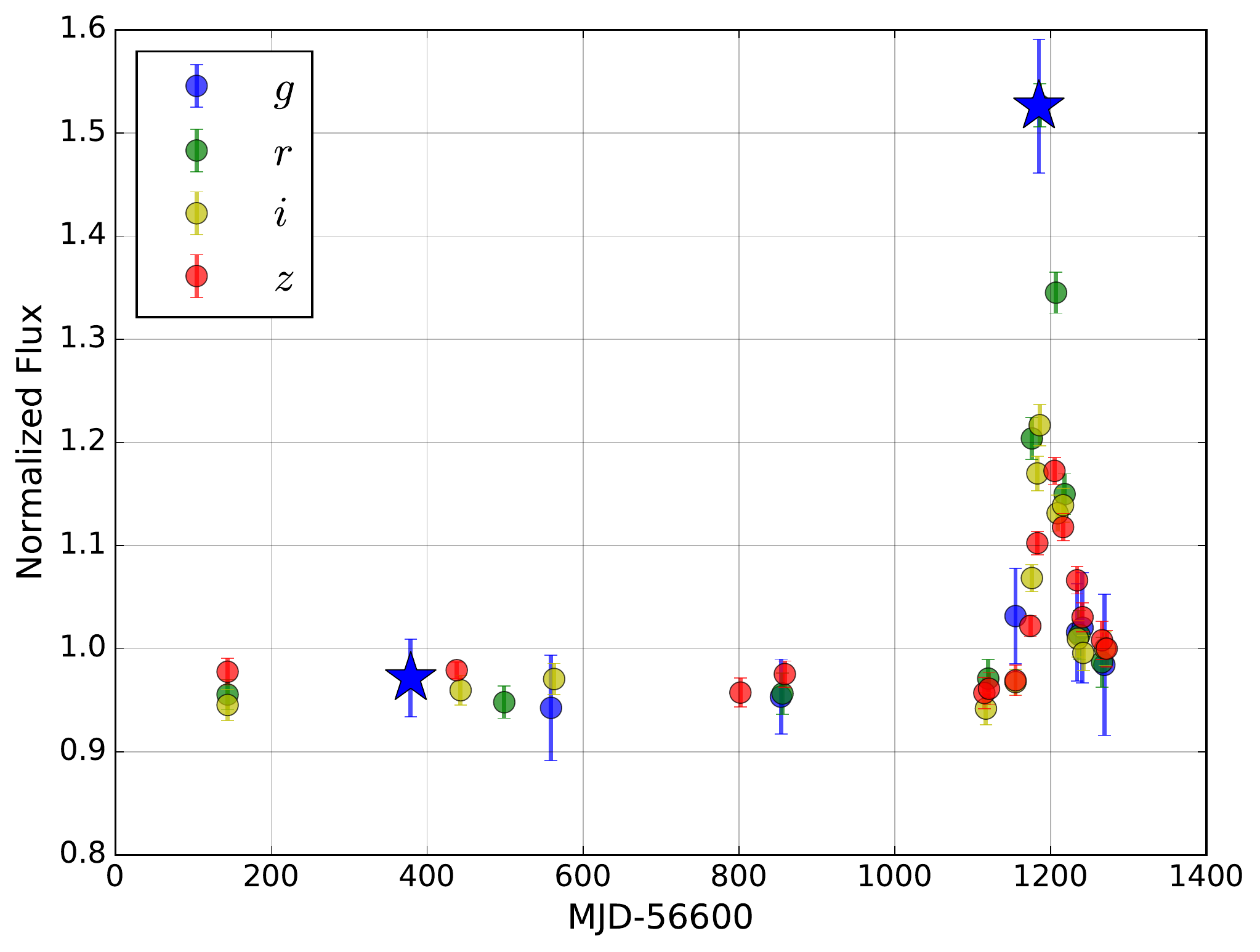}{0.31\textwidth}{light curve}
}
\gridline{
\fig{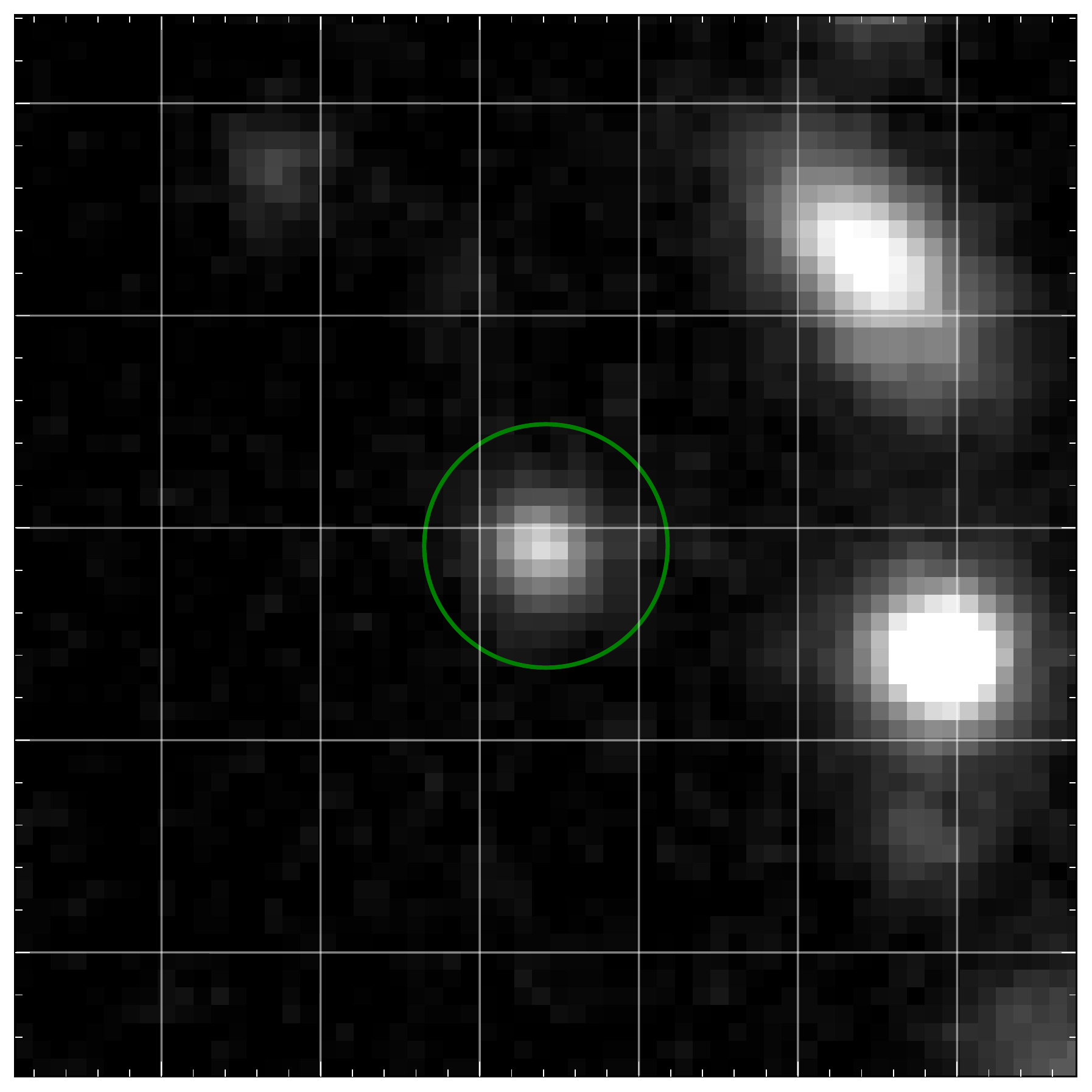}{0.23\textwidth}{$i$ band [2014-03-28]}
\fig{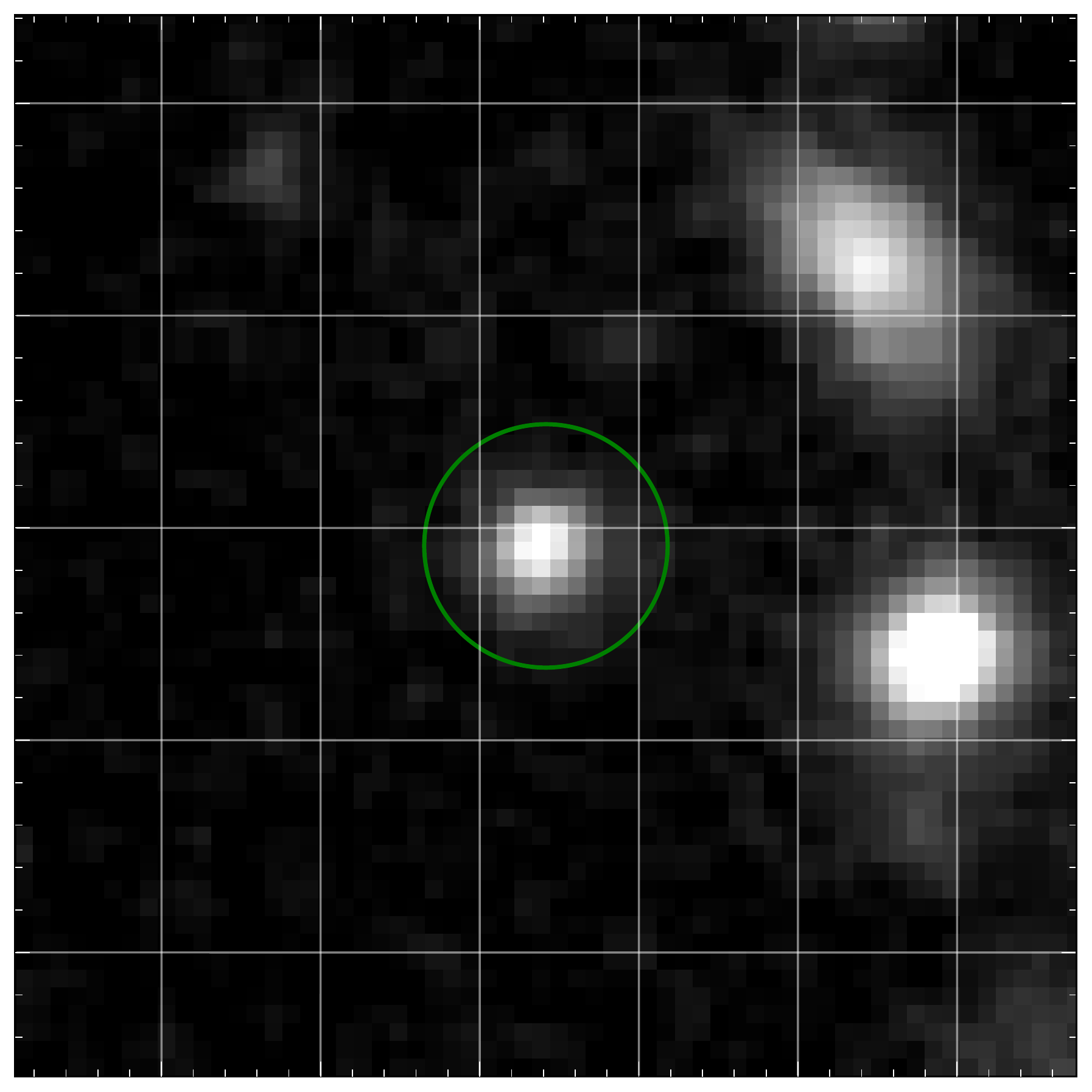}{0.23\textwidth}{$i$ band [2017-02-25]}
\fig{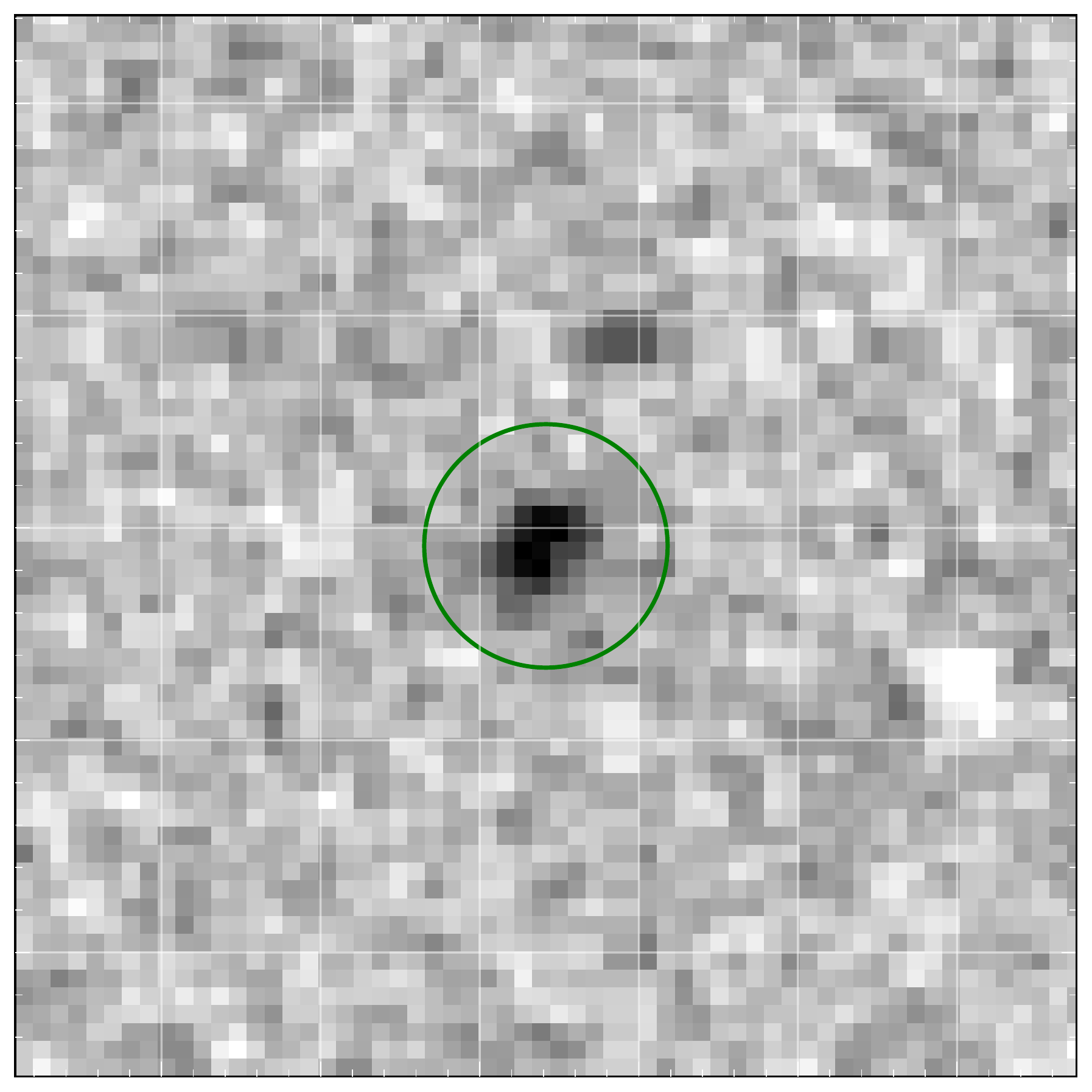}{0.23\textwidth}{subtracted}
\fig{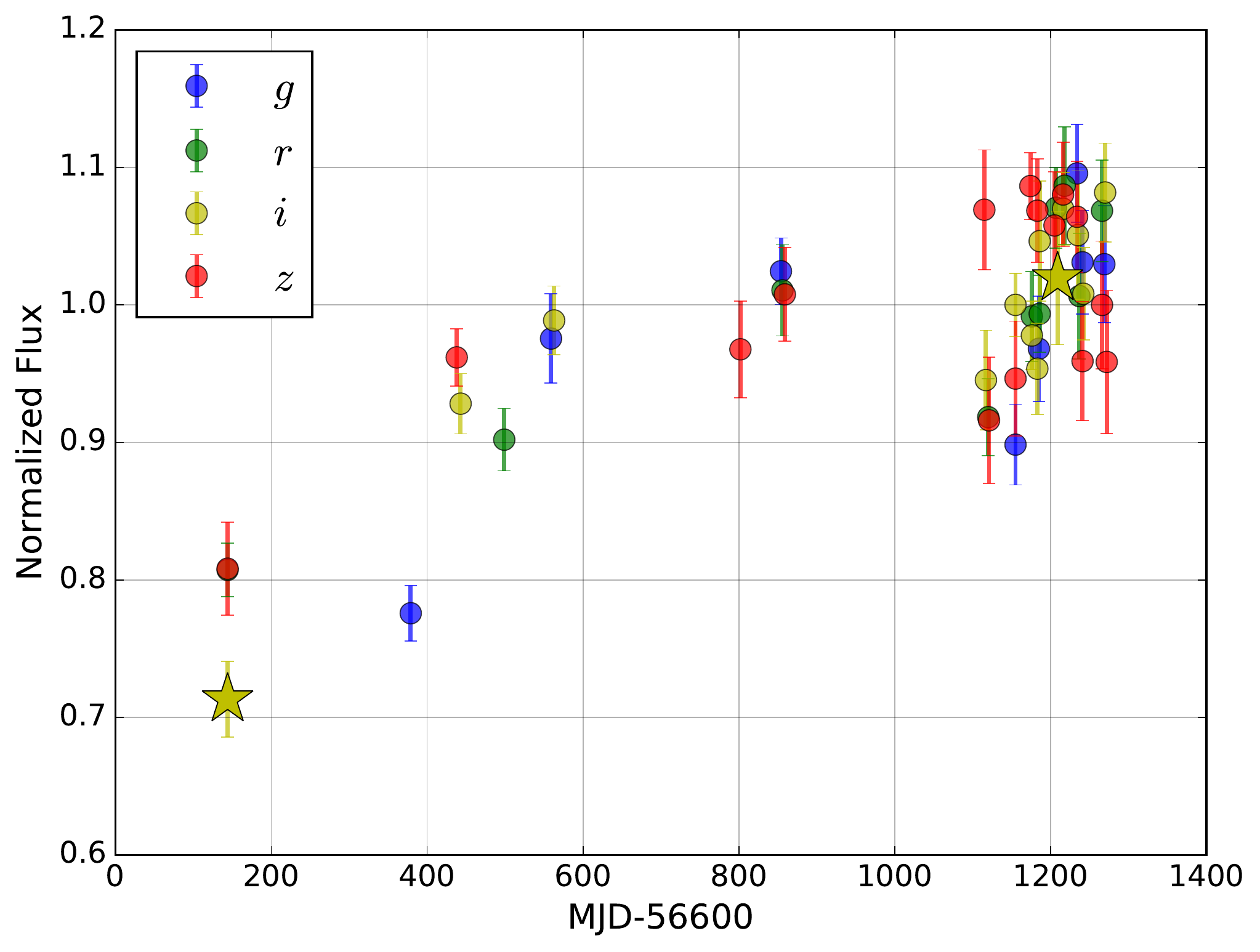}{0.31\textwidth}{light curve}
}
\gridline{
\fig{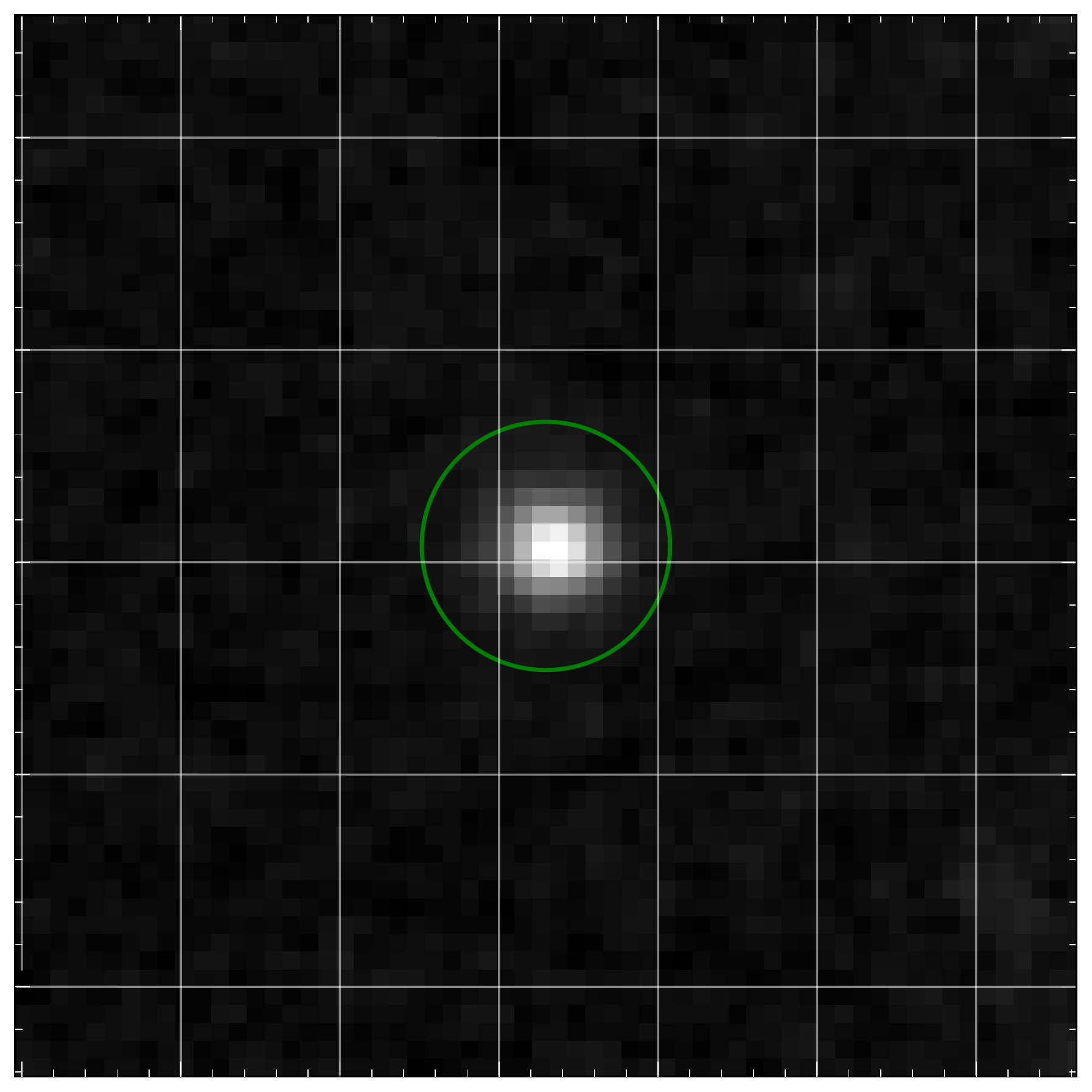}{0.23\textwidth}{$i$ band [2014-03-28]}
\fig{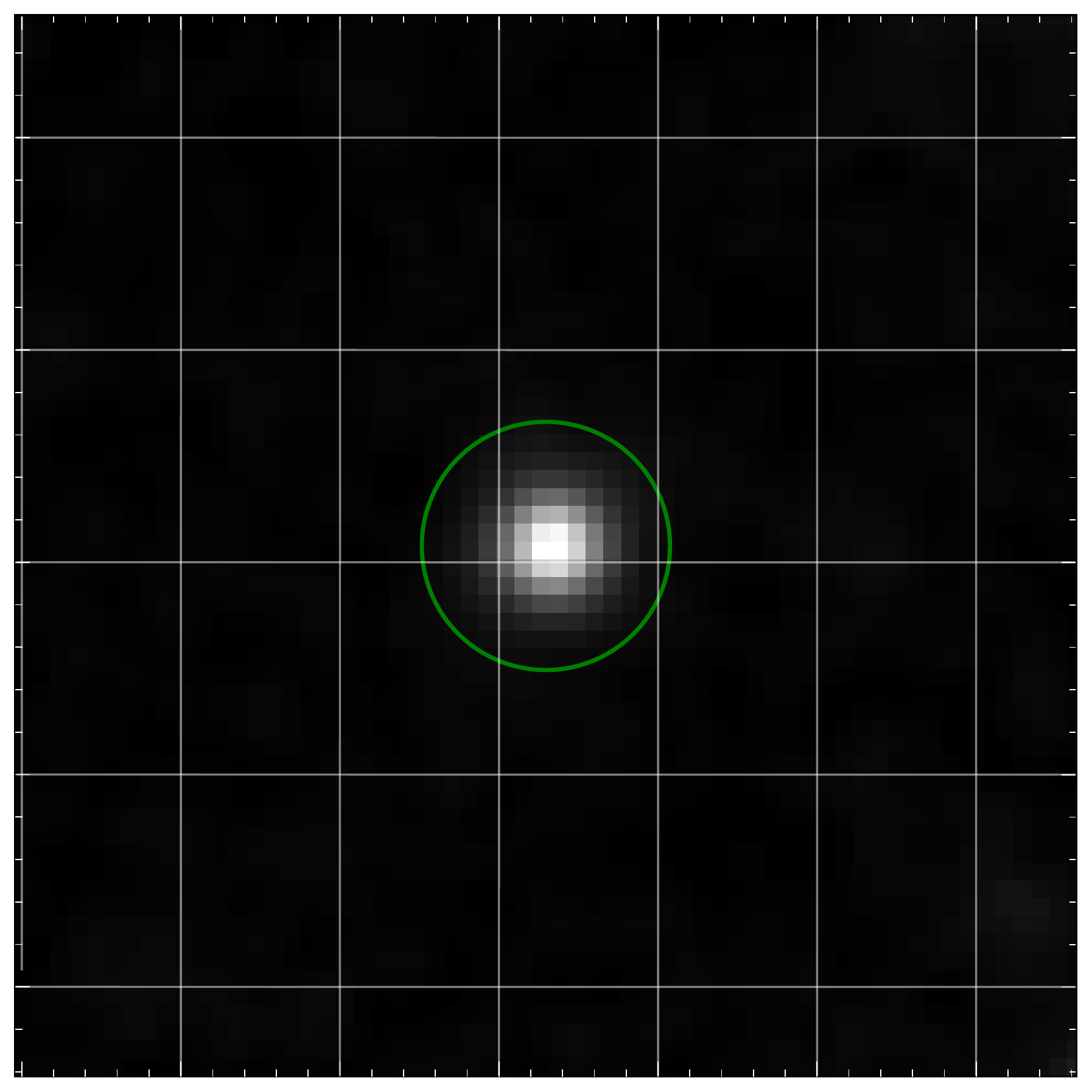}{0.23\textwidth}{$i$ band [2017-03-04]}
\fig{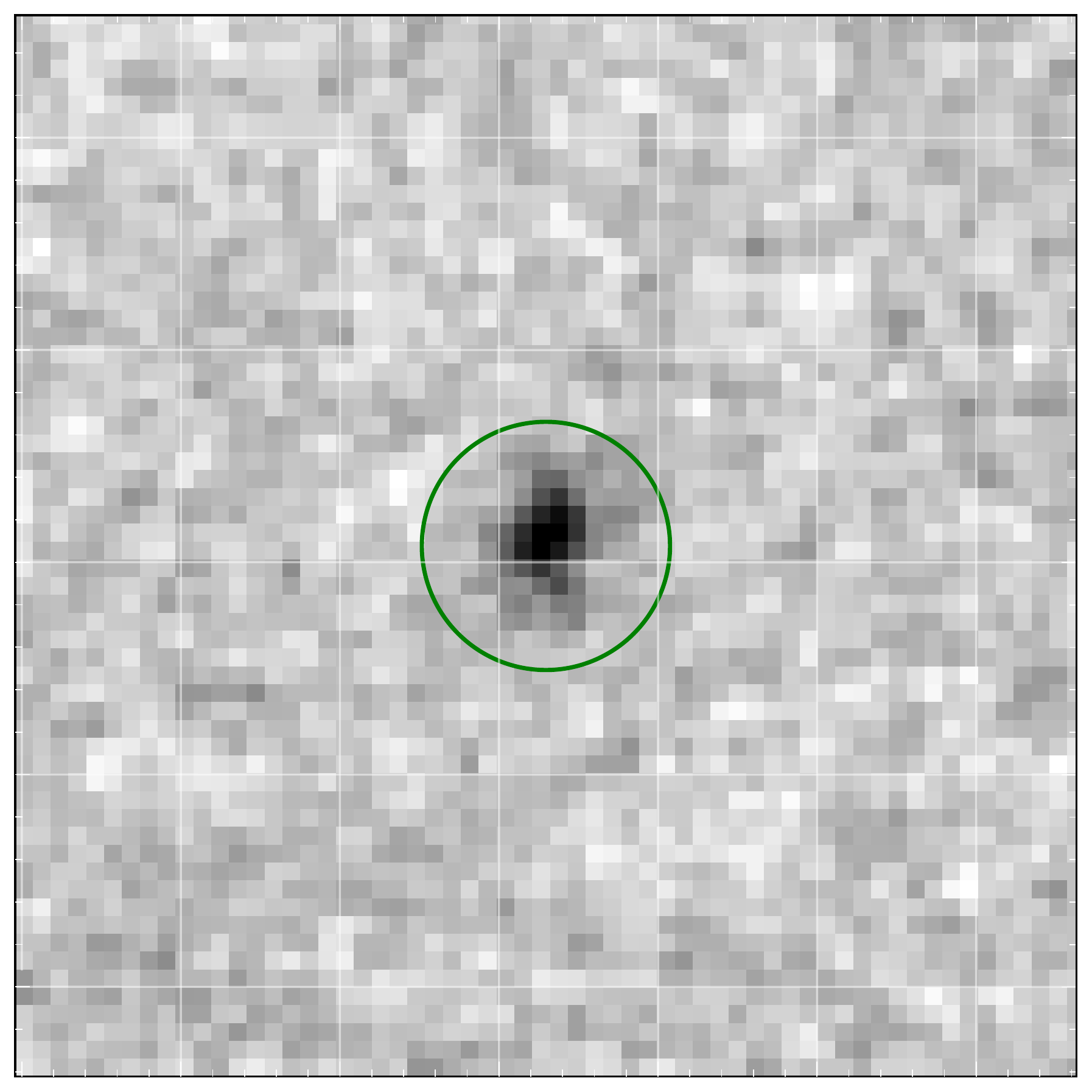}{0.23\textwidth}{subtracted}
\fig{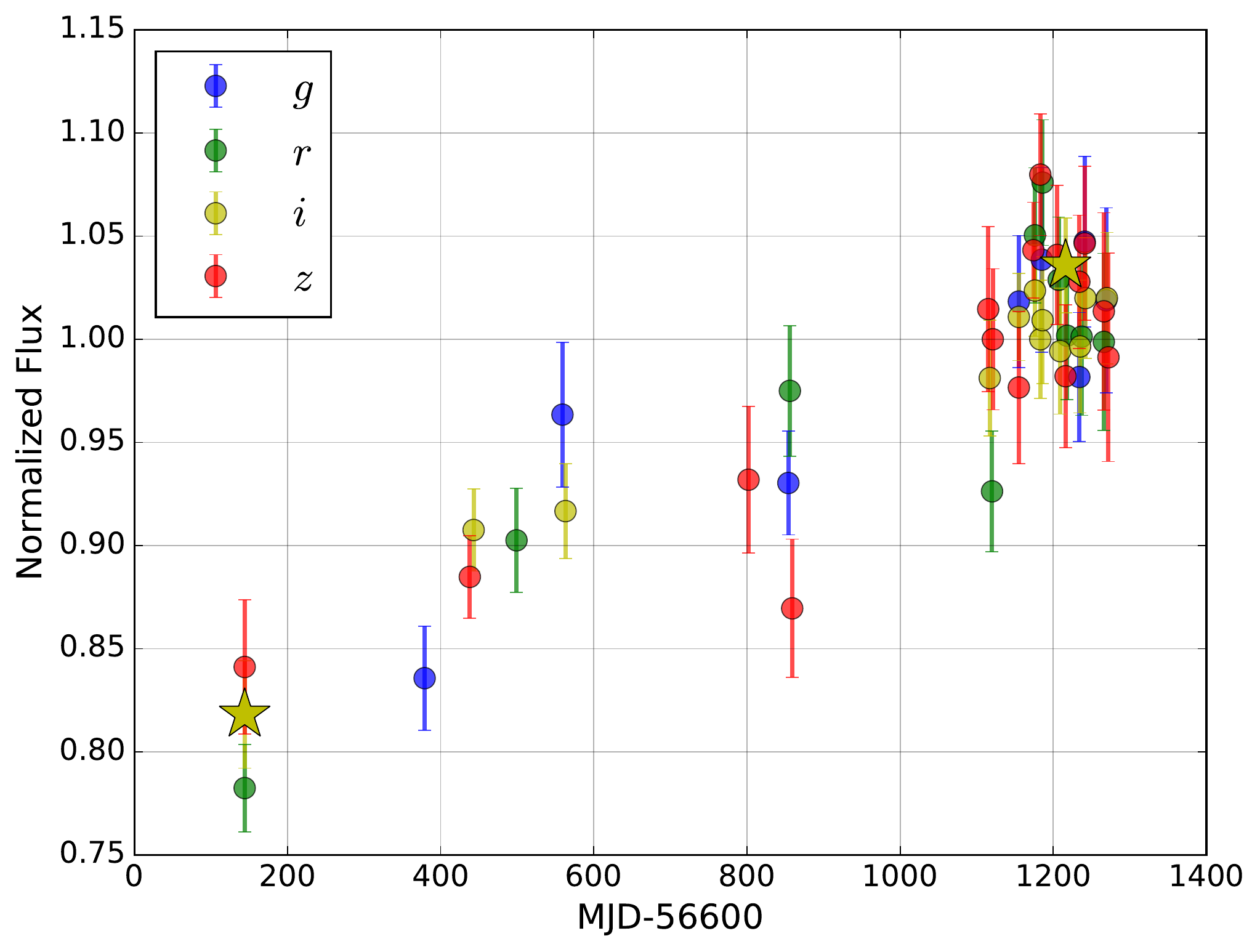}{0.31\textwidth}{light curve}
}
\caption{Postage stamp images (two arbitrary selected epochs and their difference are shown in the left three panels of each row) and light curves (right panels of each row) of the visually inspected variable objects.
The upper row shows an example of a supernova, the middle row shows an X-ray detected AGN, and the bottom row shows an X-ray undetected AGN.
The image size is $10 \times 10$~arcsec${}^{2}$.
The green circles plotted on the images represent the aperture size used in the photometry.
The star symbols in the light curves are the dates shown in the left two images.
\label{fig6}
}
\end{figure*}

\begin{deluxetable*}{ccccccccccccc}
\tablecaption{Catalog of the variable AGNs
\label{tbl4}
}
\tablehead{
\colhead{ID} & 
\colhead{R.A. J2000} & 
\colhead{Dec. J2000} &
\colhead{$i$-mag} &
\colhead{X-ray} &
\colhead{Flag-$g$} &
\colhead{Flag-$r$} &
\colhead{Flag-$i$} &
\colhead{Flag-$z$} &
\colhead{$n_{\mathrm{corr}}$} &
\colhead{Redshift} &
\colhead{Ref.} &
\colhead{ID-COSMOS2015}\\ 
(1) & (2) & (3) & (4) & (5) & (6) & (7) & (8) & (9) & (10) & (11) & (12) & (13)
}
\startdata
1 & 150.74386 & 2.20245 & 22.71 & 1 & 1 & 1 & 1 & 1 & 6 & 1.561 & 3 & 594392\\
2 & 150.73557 & 2.19957 & 20.36 & 1 & 1 & 1 & 1 & 1 & 6 & 3.499 & 1 & 592797\\
3 & 150.73353 & 2.15646 & 20.88 & 1 & 1 & 1 & 1 & 1 & 6 & 0.977 & 1 & 565402\\
4 & 150.79702 & 2.13888 & 21.01 & 1 & 1 & 1 & 1 & 1 & 6 & 0.573 & 1 & 552225\\
5 & 150.78259 & 2.19306 & 20.63 & 1 & 1 & 1 & 1 & 1 & 6 & 0.585 & 1 & 589540\\
$\vdots$ \\
491 & 150.03524 & 2.72781 & 21.04 & 0 & 1 & 0 & 0 & 1 & 1 & 0.509 & 1 & 940746
\enddata
\tablecomments{
Column (1): the identification number of each variable source.
Column (2) and (3): source coordinates (unit of degree) from the HSC catalog. 
Column (4): $i$ band cmodel magnitude from the HSC catalog. 
Column (5): X-ray detection flag (0: undetected, 1: detected). 
Column (6)$-$(9): flag of $variability$ for each band (0: unflagged, 1: flagged).  
Column (10): the number of $correlation$-flagged band pairs. 
Column (11): redshift.
Column (12): redshift reference (1: spectroscopic redshift from the HSC catalog, 2: spectroscopic redshift from the DEIMOS catalog, 3: photometric redshift from $z$\_$best$ in the Chandra catalog, 4: photometric redshift from $ZPDF$ in the COSMOS2015 catalog.
Column(13): the identification number listed in the COSMOS2015 catalog.
This table is published in its entirety in the machine-readable format.
A portion is shown here.
}
\end{deluxetable*}

\subsubsection{Visual Inspection} \label{sec2_3_3}

By applying the criteria in Equations (\ref{eq4}) and (\ref{eq6}), we obtain the sample of the 1078 variable object candidates.
These are robust candidates for variable AGNs, but some possible false-positive variables still remain. 
As we are interested in AGNs, supernovae should also be excluded from the sample.
For this purpose, we conduct a visual inspection of the images and light curves for the 1078 variable object candidates.
We visually identify
(i) objects that are clearly affected by satellites, bad pixels, adjacent objects, or bright stars,
(ii) supernova candidates that show supernova-like light curves or show off-nuclear transients, and
(iii) spurious objects that have largely extended light profiles.
In our visual inspection, 196 objects ($\sim 18\%$) are identified as case (i).
Examples of these case (i) objects are shown in Figure~\ref{fig5}.
We also identified 186 objects as case (ii).
An example of the supernova candidate is shown in the top panels of Figure~\ref{fig6}, which clearly show an off-nuclear transient and supernova-like light curves.
We then identify 134 objects as case (iii).

Finally, we identify 71 variable stars by using the star-flags (see Section~\ref{sec2_3_1}).
By removing the case (i), (ii), and (iii) objects as well as the variable stars, we obtain 491 robust variable AGN candidates, of which 441 objects ($\sim90\%$) are detected in the Chandra X-ray observations.
Examples of the X-ray detected and X-ray undetected variable AGNs are shown in Figure \ref{fig6} (middle panels: X-ray detected, bottom panels: X-ray undetected).

\subsubsection{Comparisons with the Previous Variability-selected AGNs in the COSMOS Field} \label{sec2_3_4}

We compare our sample of variability-selected AGNs with those found in the previous variability-based AGN searches in the literature.
In the COSMOS field, variability surveys were conducted by using data from the PanSTARRS1 (PS1) survey \citep{sim15} and the VLT Survey Telescope (VST) survey \citep{dec19}.

\citet{sim15} carried out optical variability analysis for X-ray-detected QSOs that have a secure optical counterpart and have pointlike light profiles, using the PS1 data in the five broad bands ($g_{\mathrm{P1}}$, $r_{\mathrm{P1}}$, $i_{\mathrm{P1}}$, $z_{\mathrm{P1}}$, $y_{\mathrm{P1}}$) covering a period of about four years from November 2009 to March 2014, obtained as a part of the $3\pi$ survey and the Medium Deep Field (MDF04) survey \citep{cha16}.
The depth ($5\sigma$ median limiting magnitude) of each survey is 22.1 ($g_{\mathrm{P1}}$), 21.9 ($r_{\mathrm{P1}}$), 21.6 ($i_{\mathrm{P1}}$), and 19.9 ($y_{\mathrm{P1}}$) for individual $3\pi$ survey data and 22.5 ($g_{\mathrm{P1}}$), 22.3 ($r_{\mathrm{P1}}$), 22.0 ($i_{\mathrm{P1}}$), and 21.3 ($y_{\mathrm{P1}}$) for individual MDF04 survey data, respectively.
90 ($g_{\mathrm{P1}}$), 54 ($r_{\mathrm{P1}}$), 14 ($i_{\mathrm{P1}}$), 37 ($z_{\mathrm{P1}}$), and 8 ($y_{\mathrm{P1}}$) sources among the 285 X-ray detected objects in the 3$\pi$ survey data and 184 ($g_{\mathrm{P1}}$), 181 ($r_{\mathrm{P1}}$), 162 ($i_{\mathrm{P1}}$), 131 ($z_{\mathrm{P1}}$), and 74 ($y_{\mathrm{P1}}$) sources among the 331 X-ray detected objects in the MDF04 survey data are identified as variable AGNs.

\citet{dec19} carried out an $r$ band variability-based AGN search using the data from the VST survey (the $5\sigma$ depth of single visits are $r\lesssim24.6$ mag) from late 2011 to early 2015 with 54 visits.
They find 299 optically variable AGN candidates ($1.3\%$ of main sample) among which 232 sources are high-confidence candidates with $r\leq23.5$ mag.

Inside of our survey field (Figure~\ref{fig1}), there are 116 PS1 variable AGNs and 235 VST variable AGNs.
We cross-match these objects with our variable AGNs and find that almost all PS1 variable AGNs (115/116) and $83\%$ of the VST sample (194/235) are matched with our variable AGNs.
If we confine to the high-confidence sample in the VST sample (190 out of the 235 VST variable AGNs; see \citealt{dec19}), $90$\% (173/190) of them are matched to our variability-selected AGNs.
These results suggest that our variable AGN sample recovers more than $90$\% of the previous robust variable AGNs at $r\lesssim23.5$.

The final catalog of our robust variability-selected AGNs are listed in Table~\ref{tbl4}, and in the following sections, we focus on these objects.

\section{Properties of the Variable AGNs} \label{sec3}
\subsection{Basic Information} \label{sec3_1}

\begin{figure}[t!]
\plotone{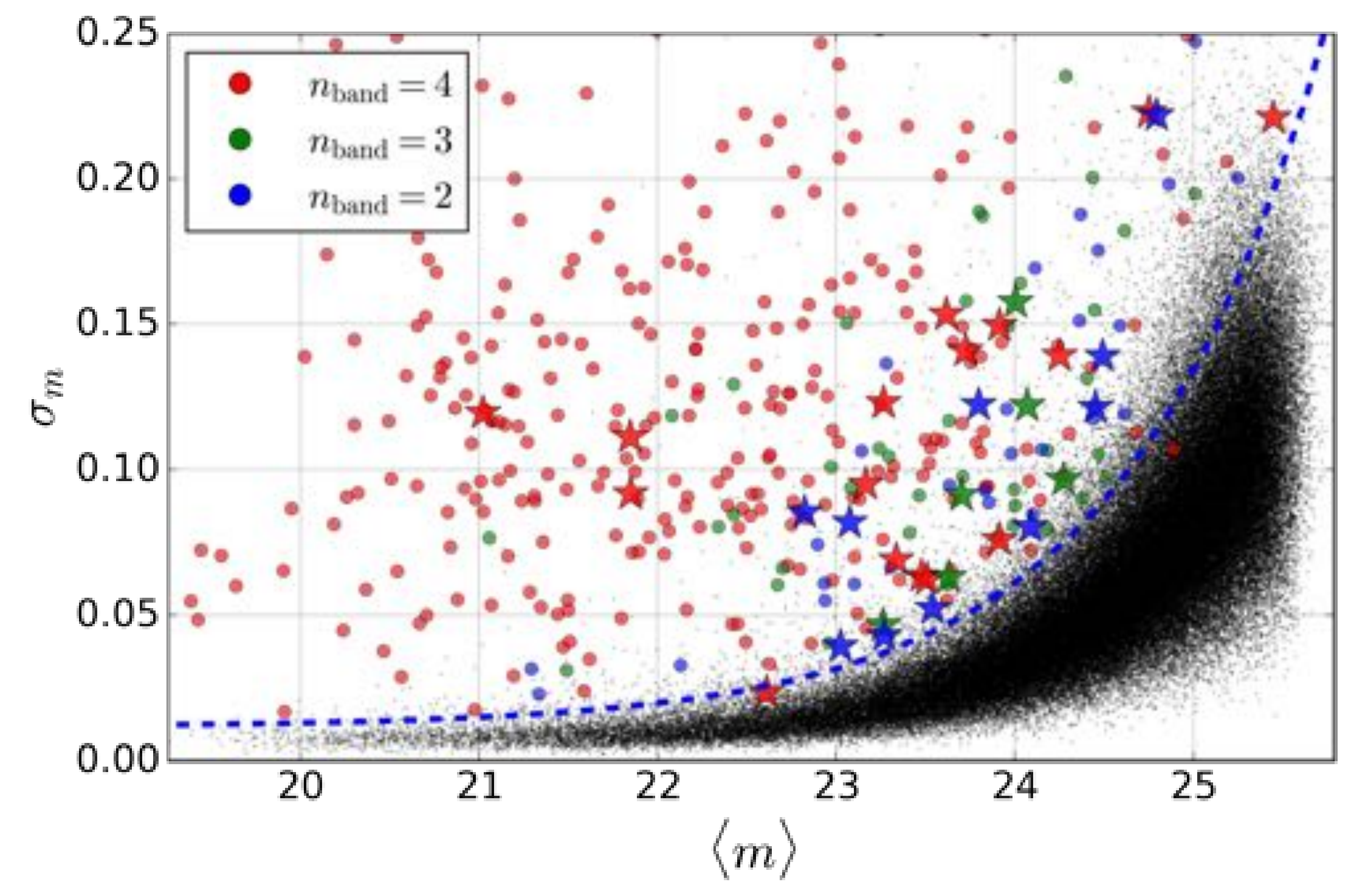}
\caption{
Significance of the variability $\sigma_{m}$ for the $g$ band light curve.
The black points are the parent sample, and the colored points are the variability AGNs with four (red), three (green), two (blue) $variability$-flagged bands.
The blue dashed line is the 95th percentile of the distribution of $\sigma_m$ as a function of mean magnitude $\langle m \rangle$.
The round symbols are the X-det sample, and the star symbols are the X-undet sample.
These objects are plotted for only the objects that are significantly detected ($S/N\geq5$) in all of the epochs. 
\label{fig7}
}
\end{figure}

\begin{figure}[htbp!]
\plotone{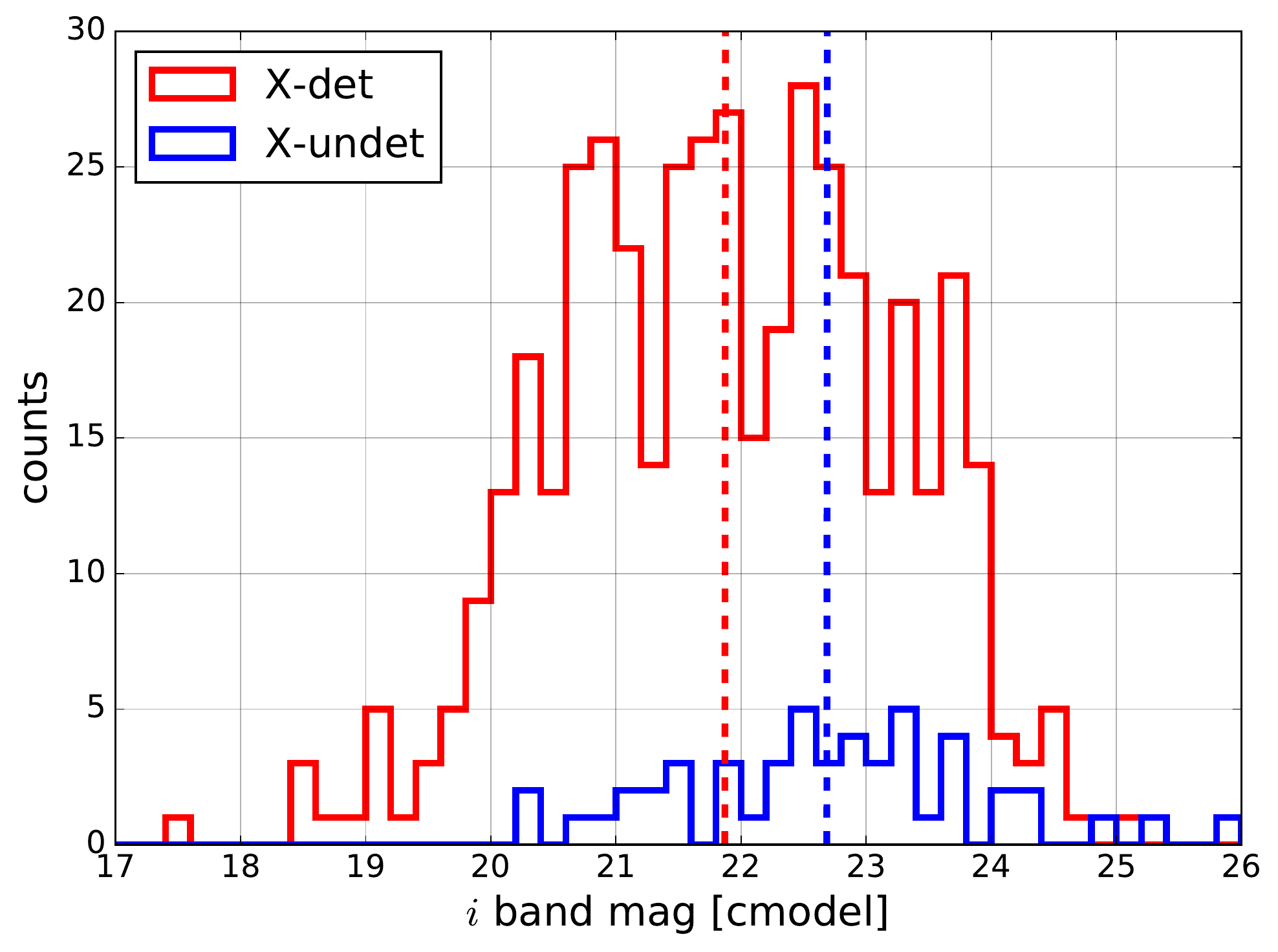}
\plotone{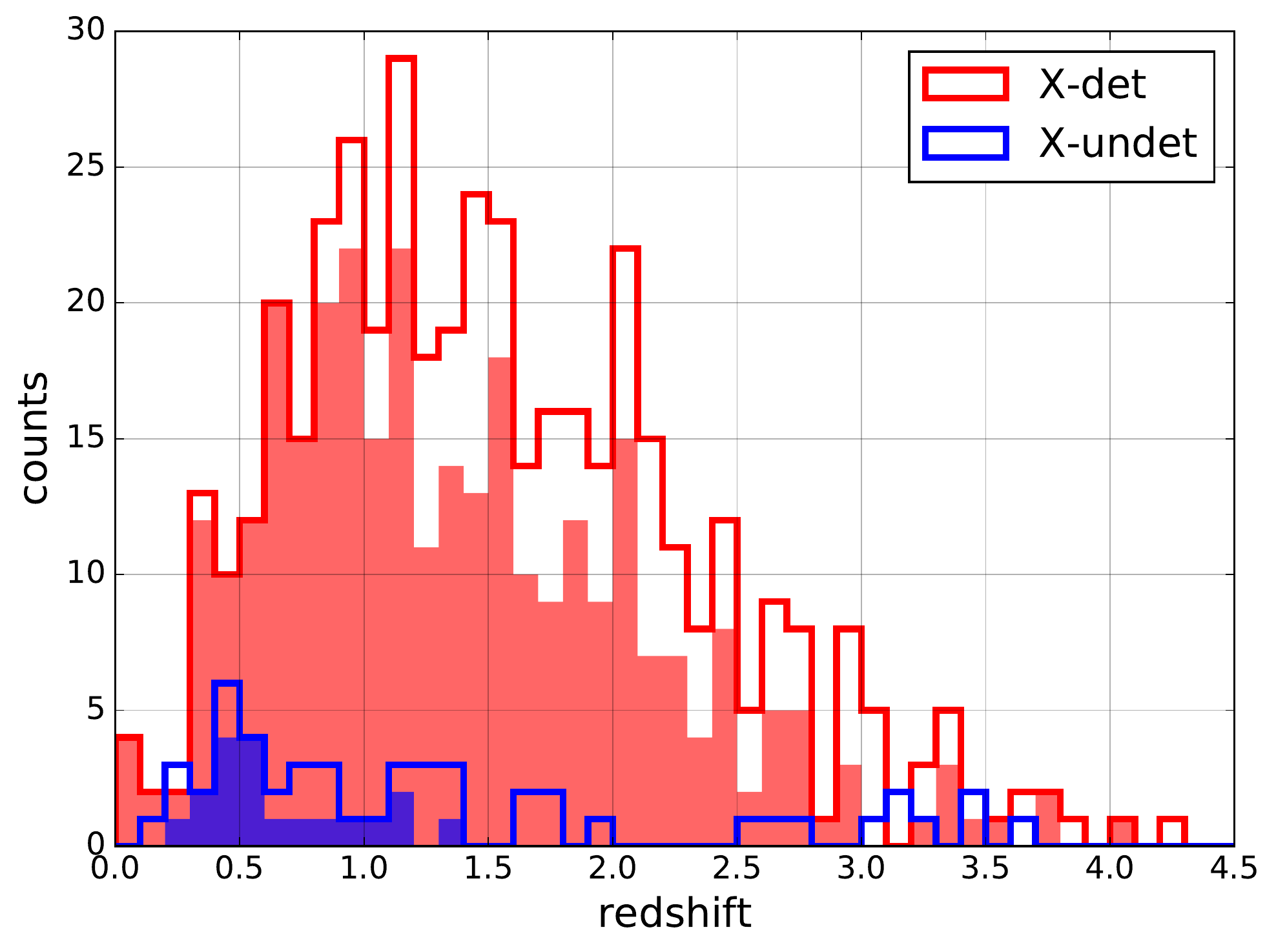}
\caption{
The histograms of the $i$ band magnitude (top panel) and redshift (bottom panel) for our variable AGN sample. 
The red (blue) histogram shows the distribution of the objects detected (undetected) in X-ray.
The vertical dashed lines in the top panel show the mean magnitudes of the X-det and X-undet samples.
The filled histograms in the bottom panel show the distributions of the objects with the spectroscopic redshifts.
\label{fig8}
}
\end{figure}
\begin{figure}[htbp!]
\plotone{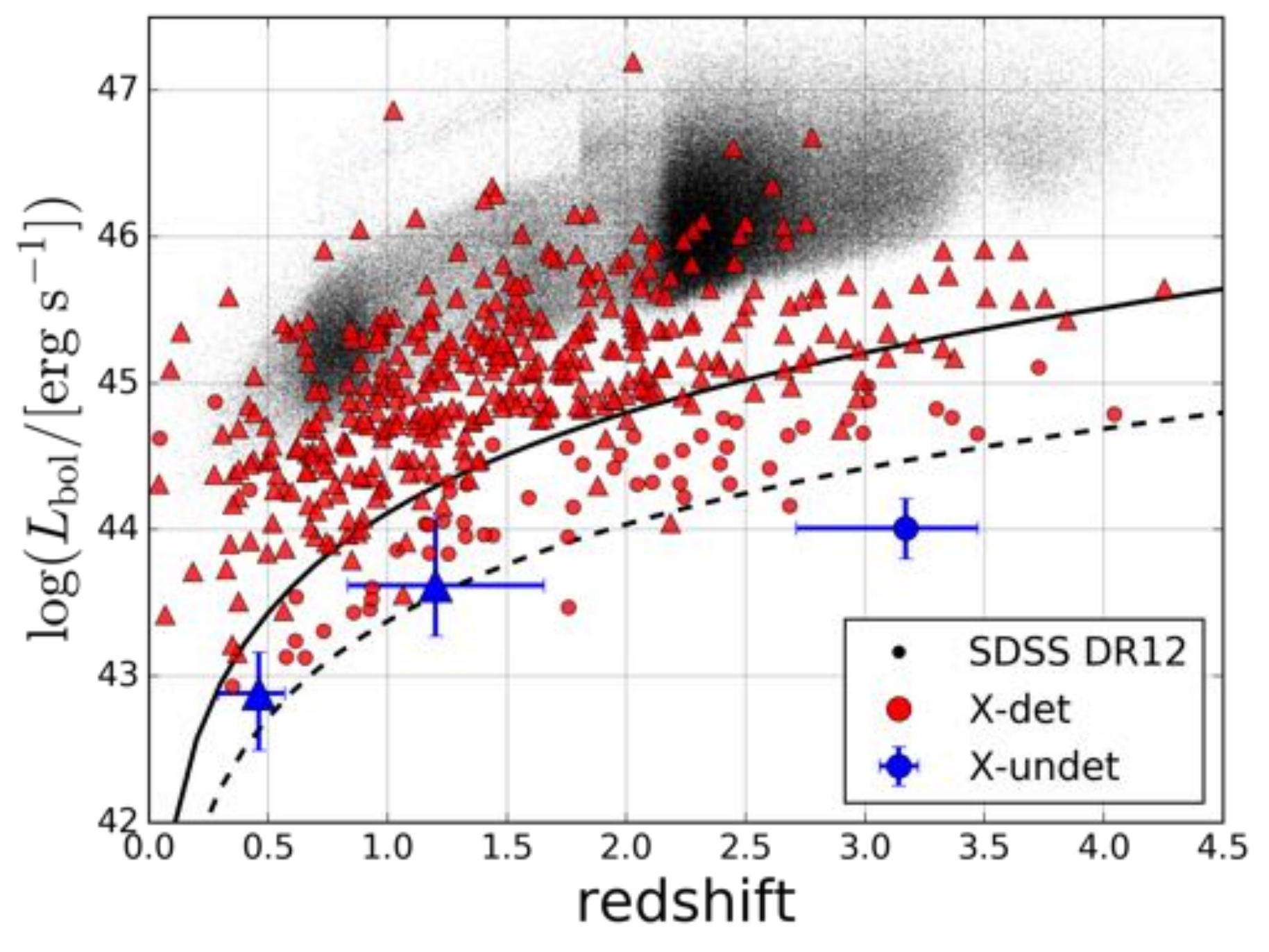}
\caption{
AGN bolometric luminosity as a function of redshift.
The red points are X-det objects in our variable AGN sample.
The blue points with error bars are X-undet objects in our variable AGN sample, which are calculated from the stacking analysis described in Section~\ref{sec3_2}.
The luminosities are calculated from the hard band (triangle symbols) or the soft band flux (circle symbols). 
The solid (dashed) line is the Chandra X-ray flux limit of hard (soft) band \citep[20\% completeness; ][]{civ16}.
The black points are a data set of the SDSS Quasar DR12 taken from \citet{koz17}.
\label{fig9}
}
\end{figure}

As mentioned in Section~\ref{sec2_3_3}, we obtained 491 variable AGN candidates, 441 ($\sim 90$\%) of which are detected in the X-ray (hereafter `X-det' sample) and the other 50 ($\sim 10$\%) are X-ray undetected (`X-undet' sample).
Figure~\ref{fig7} shows the standard deviation $\sigma_{m}$ of the $g$ band light curves of the 491 variable AGN candidates and the 271475 mostly non-variable objects in the parent sample (Section~\ref{sec2_2_1}), where $\sigma_{m}$ for each object is defined as
\begin{equation}
\sigma_{m}^2 =\frac{1}{n_{\mathrm{epoch}}} \sum^{n_{\mathrm{epoch}}}_{i} \left( m_{i} - \langle m\rangle \right)^2,
\end{equation}
where $m_{i}$ is the magnitude at the $i$th epoch, $\langle m\rangle$ is the mean magnitude over the light curve, and $n_{\mathrm{epoch}}$ is the number of epochs in which a target is not flagged as $faint$ nor $neighbor$ described in Section~\ref{sec2_2_3}.
As shown in Figure~\ref{fig7}, almost all of our variable AGNs (X-det and X-undet samples) show more than the $95$th percentile of the distribution of $\sigma_{m}$.

The $i$ band magnitude histogram of our variable AGN sample is shown in the top panel of Figure~\ref{fig8}.
The HSC survey depth enables us to identify a robust sample of variable objects down to $i \sim 25$ mag, which is more than one magnitude deeper than the previous time-domain surveys, such as the PS1 and VST surveys (Section~\ref{sec2_3_4}).
The mean values (standard deviations) of the $i$ band magnitudes for the X-det and X-undet samples are 21.89 (1.30) and 22.70 (1.21), respectively.
A Kolmogorov$-$Smirnov (KS) test rejects the null hypothesis that the $i$ band magnitude distribution of the X-undet sample is the same as that of the X-det sample (the $p$-value is 0.13\%).
This suggests that the X-undet sample is significantly fainter in the optical than the X-det sample.

We also plot the redshift distribution of our variable AGNs in the bottom panel of Figure~\ref{fig8}.
The redshift information comes from the spectroscopic redshift ($z_{\mathrm{spec}}$) in the HSC catalog, including zCOSMOS DR3 \citep{lil09}, PRIMUS DR1 \citep{coi11, coi13}, VVDS \citep{lef13}, SDSS DR12 \citep{ala15}, FMOS-COSMOS \citep{sil15}, 3D-HST \citep{mom16}, and the DEIMOS 10K Spectroscopic Survey Catalog \citep[DEIMOS catalog;][]{has18}.
If there is no spectroscopic information, for the X-det objects, we use the $z$\_$best$ values in the Chandra catalog, which are photometric redshifts obtained by the SED fitting with galaxy and AGN hybrid SED templates.
The typical uncertainty of these photometric redshifts is ${\displaystyle{ \sigma_{\Delta z/(1+z_{\mathrm{spec}})}\sim0.03}}$, and a fraction of outliers is $<8\%$ \citep{mar16}.
For X-undet objects, we use the $z$\_$PDF$ values in the COSMOS2015 catalog, which are obtained by SED fitting with only galaxy templates.
The uncertainty for these photometric redshifts is $\sigma_{\Delta z/(1+z_{\mathrm{spec}})}\lesssim0.1$ for $i<24$ mag \citep{lai16}.
Three hundred thirty-seven objects ($69\%$) have the spectroscopic redshifts.
One hundred twenty-three objects ($25$\%) have the photometric redshifts by the galaxy-AGN hybrid templates, and 31 objects ($6\%$) have the photometric redshifts by the galaxy templates.
Our sample covers a wide range of rest-frame time intervals and wavelengths, where the highest redshift object is at $z = 4.26$.

Figure~\ref{fig9} shows the AGN bolometric luminosity as a function of redshift for our variable AGN sample.
The results of the X-ray stacking for the X-undet samples are also shown with the blue points with error bars.
The method of the X-ray stacking is described in the next subsection in detail.
The bolometric luminosity is calculated from the X-ray luminosity assuming the luminosity-dependent bolometric collection factor \citep{lus12}.
To calculate the bolometric luminosity, we use the hard band (2-10~keV) luminosities if available and use the soft  band (0.5-2~keV) luminosities for the hard X-ray undetected objects.
Our variable AGNs cover a luminosity range of $10^{43.0-46.5}$~erg~s$^{-1}$

\subsection{X-Ray Undetected Variable AGNs} \label{sec3_2}
\subsubsection{X-Ray Stacking Analysis} \label{sec3_2_1}

\begin{deluxetable*}{cccccccccccc}
\tablecaption{X-Ray stacking results for X-undet samples
\label{tbl5}
}
\tablehead{
 & & & & & \multicolumn{3}{c}{Soft Band} & \multicolumn{3}{c}{Hard Band} & \\ \cmidrule(lr){6-8} \cmidrule(lr){9-11}
\colhead{Bin} & 
\colhead{$N_\mathrm{{stacked}}$} & 
\colhead{$z_{\mathrm{med}}$} &
\colhead{Exp.} &
\colhead{$\log\left(L_{\mathrm{bol}}\right)$} &
\colhead{CR (0.5-2 keV)} &
\colhead{S/N} &
\colhead{$\log\left(L_{0.5-2\mathrm{keV}}\right)$} &
\colhead{CR (2-8 keV)} &
\colhead{S/N} &
\colhead{$\log\left(L_{2-10\mathrm{keV}}\right)$} &
\colhead{HR}\\ 
 &  &  & (ks) & (erg s$^{-1}$) & ($\mu$ counts s$^{-1}$) &  & (erg s$^{-1}$) & ($\mu$ counts s$^{-1}$) &  & (erg s$^{-1}$) & \\
(1) & (2) & (3) & (4) & (5) & (6) & (7) & (8) & (9) & (10) & (11) & (12)
}
\startdata
$z\leq0.7$          & 18 & 0.46 & $1,855$ & $             42.88^{+0.28}_{-0.39}$ & $ 9.18^{+3.85}_{-4.12}$ &      2.1   & $40.59^{+0.31}_{-0.42}$ & $30.8^{+7.75}_{-9.93} $ & 3.5 & $41.75^{+0.28}_{-0.39}$ & $0.54^{+0.21}_{-0.16}$\\
(low-mass)          &   9 & 0.34 & $    915$ & $             42.62^{+0.48}_{-0.35}$ & $ 4.54^{+4.99}_{-4.73}$ & $<1.0$ & $<40.32$                             & $32.6^{+13.2}_{-17.1} $ & 2.2 & $41.49^{+0.48}_{-0.35}$ & $>0.68$ \\
(high-mass)        &   9 & 0.51 & $    940$ & $             42.97^{+0.23}_{-0.21}$ & $ 13.1^{+7.17}_{-6.22}$ &      1.9    & $<41.04$                             & $29.9^{+11.4}_{-10.2} $ & 2.9 & $41.84^{+0.23}_{-0.21}$ & $0.37^{+0.28}_{-0.30}$\\ \hline
$0.7<z\leq2.0$ & 22 & 1.20 & $2,330$ & $             43.62^{+0.45}_{-0.35}$ & $ 6.09^{+4.84}_{-5.05}$ &      1.2    & $<41.58$                             & $20.4^{+6.58}_{-6.43} $ & 3.1 & $42.48^{+0.45}_{-0.35}$ & $0.59^{+0.22}_{-0.21}$\\ 
(low-mass)         & 11 & 1.29 & $ 1,225$ & $            43.72^{+0.41}_{-0.44}$ & $-4.23^{+3.40}_{-3.11}$ & $<1.0$  & $<41.15$                             & $21.9^{+11.3}_{-9.80} $ & 2.1 & $42.58^{+0.41}_{-0.44}$ & $>0.90$ \\
(high-mass)       & 11 & 1.18 & $ 1,105$ & $             43.54^{+0.35}_{-0.36}$ & $ 18.1^{+8.54}_{-6.61}$ &      2.4    & $41.79^{+0.35}_{-0.34}$ & $18.1^{+8.48}_{-8.69} $ & 2.2 & $42.41^{+0.35}_{-0.36}$ & $0.05^{+0.32}_{-0.22}$\\ \hline
$z>2.0$              & 10 & 3.17 & $     967$ & $^{\ast}44.01^{+0.20}_{-0.21}$ & $ 23.0^{+8.98}_{-8.00}$ &     2.6    & $42.77^{+0.20}_{-0.21}$ & $15.8^{+8.61}_{-7.10} $ & 1.9 & $<43.43$                             & $ -0.14^{+0.33}_{-0.26}$
\enddata
\tablecomments{
Column (1): redshift bin. 
Column (2): the number of the stacked objects. 
Column (3): the median redshift.  
Column (4): total effective exposure time. 
Column (5): bolometric luminosity calculated from the stacked hard band X-ray flux. 
Column (6) and (9): the median net source count rates in the resampled dataset for each band (0.5-2 keV, 2-8 keV). The errors represent the upper and lower values at a $68\%$ confidence level. 
Column (7) and (10): significances above the photometric noise for each band (0.5-2 keV, 2-8 keV). 
Column (8) and (11): X-ray luminosities for each band (0.5-2 keV, 2-10 keV). 
Column (12): hardness ratio.
For the lower S/N data (${\mathrm{S/N}}<2$), the luminosities are calculated from the $1\sigma$ variation of the noise value.
The asterisk symbol in the column (5) means that the bolometric luminosity is calculated from the soft band luminosity due to the low S/N of the hard band X-ray luminosity.
}
\end{deluxetable*}


\begin{figure}[htbp!]
\plotone{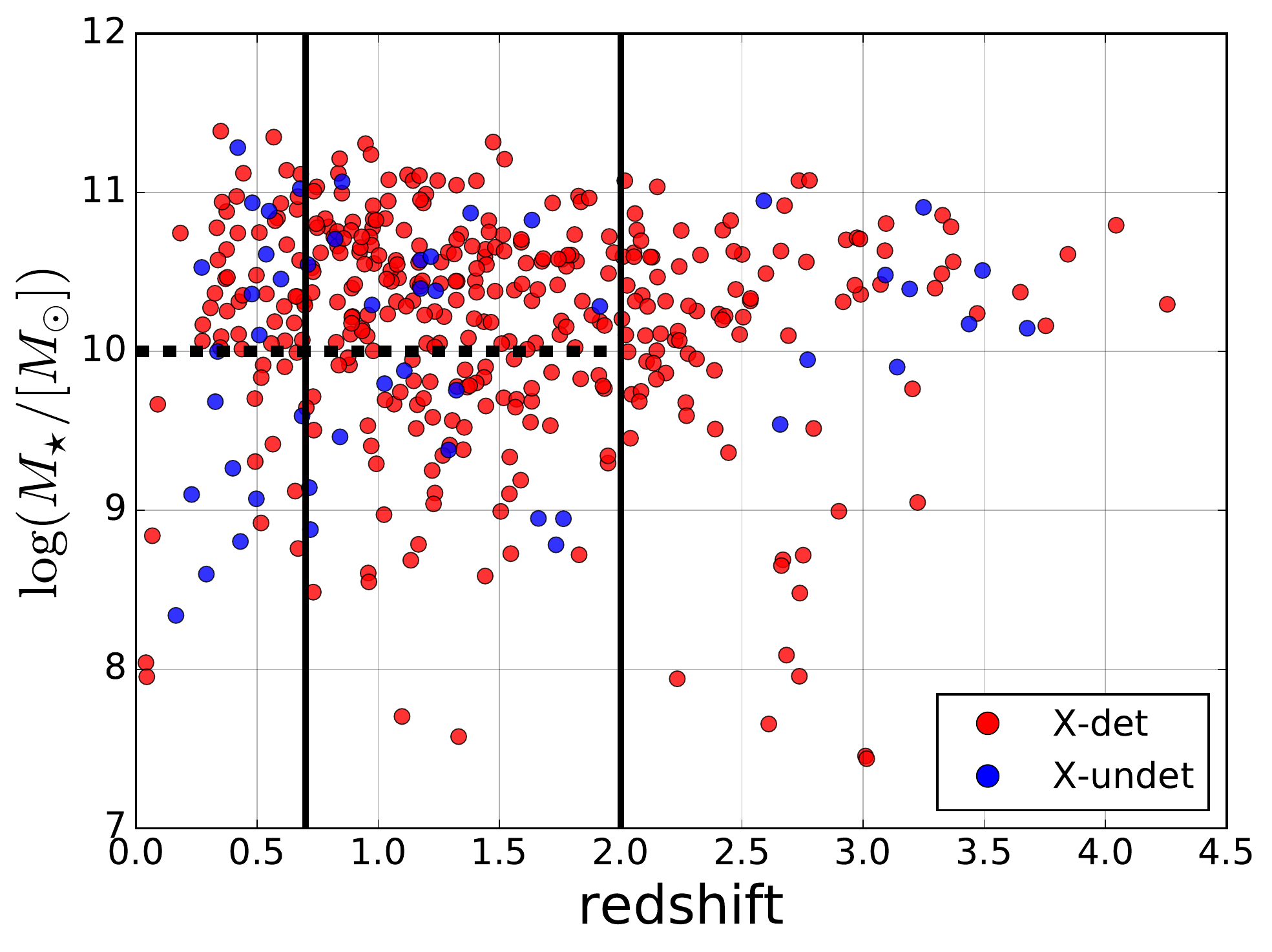}
\caption{
Stellar mass and redshift distribution for our variable AGN sample.
We use the stellar mass from $MASS\_BEST$ in the COSMOS2015 catalog, which is estimated from SED fitting~\citep{lai16}.
The red (blue) points are X-det (X-undet) objects.
The black lines are the boundaries of each bin for the stacking analysis (solid: redshift bin, dashed: stellar mass bin) described in Section~\ref{sec3_2}.
\label{fig10}
}
\end{figure}

\begin{figure}[t!]
\plottwo{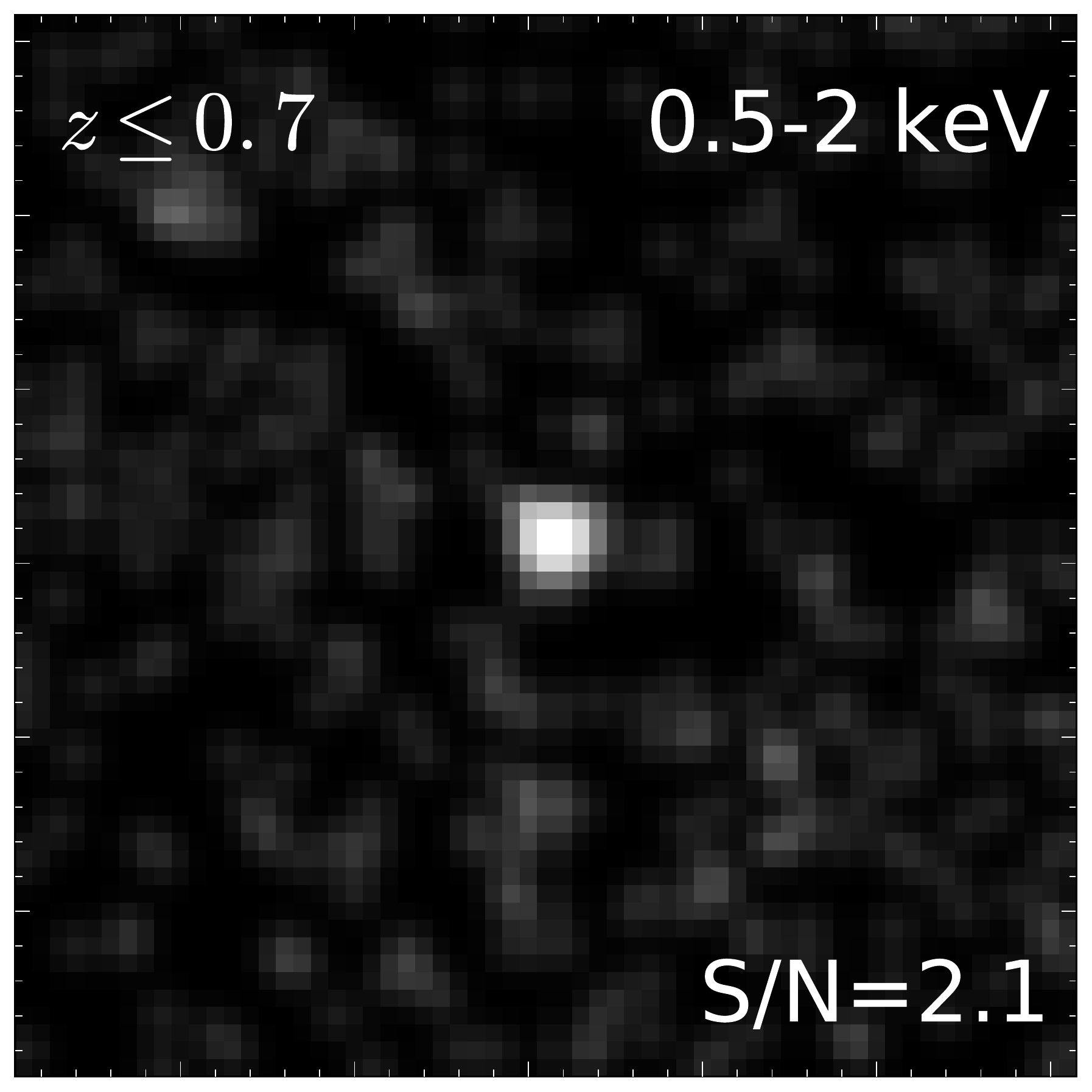}{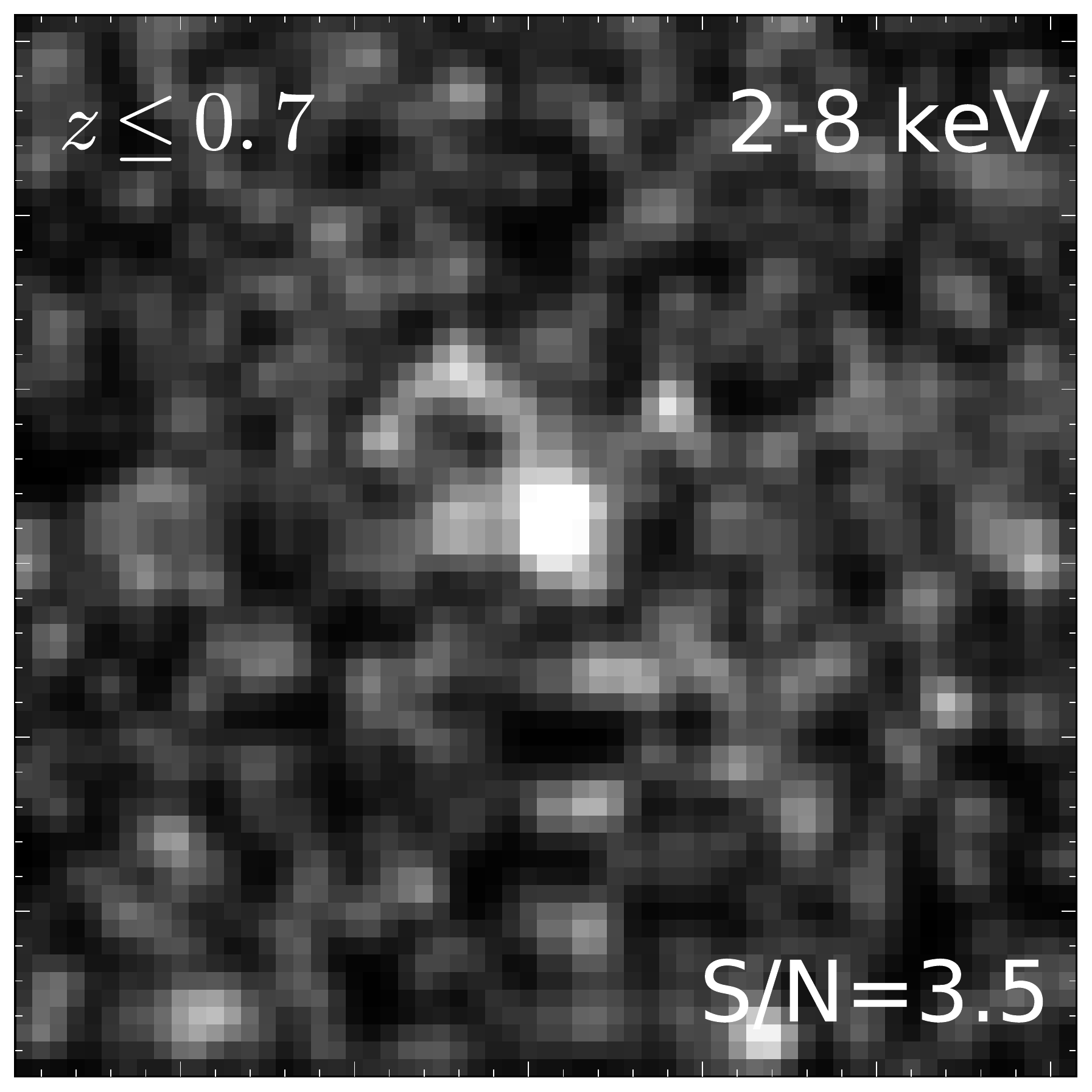}
\plottwo{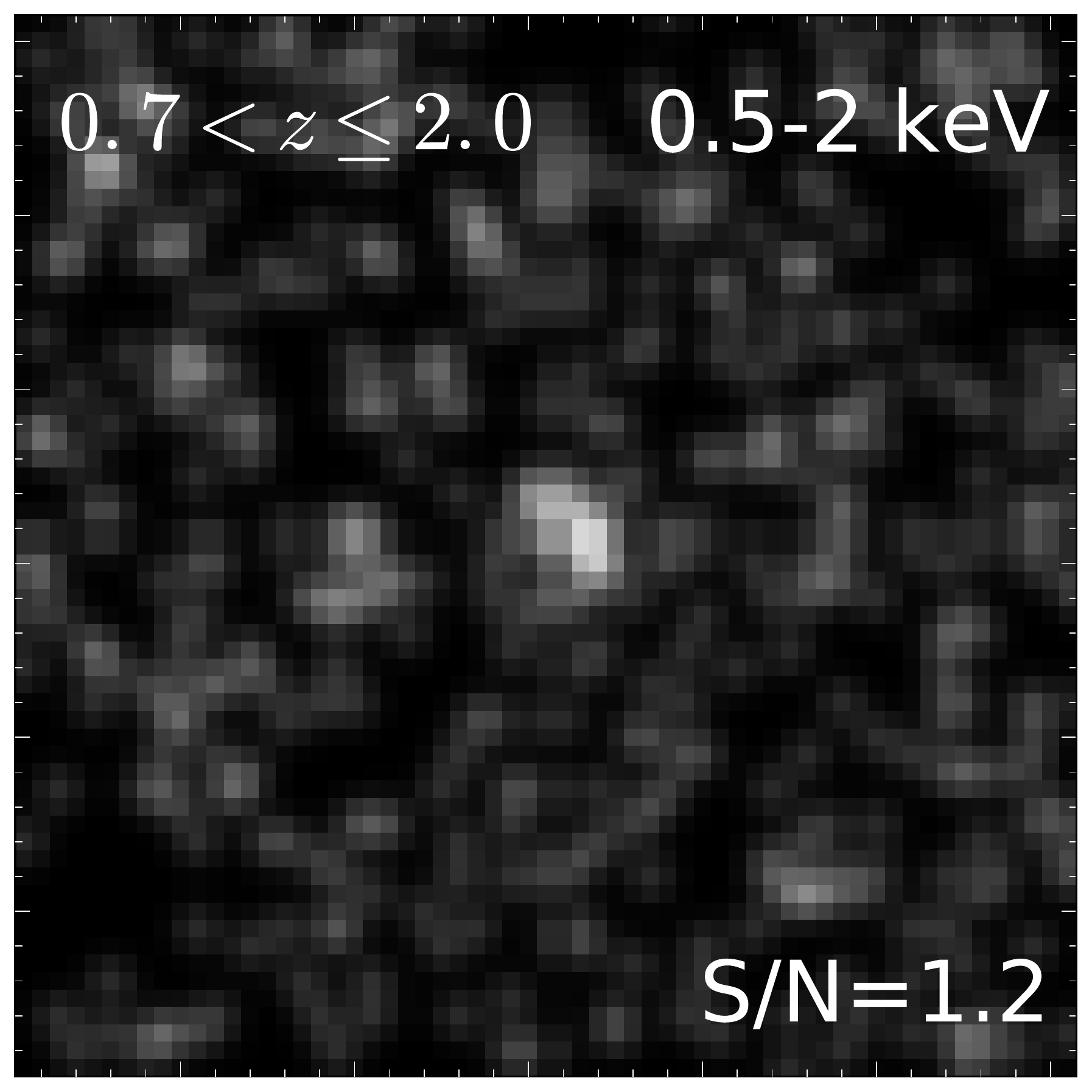}{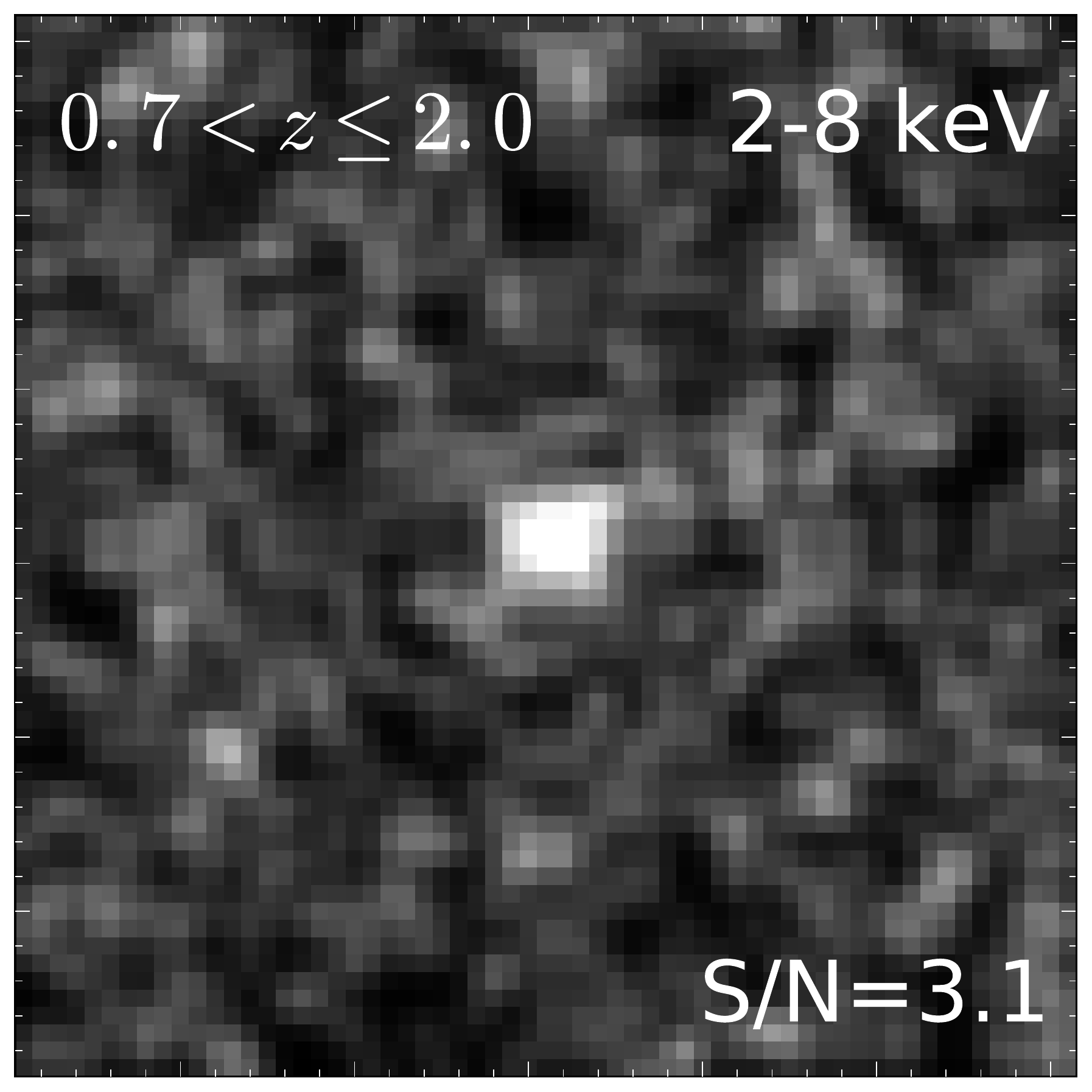}
\plottwo{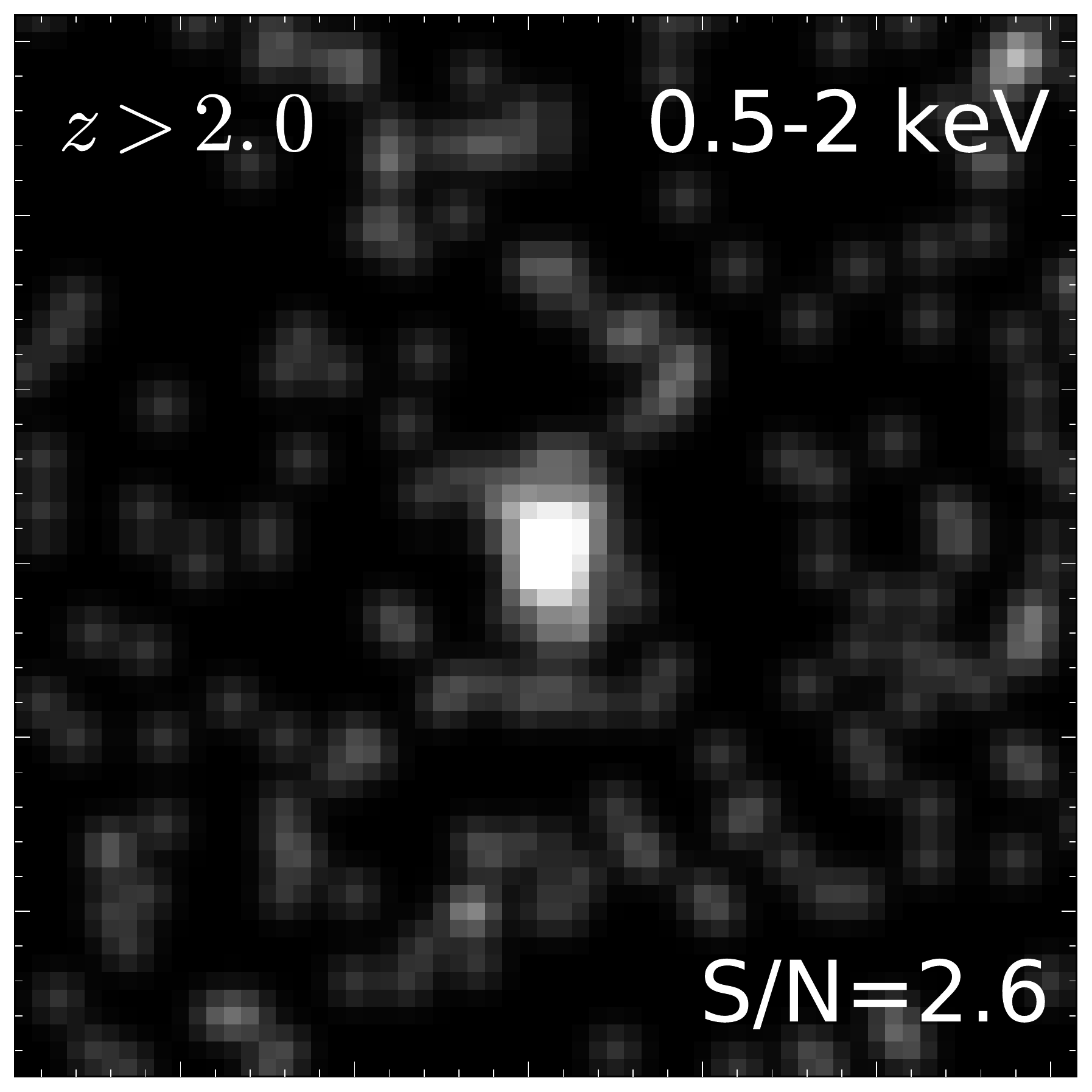}{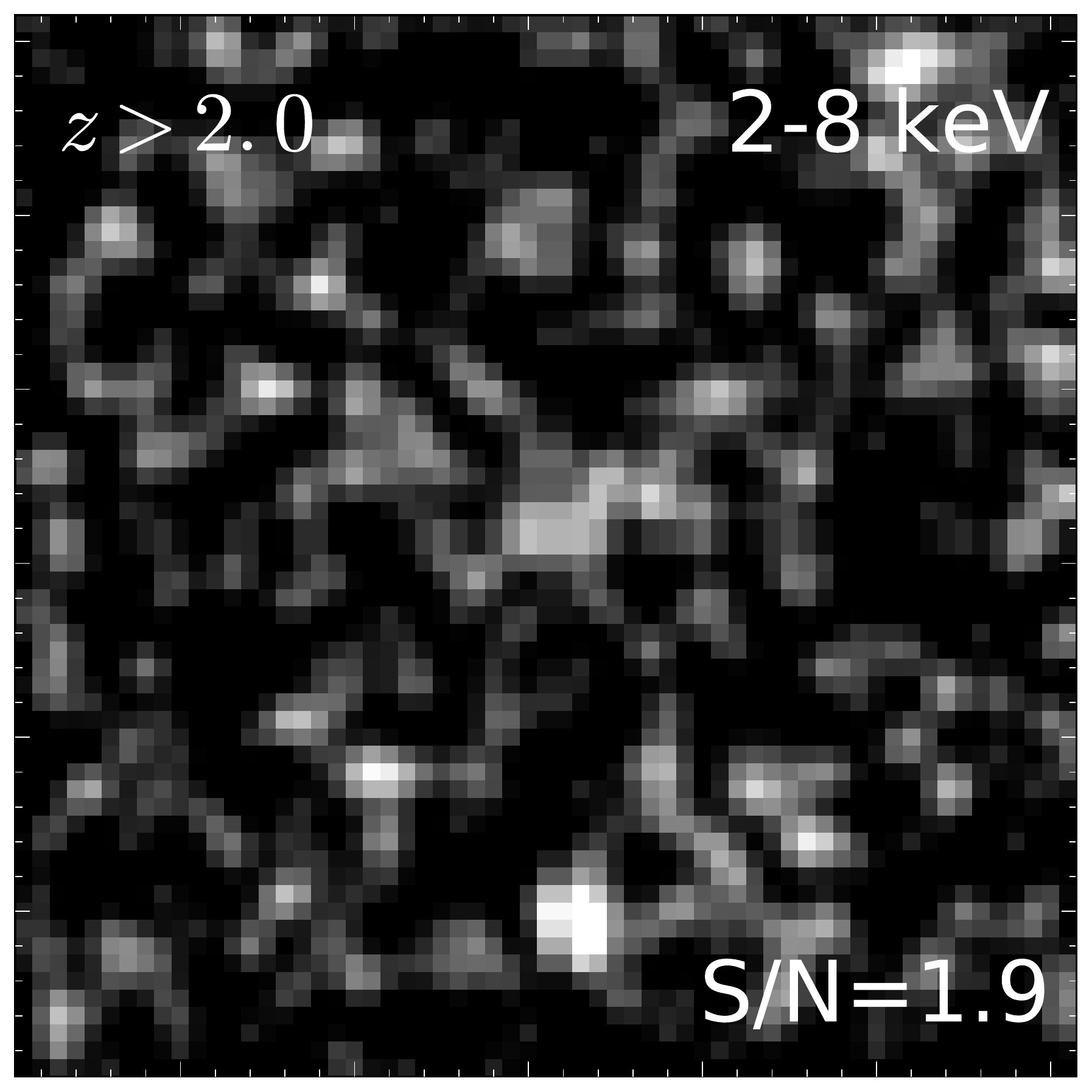}
\caption{
Stacked X-ray images for each redshift bin (top panels: $z\leq0.7$, middle panels: $0.7<z\leq2.0$, bottom panels: $z\geq2$) in the soft (left panels) and hard (right panels) bands.
Images are smoothed by a Gaussian filter with the standard deviation of 1 pixel.
\label{fig11}
}
\end{figure}

Here, we focus on the X-undet objects (50 out of 491 our variable AGN objects).
They are likely to be AGNs with lower X-ray flux than the Chandra detection limit.
We checked their statistical X-ray properties by using the Chandra X-ray stacking analysis tool, CSTACK v4.32 \citep{miy08}\footnote{
\url{http://cstack.ucsd.edu/} or \url{http://lambic.astrosen.unam.mx/cstack/}
}.
Using 117 observations from the Chandra COSMOS Legacy survey data \citep[the total exposure time is 4.6~Ms;][]{civ16}, CSTACK can calculate exposure-weighted mean X-ray count rates in the soft (0.5-2~keV) and hard (2-8~keV) bands by stacking Chandra images centered at given sky positions.

Since the number of the X-undet objects is limited, we divide the X-undet objects into three redshift bins: $z\leq0.7$, $0.7<z\leq2.0$, and $z>2.0$, where the median redshifts of each bin are $z_{\mathrm{med}} = 0.46$, $1.20$, and $3.17$, respectively, as shown in Figure~\ref{fig10}.
The results of the CSTACK X-ray stacking analysis are summarized in Table~\ref{tbl5}, and postage stamps of the stacked X-ray images are shown in Figure~\ref{fig11}.
To check the significance of the stacked count rates, CSTACK conducts a bootstrap resampling analysis that provides the distribution of the stacked count rates for 500 resampled datasets, each of which consists of the same number of objects as the input ones selected at random allowing for duplicates.
X-ray emissions are statistically detected ($S/N\geq2$) in the lowest redshift bin sample in both bands, $0.7<z\leq2.0$ bin sample in the hard band, and $z>2.0$ bin sample in the soft band.
We then obtain the X-ray flux from the stacked X-ray count rates by adopting the conversion factor from the PIMMS\footnote{
\url{http://cxc.harvard.edu/toolkit/pimms.jsp}
} utility.
The conversion factor\footnote{
We use the ACIS-I response for Chandra Cycle 14.
} from 0.5-2~keV (2-8~keV) count rate to 0.5-2~keV (2-10~keV) X-ray band flux is $6.563\times10^{-12}$~erg~cm$^{-2}$~count$^{-1}$ ($2.784\times10^{-11}$~erg~cm$^{-2}$~count$^{-1}$), where a power-law photon index $\Gamma=1.4$ and a Galactic column density of $N_{\mathrm{H}}=2.6\times10^{20}$~cm$^{-2}$ \citep{kal05} are assumed, as used in Chandra COSMOS Legacy survey (see \citealt{civ16}).
For each redshift bin, the observed X-ray flux is converted to the rest-frame flux by using a k-correction factor of $(1+z_{\mathrm{med}})^{\Gamma-2}$.
We finally derive the X-ray luminosity for each band and the bolometric luminosity by using the luminosity distance at $z_{\mathrm{med}}$ and the luminosity-dependent bolometric collection factor (\citealt{lus12}).
The results are summarized in Table~\ref{tbl5} and are also plotted as the blue points in Figure~\ref{fig9}.
As shown in Figure~\ref{fig9}, we conclude that the X-undet sample indeed has a lower flux than the Chandra detection limit.

\subsubsection{Non-AGN X-Ray Flux Contribution} \label{sec3_2_2}

\begin{deluxetable*}{cccccccc}[t!]
\tablecaption{Non-AGN X-Ray contributions
\label{tbl6}
}
\tablehead{
\colhead{Bin} & 
\colhead{$z_{\mathrm{med}}$} & 
\colhead{$\log\left(M_{\star,\mathrm{med}}\right)$} & 
\colhead{$\log\left(\mathrm{SFR}_{\mathrm{med}}\right)$} &  
\multicolumn{2}{c}{$\log\left(L_{2-10\mathrm{keV}}^{\mathrm{XRB}}\right)$} & 
\multicolumn{2}{c}{$\log\left(L_{0.5-2\mathrm{keV}}^{\mathrm{Hot}}\right)$} \\ 
\cmidrule(lr){5-6} \cmidrule(lr){7-8}
 &  &  ($M_{\odot}$) & ($M_{\odot}\ \mathrm{yr}^{-1}$) & (erg s$^{-1}$) & (\%) & (erg s$^{-1}$) & (\%)\\
 (1) & (2) & (3) & (4) & (5) & (6) & (7) & (8)
}
\startdata
$z\leq 0.7$             & 0.46 & 10.05 & 0.53 & 40.2  & $3.2$   & 39.4 & $7.2$\\
$0.7 < z \leq 2.0$ & 1.20 & 10.08 & 1.18 & 41.0  &  $3.2$   & 40.1 & $>3.3$\\
$ z > 2.0 $               & 3.17 & 10.28 & 1.89 & 42.0  & $>3.5$ & 40.8 & $1.1$
\enddata
\tablecomments{
Column (1): redshift bin. 
Column (2): the median redshift. 
Column (3): the median stellar mass of host galaxy.
Column (4): the median star formation rate.
Column (5): hard band (2-10~keV) X-ray contributions from XRBs.
Column (6): fraction of the luminosities of XRBs to the overall measured hard band luminosities.
Column (7): soft band (0.5-2~keV) X-ray contributions from ISM diffuse X-ray emissions.
Column (8): fraction of the luminosities of ISM diffuse X-ray emissions to the overall measured soft band luminosities.
}
\end{deluxetable*}

In addition to the AGN emission, X-ray binaries (XRBs) in the AGN host galaxies can also be the sources of the observed X-ray emission.
The integrated X-ray emission from XRBs consists of radiation from low-mass XRBs and high-mass XRBs, and their total luminosities are proportional to the stellar mass ($M_{\star}$) and the star formation rate (SFR) of the host galaxy.
\citet{leh16} investigate the redshift dependence of the contributions of the XRBs in normal galaxies and provide the following empirical relation:
\begin{eqnarray}
L_{\mathrm{2-10keV}}^{\mathrm{XRB}} = \alpha(1&+&z)^{\gamma}\left(\frac{M_{\star}}{M_{\odot}}\right)\nonumber\\
                                                            &+& \beta(1+z)^{\delta}\left(\frac{\mathrm{SFR}}{M_{\odot}\ {\mathrm{yr}}^{-1}}\right)\ \ [\mathrm{erg\ s^{-1}}] \label{eq8},
\end{eqnarray}
where $\log\alpha = 29.30\pm0.28$, $\log\beta=39.40\pm0.08$, $\gamma=2.19\pm0.99$, $\delta=1.02\pm0.22$, and the scatter is 0.17~dex \citep[the best-fit values for the 6~Ms Chandra Deep Field South data, see ][]{leh16}.
We adopt the median values of redshift, stellar mass, and SFR in each redshift bin to evaluate the contributions of the XRBs to the X-undet samples.
We use the stellar mass and SFR from $MASS\_BEST$ and $SFR\_BEST$ values listed in the COSMOS2015 catalog, which are products of an SED fitting \citep{lai16}.
The X-ray contributions from XRBs are listed in Table~\ref{tbl6}.
For all of the stacked samples, the hard band luminosities are about 1.5~dex brighter than the contributions from XRBs.

Another possible source of the X-ray emission is thermal plasma ($\sim$sub-keV temperature) in the galaxy interstellar medium (ISM), which mainly contributes to the soft band X-ray luminosity.
Since this diffuse X-ray emission arises from collective effects of supernova remnants and winds from massive stars, the soft band X-ray luminosity depends on the SFR.
\citet{min12} derive an empirical relationship between the diffuse X-ray luminosity and SFR for nearby late-type galaxies as
\begin{eqnarray}
L_{\mathrm{0.5-2keV}}^{\mathrm{Hot}} = (8.3\pm0.1)\times10^{38} \left(\frac{\mathrm{SFR}}{M_{\odot}\ {\mathrm{yr}}^{-1}}\right)\ \ [\mathrm{erg\ s^{-1}}] \label{eq9},
\end{eqnarray}
with an intrinsic  scatter of 0.34~dex.
From this equation, we calculate the effect of the ISM diffuse X-ray emission in the soft band X-ray luminosity for our stacked samples.
The results are also listed in Table~\ref{tbl6}, and we find that the stacked soft band X-ray luminosities are more than 1~dex brighter than the contributions from the ISM diffuse X-ray emissions.

It should be noted that ultra luminous X-ray sources (ULXs) may also contribute to the stacked X-ray emission in the X-undet samples.
ULXs are usually defined as off-nuclear pointlike X-ray sources and typically have X-ray luminosities $> 10^{39}$~erg s$^{-1}$ \citep[e.g. ][]{fen11}.
The X-ray luminosity of ULXs in elliptical galaxies are weak ($<10^{40}$~erg s$^{-1}$), while one-third of spiral galaxies have luminosities $\geq5\times10^{39}$~erg~s$^{-1}$ and about 10 \% of ULXs have luminosities $> 10^{40}$~erg~s$^{-1}$ \citep{swa04,wal11}.
The X-ray luminosities of our samples, however, are still significantly higher than that of the expected ULX emission.

Thus, the X-ray luminosities from XRBs, hot ISM, and ULXs are too weak to explain the stacked X-ray luminosities, and we find that X-ray emission of the stacked samples is dominated by the emission from AGNs.

\subsubsection{Hardness Ratio} \label{sec3_2_3}

\begin{figure}[htbp!]
\plotone{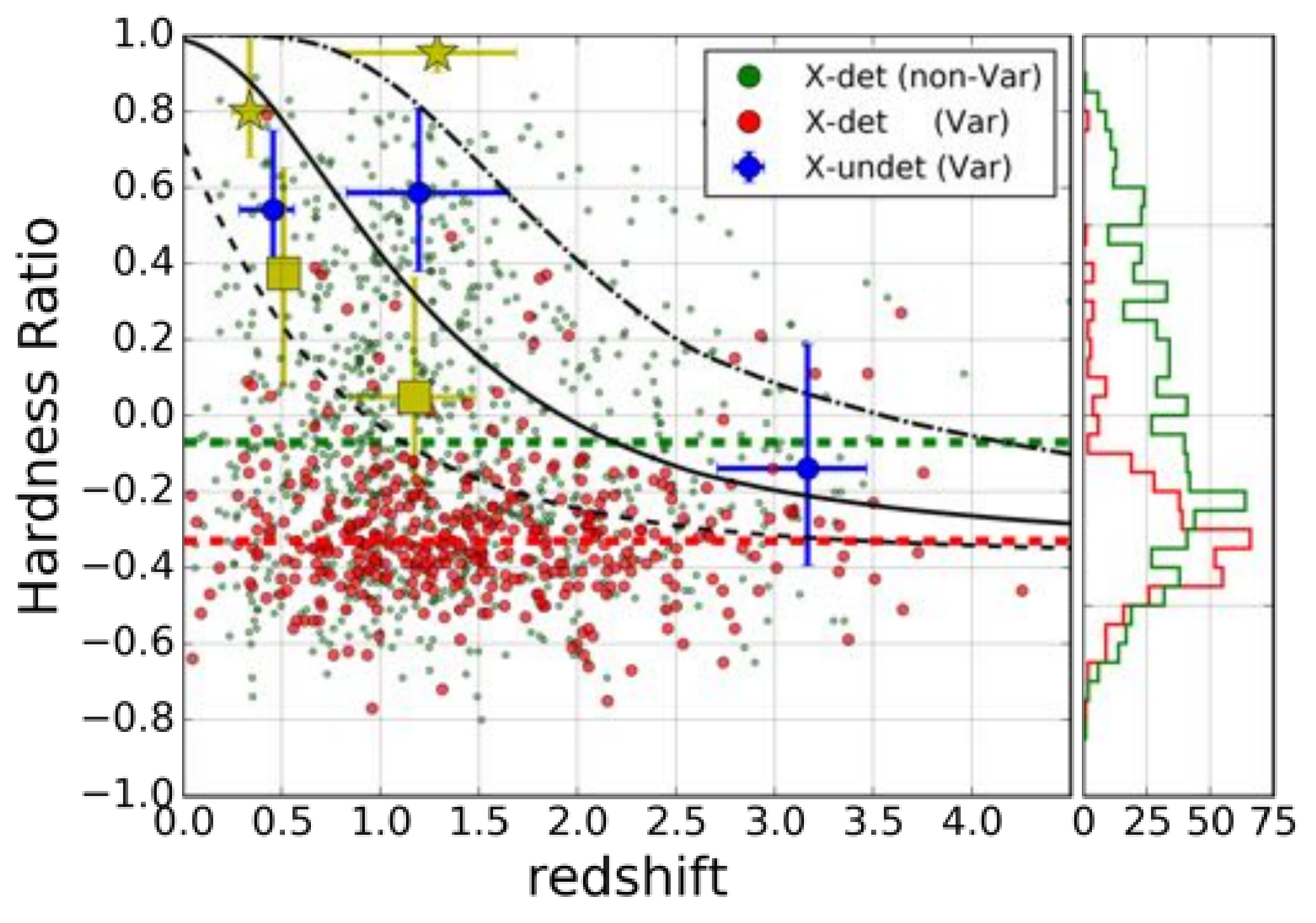}
\caption{
Hardness ratio as a function of redshift.
The red (green) points are variability-detected (undetected) X-ray detected objects.
The dashed lines are the median values for each sample.
The histograms of the hardness ratios are shown in the right side.
The blue points with error bars show the results of stacking analysis for the X-undet samples in our variable AGNs (three redshift bins).
The yellow points are also the results of stacking analysis for the low mass bin (square symbols) and the high mass bin (star symbols). 
The black curves are model predictions assuming a power-law spectrum with photon index $\Gamma=1.8$ and gas column densities of $N_{\mathrm{H}}=10^{22.5}$ (dashed), $10^{23}$ (solid), and $10^{23.5}$ (dashed-dotted)~cm$^{-2}$, respectively.
\label{fig12}
}
\end{figure}

\begin{figure}[t!]
\plottwo{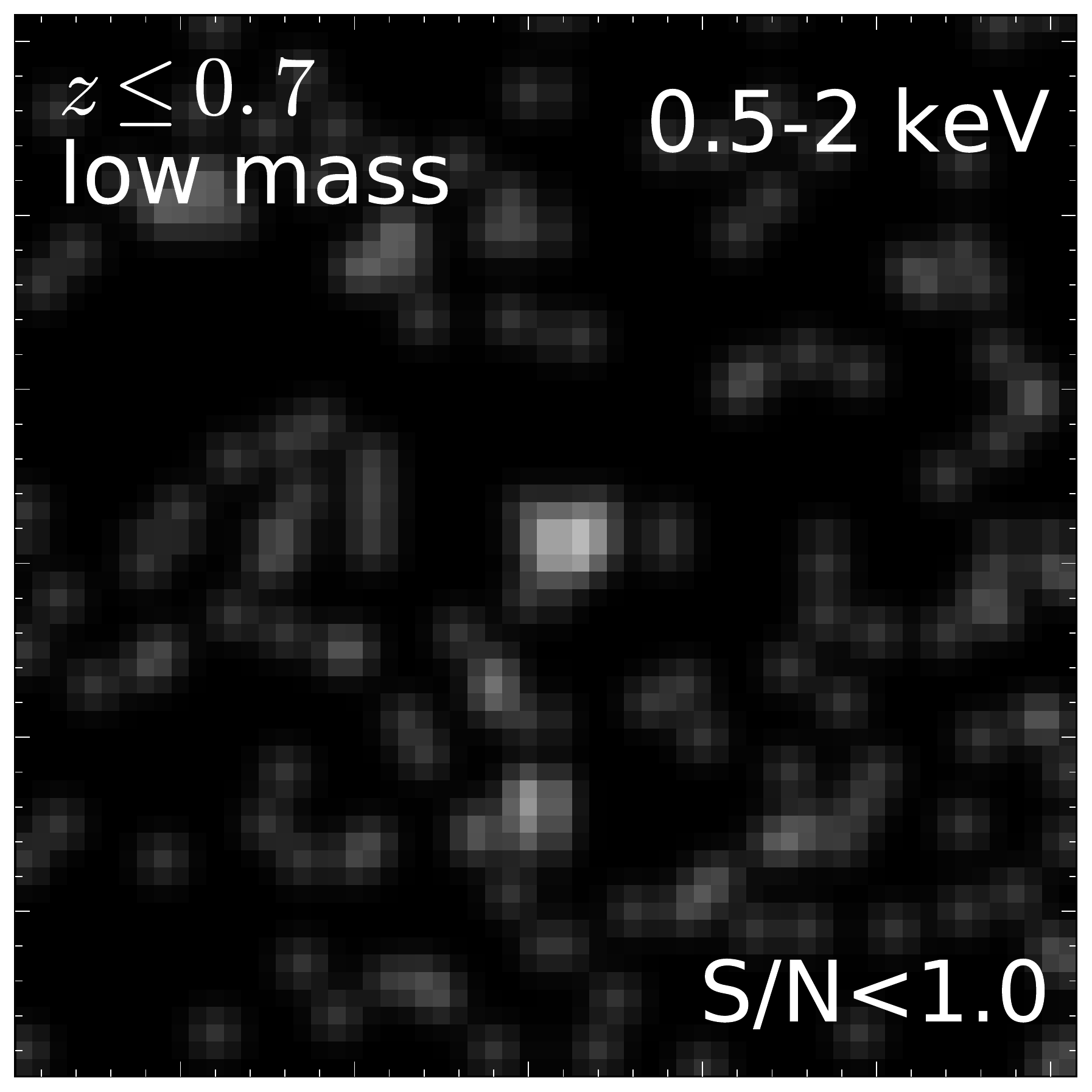}{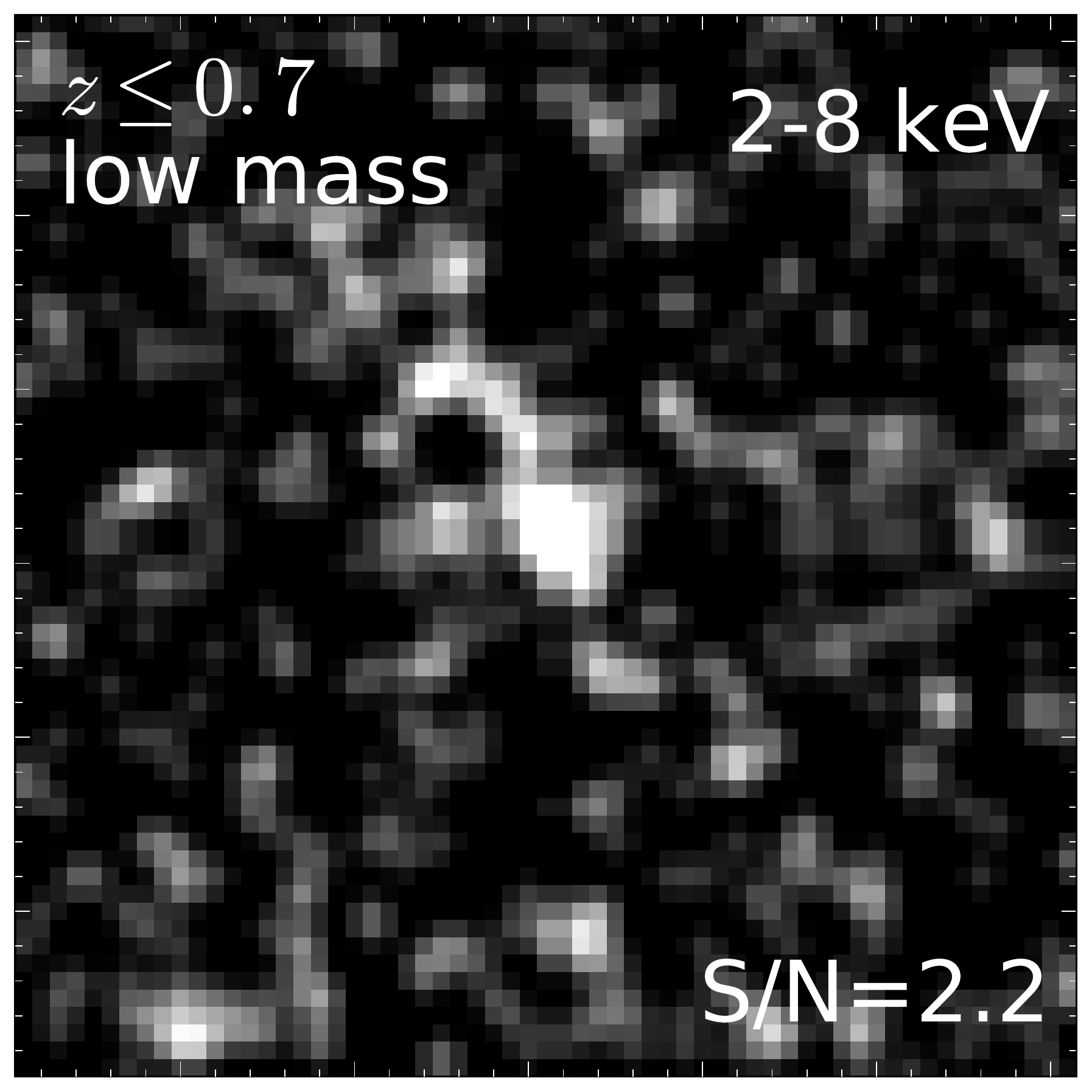}
\plottwo{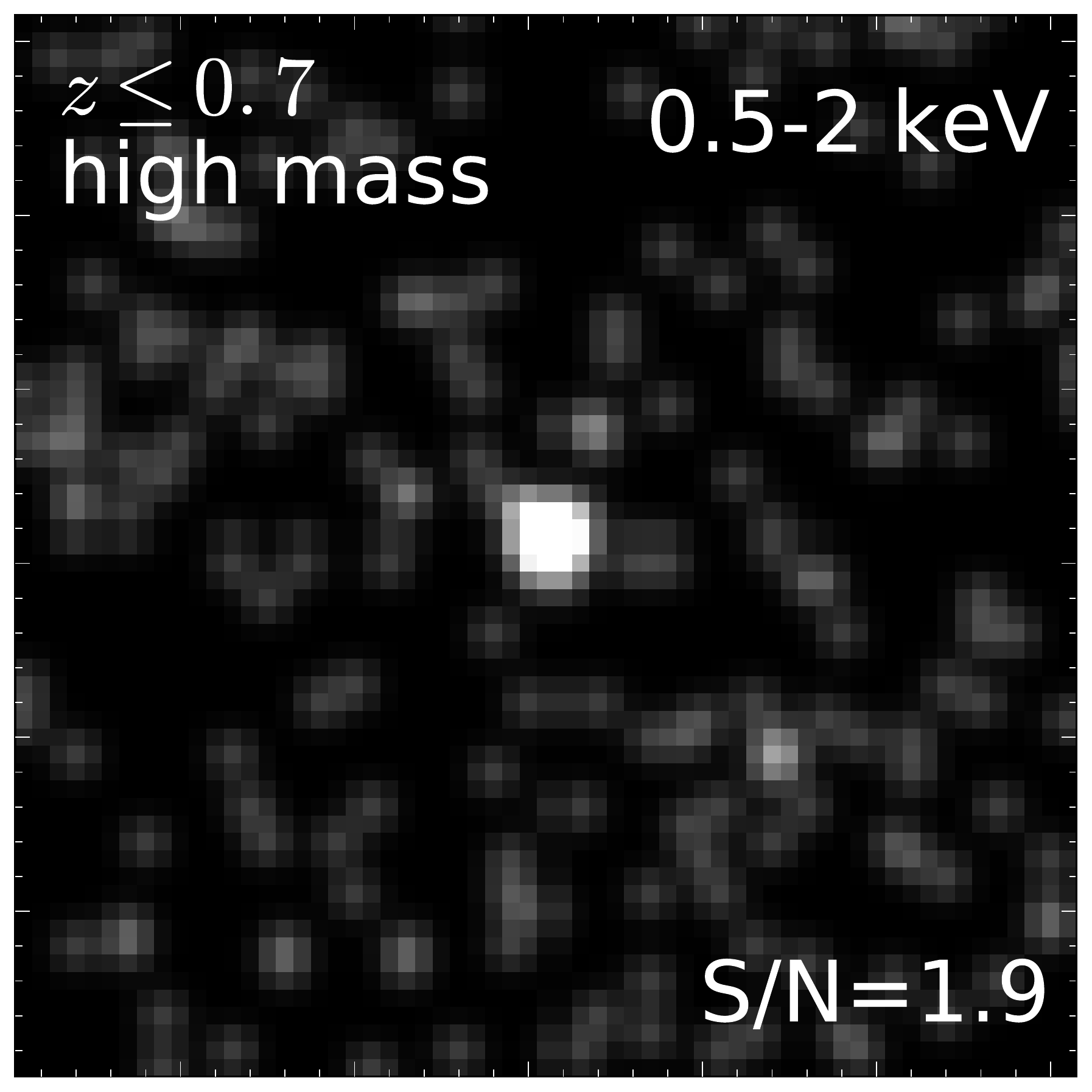}{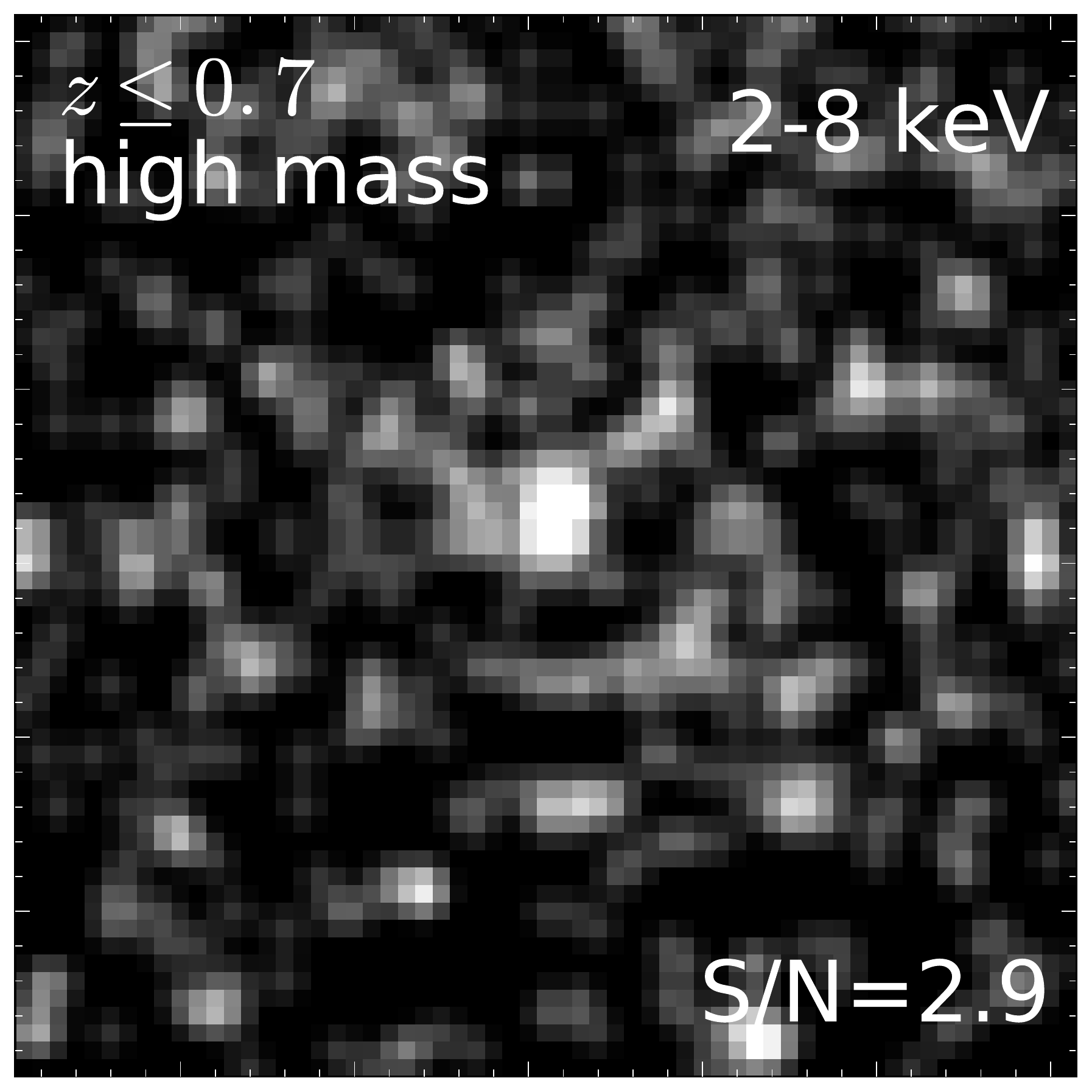}
\plottwo{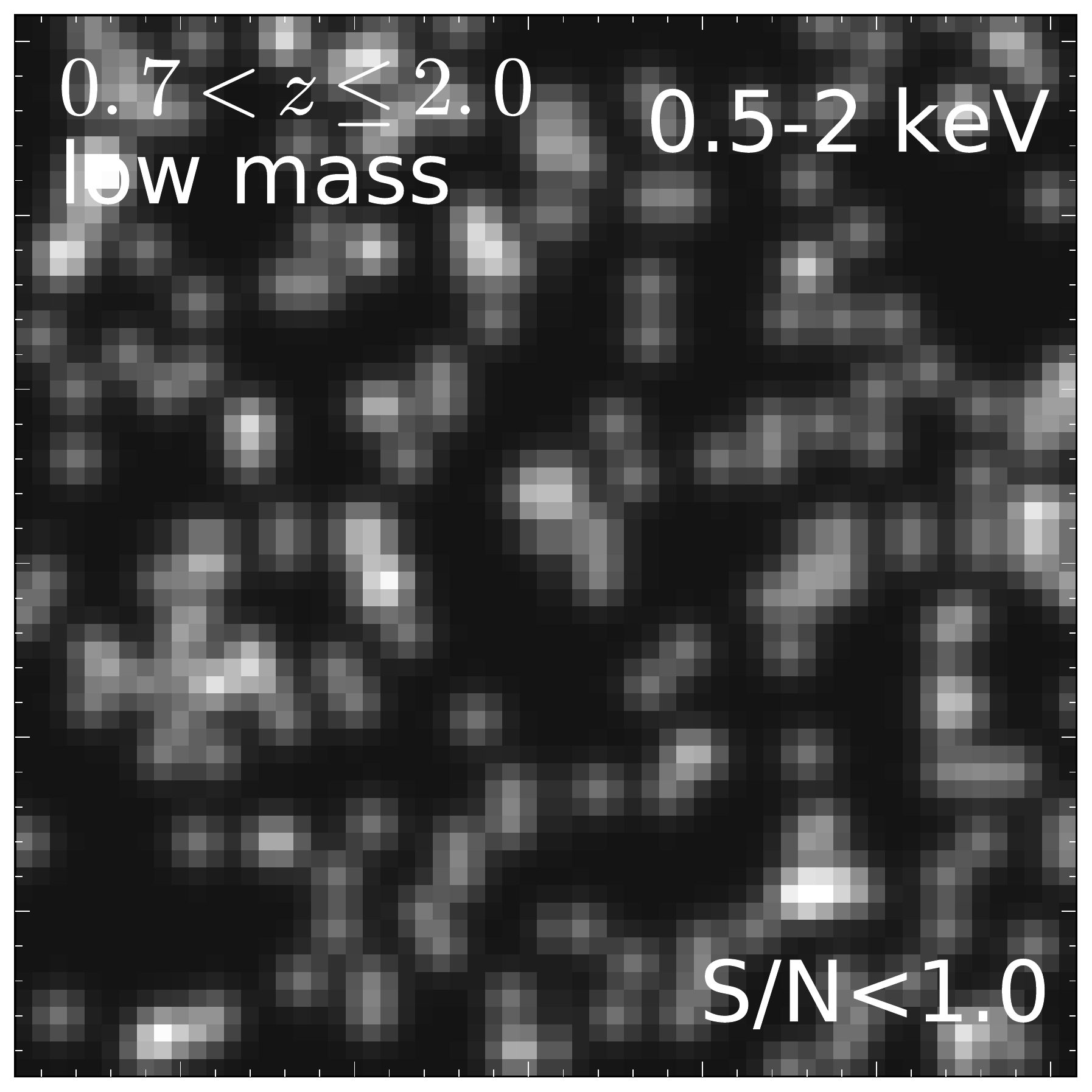}{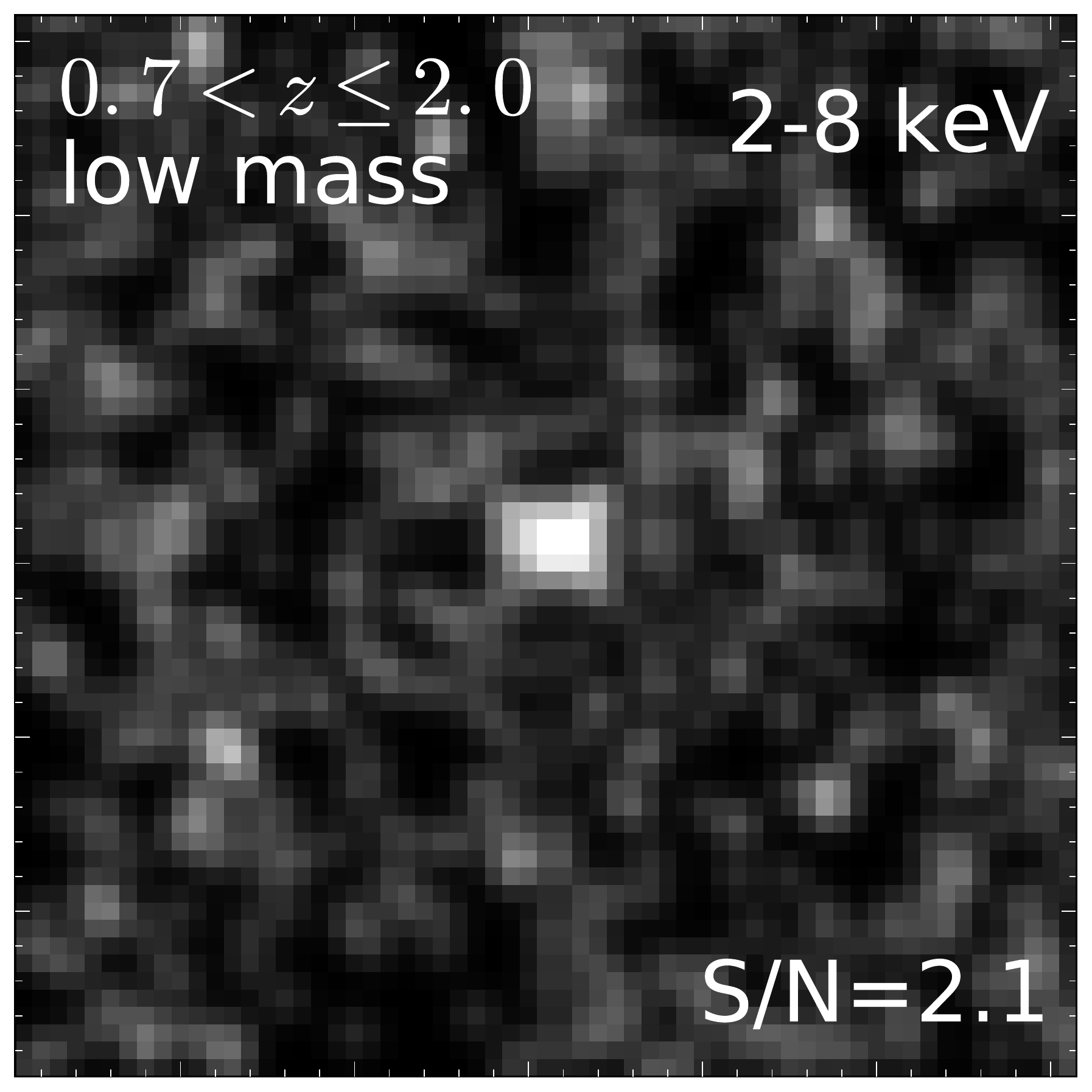}
\plottwo{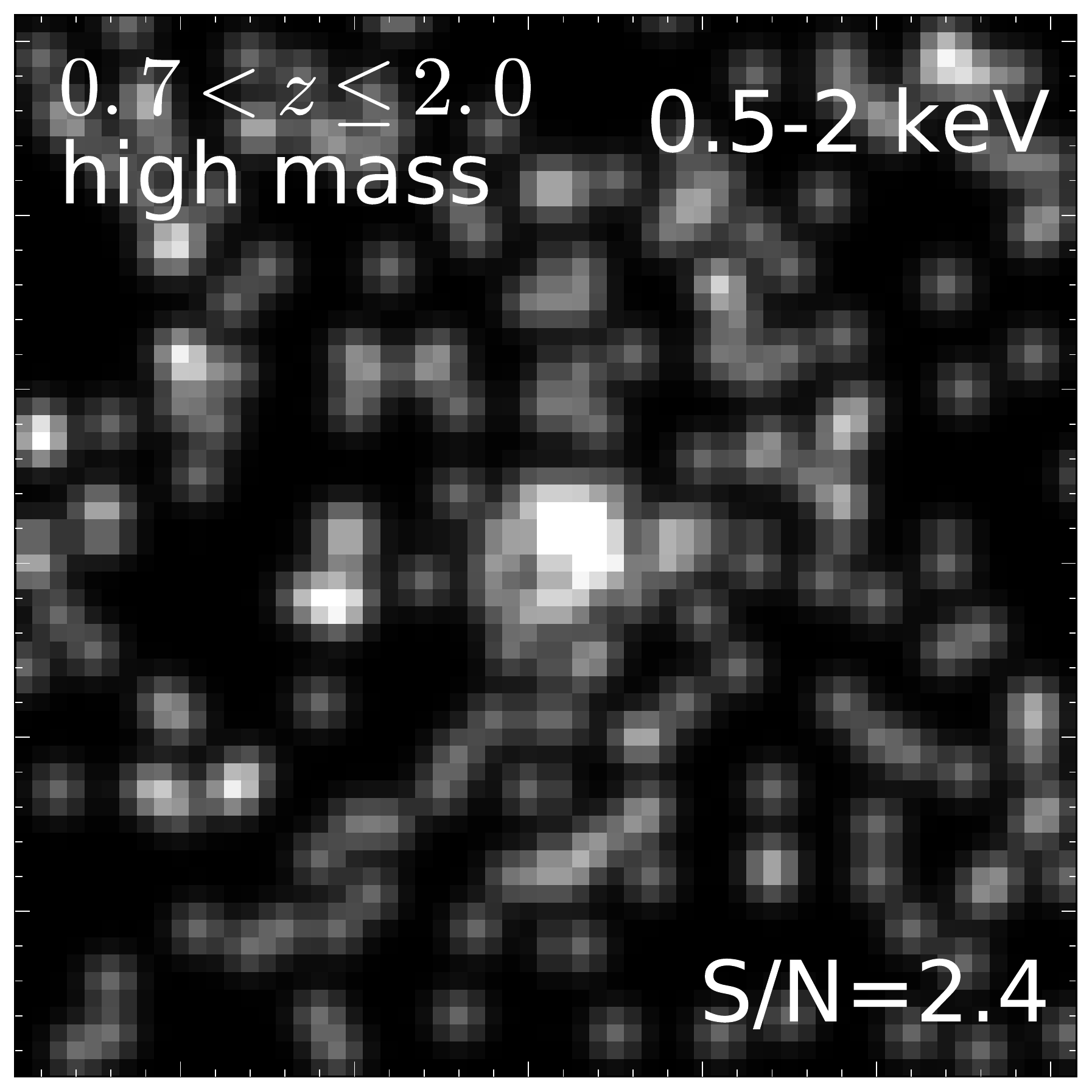}{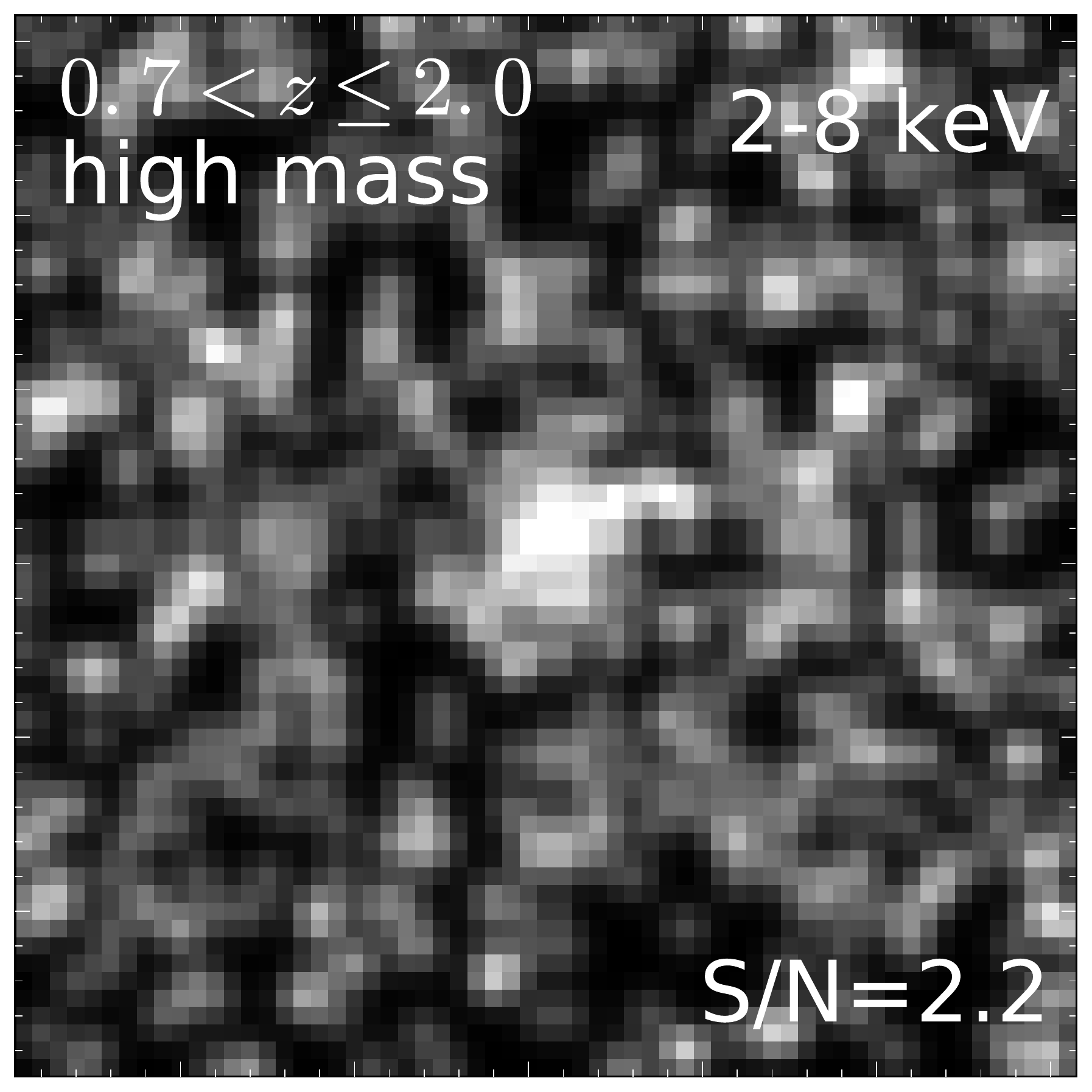}
\caption{
Stacked X-ray imaging for each redshift and stellar mass bin (upper four panels: $z\leq0.7$; lower four panels: $0.7<z\leq2.0$) in the soft (left panels) and hard (right panels) band.
Images are smoothed by a Gaussian filter with the standard deviation of 1 pixel.
\label{fig13}
}
\end{figure}

For the stacked samples, we calculate the Hardness Ratio defined as $\mathrm{HR=(H-S)/(H+S)}$, where H and S are the net source counts in the hard band (2-8~keV) and the soft band (0.5-2~keV), respectively.
The HR represents the shape of the X-ray spectrum that can be characterized by the intrinsic power-law photon index $\Gamma$, absorption column density $N_{\mathrm{H}}$, and redshift.
To calculate HR, we use the Bayesian Estimation of Hardness Ratios (BEHR) tool \citep{par06}.
BEHR calculates the HR from the input parameters (source/background counts and effective areas of source/background in both the soft and hard bands), all of which are given by the CSTACK analysis.

The results are summarized in Table~\ref{tbl5}.
Figure~\ref{fig12} shows the HR distribution as a function of redshift for the stacked samples, X-ray detected variable objects, and the X-ray detected non-variable objects in the Chandra catalog.
The median value of HR for the X-ray detected variable objects is $-0.33$, which is softer than the median value of HR for the X-ray detected non-variable objects ($-0.07$).

It is interesting to note that the stacked X-undet samples have higher HR values than most of the X-det objects in our variable AGN sample and are comparable to the X-ray detected non-variable objects.
This can be interpreted as there is a significant amount of X-ray absorbing materials in the line of sight, which absorb the X-ray photons in a low energy band ($\lesssim5$ keV).

We also divide the X-undet variable samples into two subsamples by the stellar mass.
Since the number of X-undet objects is limited, we here use the X-undet objects at $z\leq2$ (see Figure~\ref{fig10}).
The CSTACK X-ray stacking results for the low-mass ($M_{\star}<10^{10}~M_{\odot}$) and high-mass ($M_{\star}\geq10^{10}~M_{\odot}$) subsamples are shown in Figure~\ref{fig13}. 
Statistically significant X-ray signals are still detected ($S/N>2$) for all the subsamples in the hard band, but the X-ray signals in the soft band are low especially for the low mass subsamples in both redshift bins.
The HRs of these subsamples are also plotted in Figure~\ref{fig12} (yellow points with error bars).
It is found that the low-mass subsamples show a harder X-ray spectrum than the high-mass subsamples in both of the redshift bins. 
The PIMMS utility shows that if we assume the Galactic column density $N_{\mathrm{H}}=2.6\times10^{20}$~cm$^{-2}$ and the intrinsic photon index $\Gamma=1.8$ with the source column density $N_{\mathrm{H}}=10^{22.5}$, $10^{23.0}$, and $10^{23.5}$~cm$^{-2}$, the observed HRs are $0.268$, $0.805$, and $0.996$ at $z = 0.46$ and $-0.096$, $0.303$, and $0.812$ at $z=1.20$, respectively. 
These HR values are similar to our results for the X-undet samples (see Table~\ref{tbl5} and Figure~\ref{fig12}).
From these results, we argue that our variable AGN sample contains a significant fraction of the X-ray obscured but optically unobscured (variable) type-I AGNs.
The physical interpretation of these X-ray obscured optically variable objects is discussed in Section~\ref{sec5_2}.

\subsection{Dust Covering Factor}\label{sec3_3}

\begin{figure}[t]
\plotone{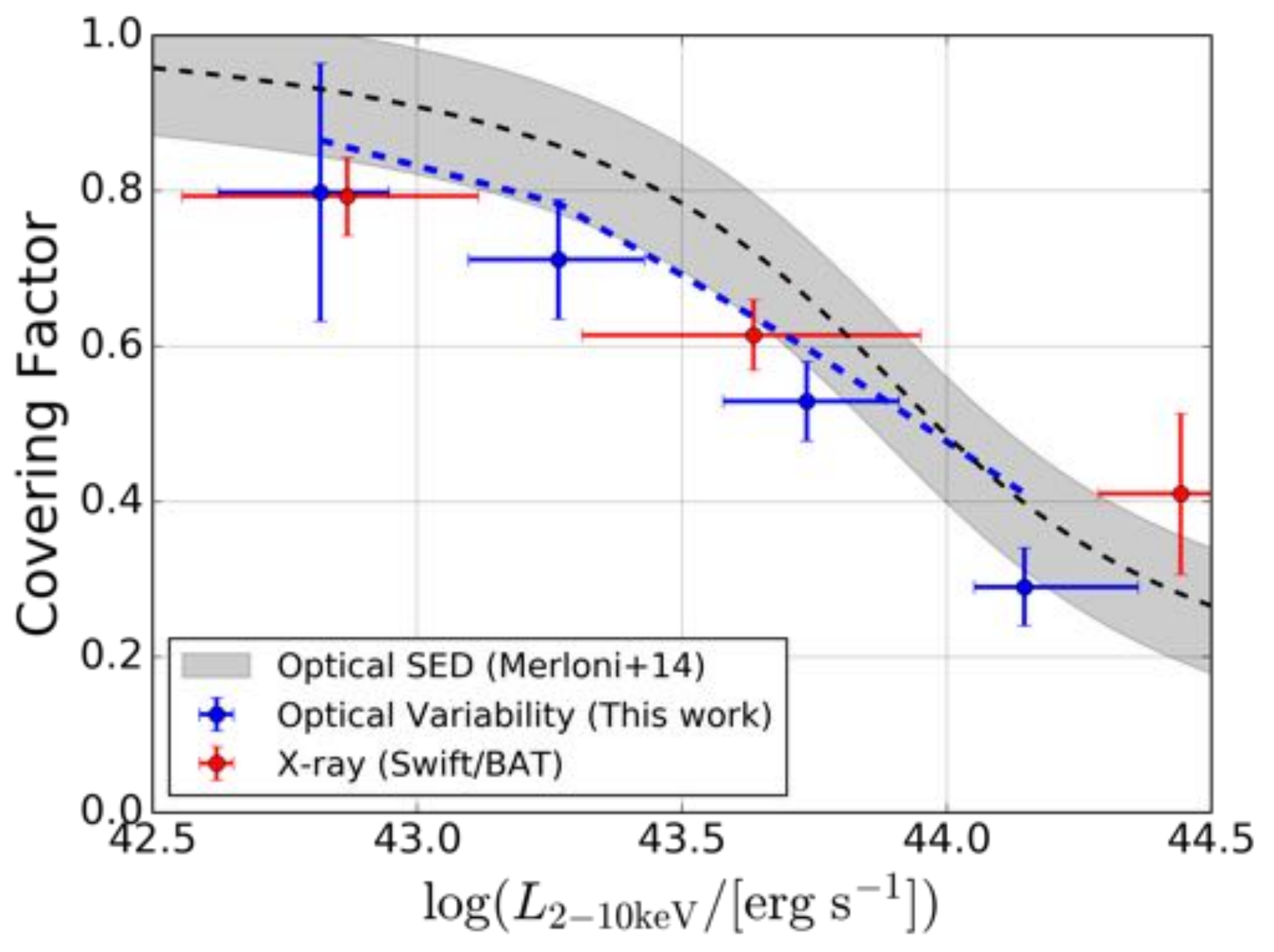}
\caption{
Covering factor for obscured material as a function of hard band (2-10~keV) X-ray luminosity.
The blue dashed line is a result obtained for our variable AGN sample, and the blue points with error bars are the values corrected for the detection rate of the known BLAGNs in each luminosity bin. 
The black dashed line is the obscured fraction from the optical diagnostics \citep[Equation (1) in][]{mer14}.
The error range of the \citet{mer14} result ($\sim0.8\%$) is shown as the black shaded region.
The red points with error bars are the results of the X-ray absorbed fraction from the Swift-BAT observations \citep{ric17a, ric17b, ich19}.
\label{fig14}
}
\end{figure}

The detection of the variability in AGNs means that we directly see UV/optical emission from the accretion disk, which is not obscured by the AGN dust tori.
The dust covering factor of AGNs, which is defined as the ratio of optically obscured AGNs to the entire AGN population, can be constrained by the variability detection fraction among the X-ray detected AGNs.

Within our HSC variability survey area in the COSMOS field, 743 X-ray detected AGNs at $0.5\leq z\leq2.5$ (the redshift information is obtained in the same manner as described in Section~\ref{sec3_1}), which are hard band X-ray detected sources listed in the Chandra catalog (brighter than the detection limit of $f_{\mathrm{2-10keV}}=3.1\times10^{15}$~erg~s$^{-1}$~cm$^{-2}$ \citealt{civ16}), have counterparts in our HSC parent sample.
Among them, 284 are variable objects, and the other 459 objects are non-variable.
We then calculate the non-variable fraction in each hard band X-ray luminosity bin of $\log(L_{\mathrm{2-10keV}}/(\mathrm{erg~s^{-1}}))=42.5-43.0$, $43.0-43.5$, $43.5-44.0$, and $44.0-44.5$.
The results are indicated by the blue dashed line in Figure~\ref{fig14}.
As described in the Section~\ref{sec2_3_2}, our variability selection method misses about $18\%$ of the known BLAGNs.
To compensate for the number of the type-I AGNs not selected by our variability selection method, we calculate the fraction $f_{\mathrm{corr}}$, defined as the fraction of non-variable BLAGNs among the known BLAGNs in each luminosity bin, and multiply the number of the variable objects in each luminosity bin by a factor of $100/(100-f_{\mathrm{corr}})$.
The corrected covering factors are shown as blue points with error bars in Figure~\ref{fig14}.
The error bars are calculated based on the Poisson statistics of the number of samples.

Figure~\ref{fig14} shows that our results have a similar trend but slightly lower values compared to the results of \citet{mer14} who select optically unobscured AGNs by spectroscopic detection of broad emission lines (FWHM$\ \geq\ $2000~km~s$^{-1}$) or by SED fitting (see also \citealt{sal11}).
This difference indicates that the our variability-based AGN selection method can select unobscured AGNs more efficiently than the optical spectroscopic or photometric AGN selection method.
We note that low black hole mass BLAGNs can have broad emission lines with a line width of $\mathrm{FWHM}<2000$~km~s$^{-1}$ and such objects can be misclassified as optically obscured type-II AGNs in optical spectroscopy-based classification (see Section~\ref{sec1}).
The variability selection is independent of the broad line width distribution of the unobscured AGNs and can select these narrow line unobscured AGNs; thus, we obtain a lower covering factor.

Figure~\ref{fig14} also shows that our results have a similar luminosity dependence to that of the X-ray absorbed fraction \citep[Swift/BAT survey; see][]{ric17a, ric17b, ich19}, the classification of the X-ray absorbed sources of which is based on gas column densities from X-ray spectral fitting.
The similar luminosity dependences indicate that the optical absorber and the X-ray absorber are regulated by the same physical mechanism, and the geometry of these absorbers gradually changes with increasing AGN luminosity.
\citet{ric17b} suggest that the main physical mechanism that regulates the covering factor of the X-ray absorption material is the Eddington ratio, and the radiation pressure affects the dusty gas.
Although it is not clear that our sample with higher luminosities has a higher Eddington ratio compared to the sample with low luminosities (because we do not know the black hole masses), our results suggest that the dust covering factor responsible for the optical absorption is also regulated by the radiation pressure.

\subsection{Comparisons with Mid-infrared Color-based AGN Selection} \label{sec3_4}

\begin{figure*}[htbp!]
\plottwo{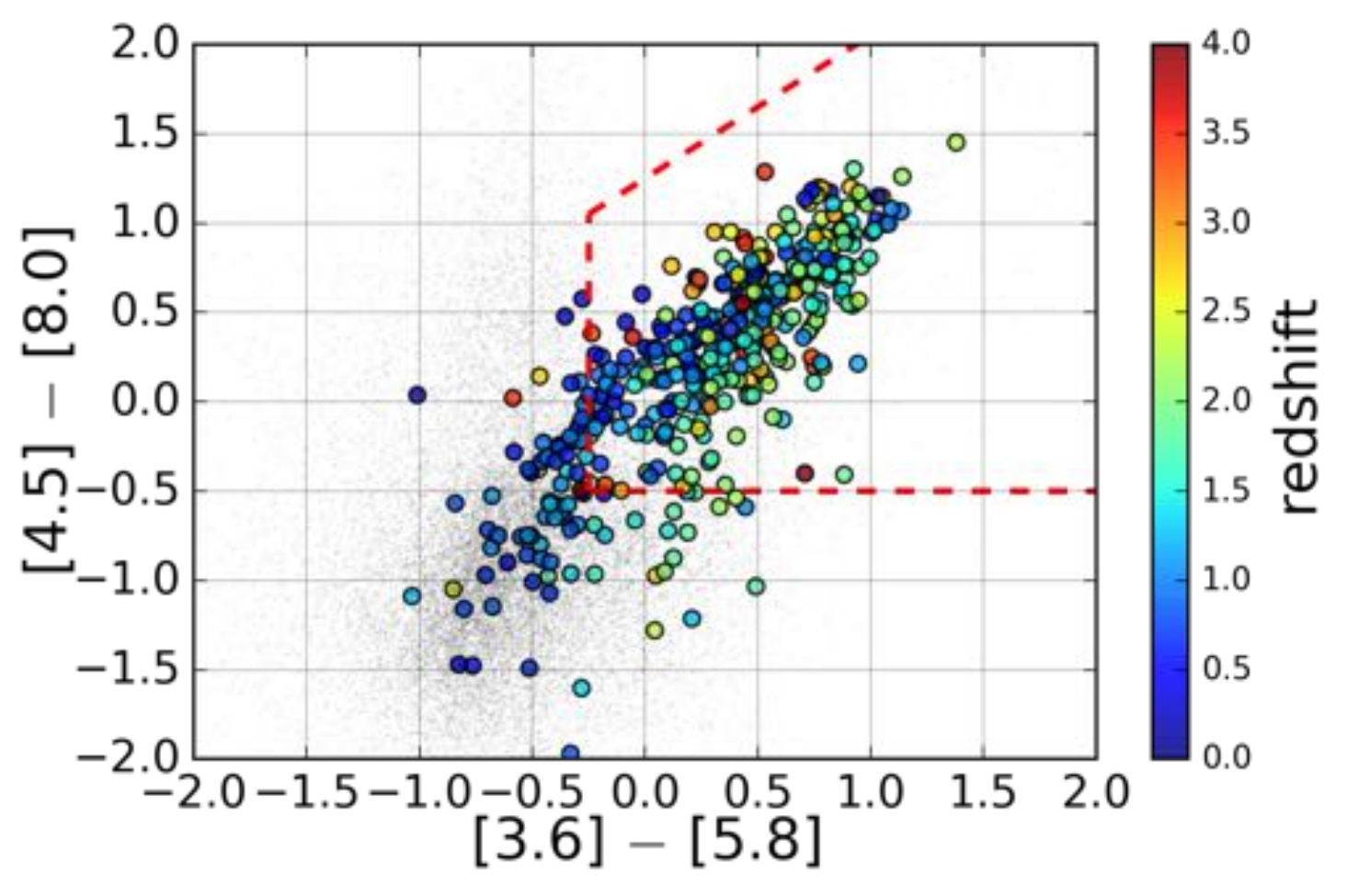}{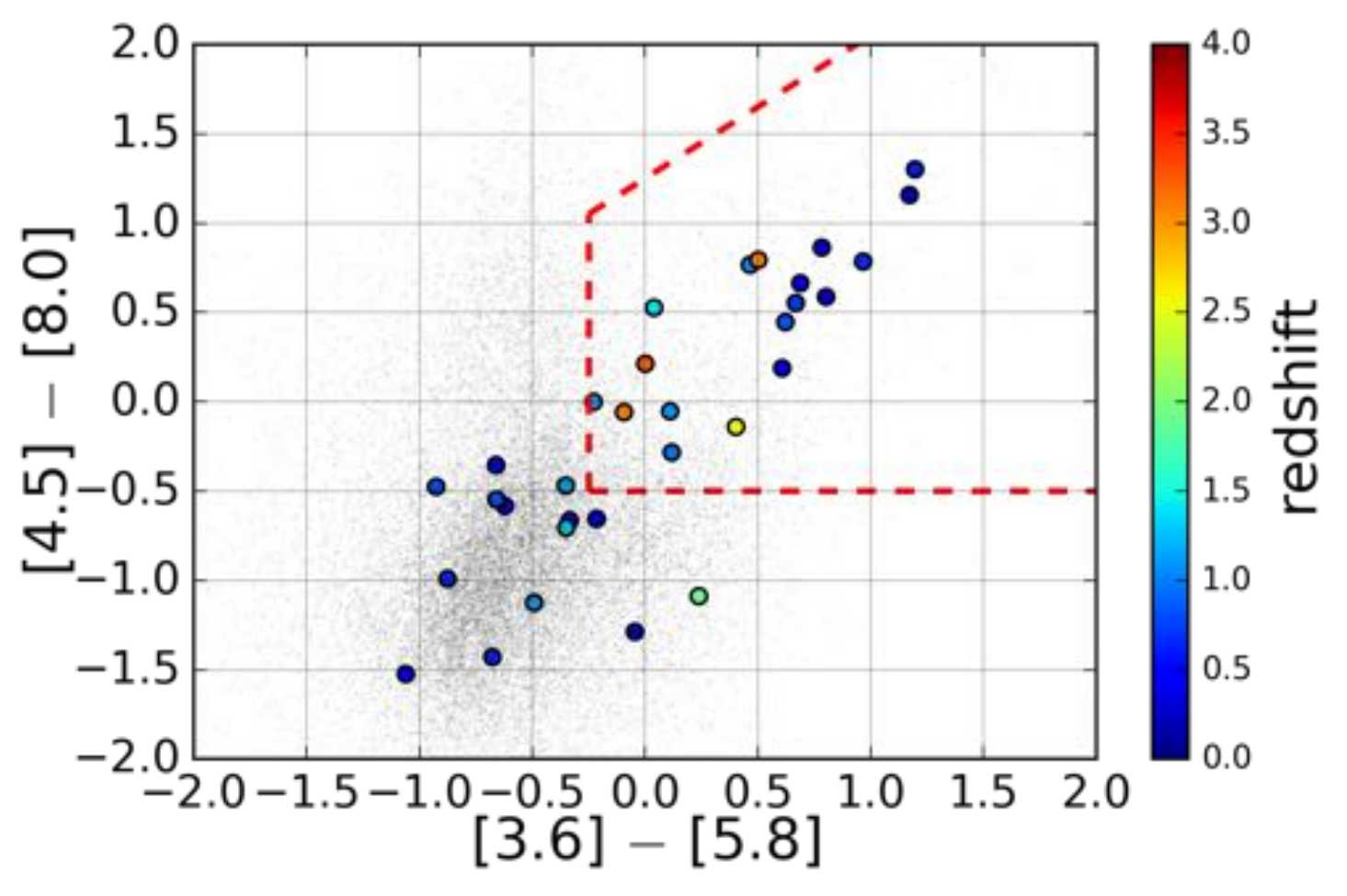}
\plottwo{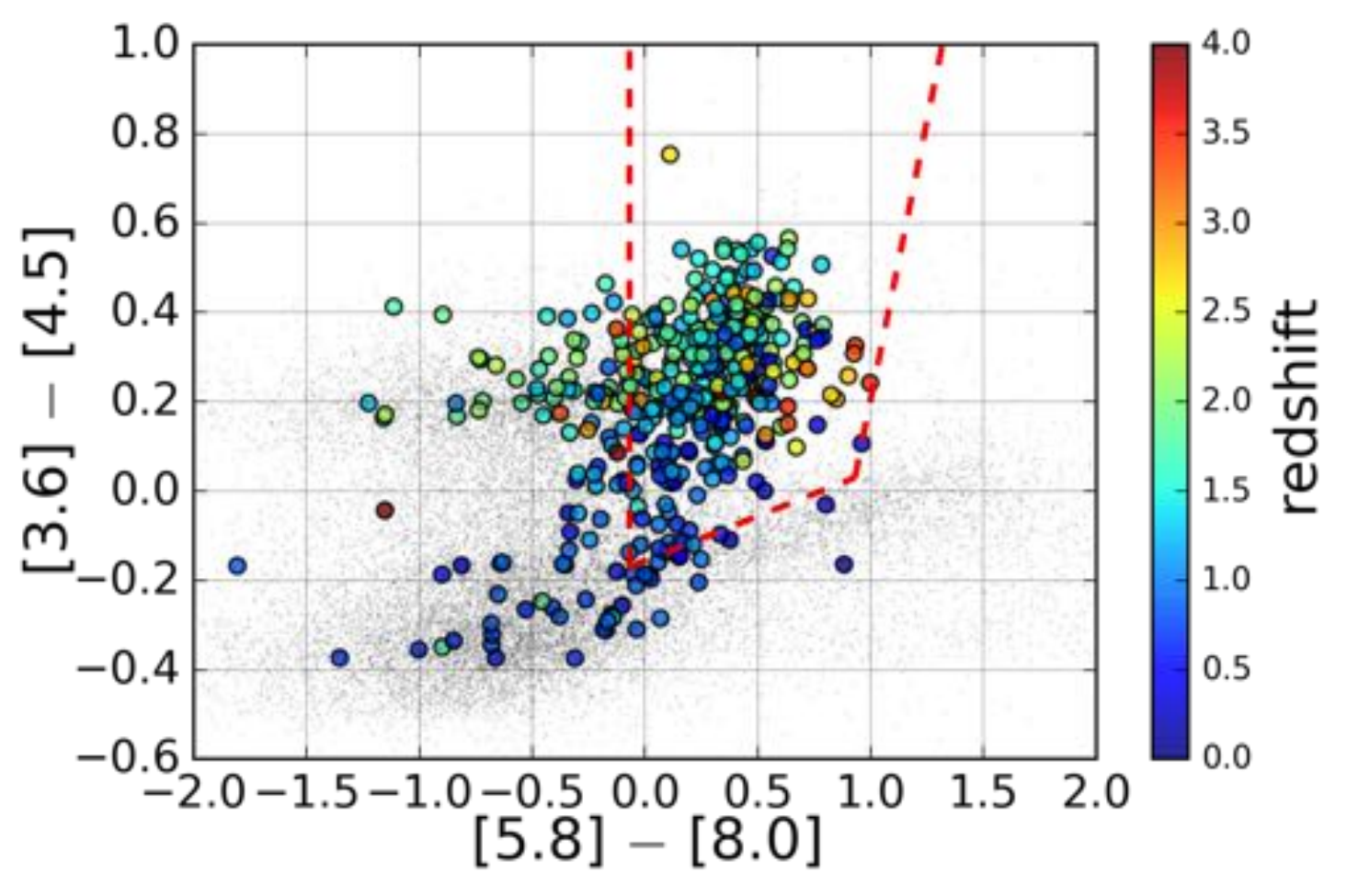}{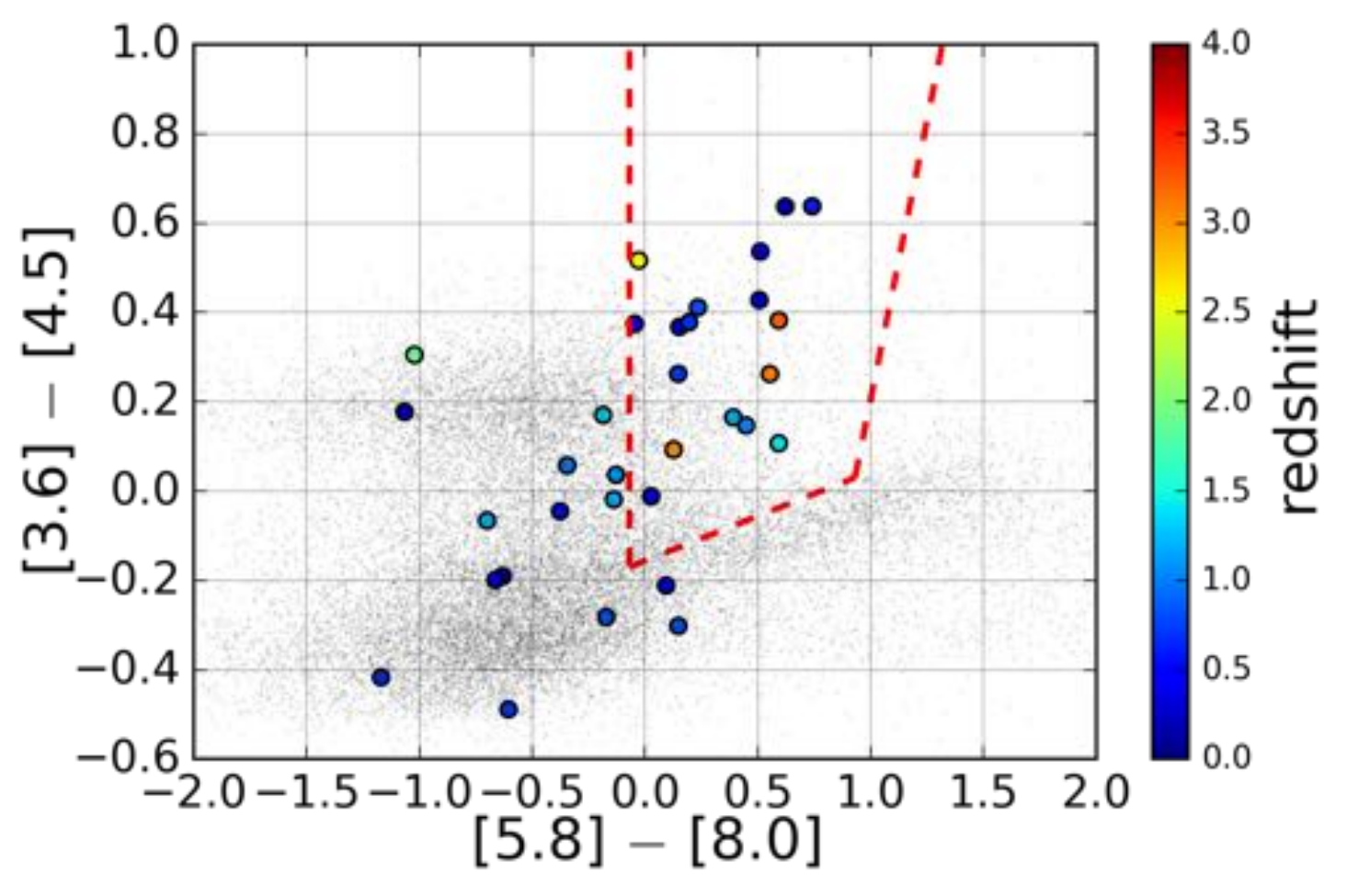}
\caption{
MIR color-color magnitude diagrams from \cite{lac07} (top panels; $3.6-5.8~\mu$m vs. $4.5-8.0~\mu$m) and \cite{ste05} (bottom panels; $5.8-8.0~\mu$m vs. $3.6-4.5~\mu$m).
The X-det (X-undet) sample in our variability-selected AGNs is shown in the left (right) panels.
The small gray dots are the parent sample.
The wedge of Lacy et al. and the wedge of Stern et al. are shown by the red dashed lines.
\label{fig15}
}
\end{figure*}

In AGNs, radiation from the hot/warm dust produces a MIR bump in the SED, while in normal galaxies, stellar continuum (e.g., $1.6~\mu$m bump), PAH emissions, and some warm dust radiation heated by the star-forming regions are the dominant components in the MIR wavelength range.
The difference of the MIR SED shape between AGNs and non-AGNs can be clearly seen in the MIR color-color space.

\citet{lac04,lac07} propose a MIR color method used to identify AGNs by using four channel data ($3.6$, $4.5$, $5.8$, and $8.0~\mu$m for ch1, ch2, ch3, and ch4, respectively) of the Infrared Array Camera \citep[IRAC; ][]{faz04} of the Spitzer Space Telescope.
We first investigate whether our variability-selected AGN sample satisfies this MIR color criteria expected for AGNs or not.
The top panels in Figure~\ref{fig15} show the MIR color-color diagram for our variability-selected AGN samples (left panel: X-det sample, right panel: X-undet sample). 
Here, we plot objects with statistically significant MIR detections \citep[large circles; $S/N\geq3$, corresponding to magnitude limits for 3\arcsec aperture photometry of 25.5, 25.5, 23.0, and 22.9~mag for Spitzer/IRAC ch1, ch2, ch3, and ch4, respectively;][]{lai16}.
Four hundred thirteen out of 441 candidates from the X-det sample and 32 out of 50 candidates from the X-undet sample are plotted in the figure.
It is clear that more than half of our samples satisfy Lacy's selection criteria, but a fraction of objects is located outside of the AGN wedge.
Ninety objects in the X-det sample (21.8\%) and 14 objects in the X-undet sample (43.8\%) do not satisfy the criteria of Lacy et al.
The MIR color of the variability-selected AGNs outside the AGN wedge is consistent with that of normal galaxies, suggesting that their MIR SEDs are dominated by the host galaxy emissions.

\citet{ste05} define another MIR color-color selection criterion to select AGNs, and we plot this in the bottom panels of Figure~\ref{fig15}. 
In this diagnostics, 101 objects in the X-det sample (24.5\%) and 13 objects in the X-undet sample (40.6\%) do not satisfy the criteria of Stern et al.
These results confirm that optical variability-based AGN selection is indeed a complementary tool to identify AGNs.

\section{Structure Function Analysis} \label{sec4}

In this section, we investigate the variability properties of our variability-selected AGNs.
Previous quasar studies show that the optical variability amplitude depends on the AGN luminosity, rest-frame wavelength, and  rest-frame time interval, i.e., the variability amplitude is larger at a lower luminosity, shorter wavelength, and longer time interval \citep[e.g.,][]{van04}.
However, it is unclear whether this trend still holds or not for low-luminosity AGNs ($L_{\mathrm{bol}} < 10^{45}$~erg~s$^{-1}$).
Here, we study the variability properties of the low-luminosity variability-selected AGNs by the structure function analysis. 
Since the light-curve sampling is limited (Table~\ref{tbl1}), in this paper, we study their ensemble structure functions, which represent typical variability amplitudes of the sample as a function of the rest-frame time interval.

\begin{figure}[t!]
\plotone{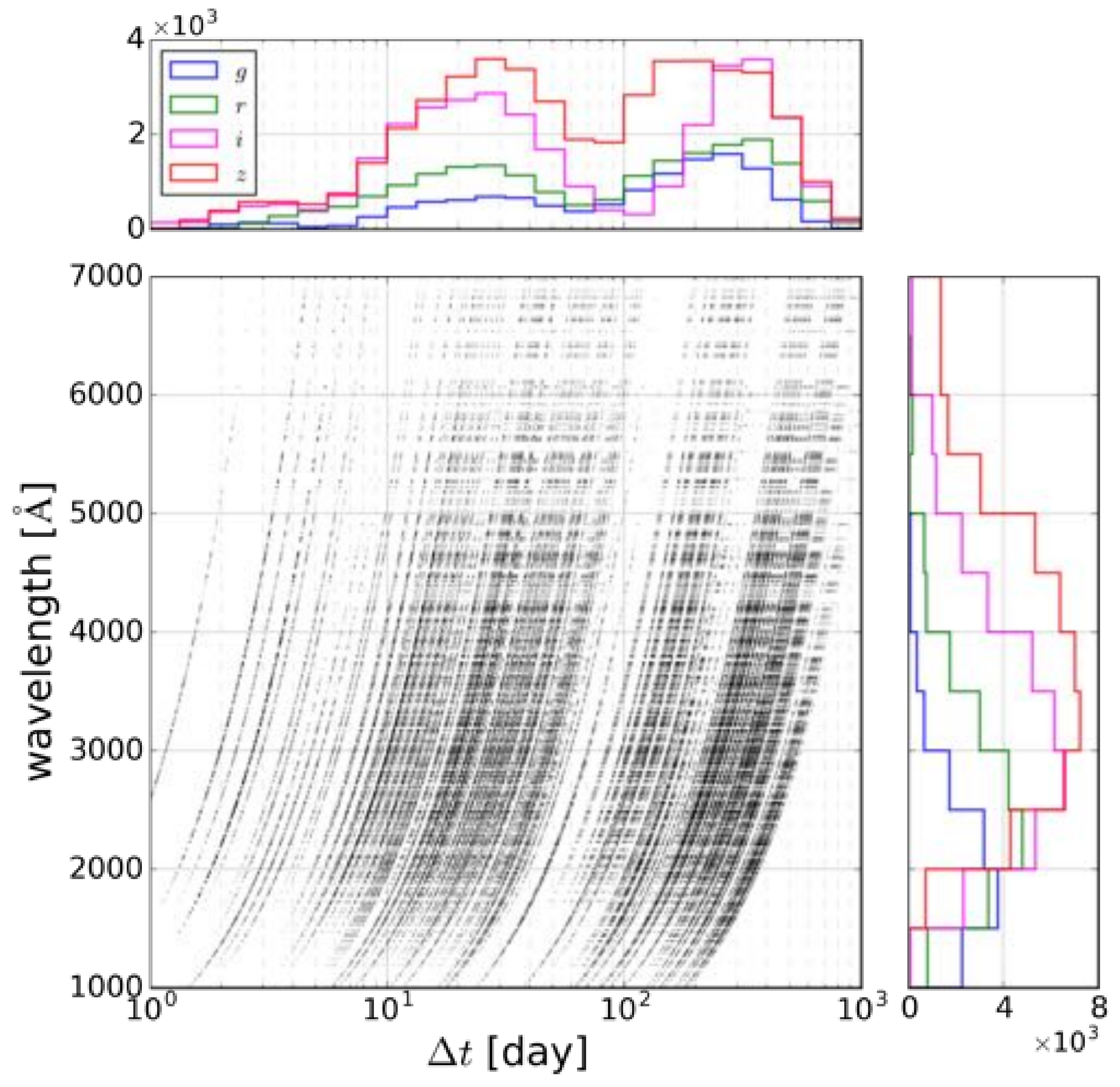}
\caption{
Data points of time interval and wavelength for the X-det sample.
The histogram for each parameter is shown on each side (blue, green, magenta, and red histograms correspond to the $g$, $r$, $i$, and $z$ bands, respectively).
\label{fig16}
}
\end{figure}

In the ensemble structure function analysis, we use the X-det sample (441 objects in our variability-selected AGNs, see Section~\ref{sec3_1}), which has redshift ($72\%$ of them have spectroscopic redshift and the other 28\% have photometric redshift) and bolometric luminosity information.
To decrease the photometric noise and the effect of host galaxy light (especially for extended sources) in the structure function analysis, we use the magnitude that was calculated with the PSF size aperture after all PSFs are matched to the worst size in each filter.
The overall rest-frame time interval and rest-frame wavelength coverages of the sample are shown in Figure~\ref{fig16}.
The rest-frame time interval $\Delta t$ is calculated from the observed-frame time interval $\Delta t_{\mathrm{obs}}$ and the redshift $z$ with $\Delta t = \Delta t_{\mathrm{obs}}/(1+z)$, and the rest-frame wavelength $\lambda$ is calculated from the effective wavelength of the HSC filters ($\lambda_{\mathrm{eff}}=4816$, $6264$, $7740$, and $9125~\mbox{\AA}$ for the $g$, $r$, $i$, and $z$ bands, respectively) with $\lambda=\lambda_{\mathrm{eff}}/(1+z)$.
Hereafter, we refer to rest-frame time interval and rest-frame wavelength as time interval and wavelength, respectively, unless otherwise noted.

\subsection{Ensemble Structure Function}\label{sec4_1}

\begin{figure}[t!]
\plotone{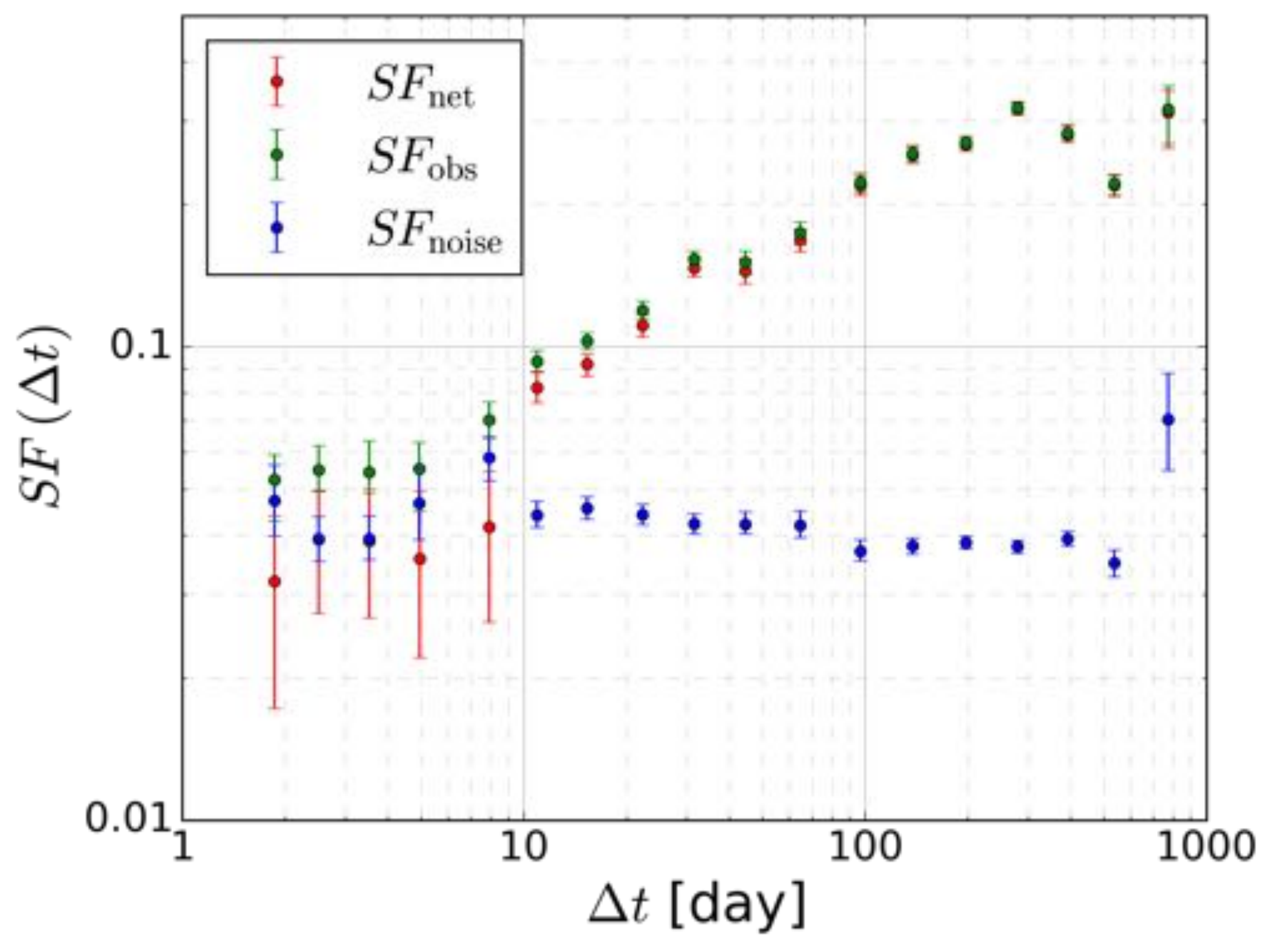}
\caption{
The $g$ band structure function for the X-det sample.
The red points are the net SF values and the green points are the observed SF values.
The photometric noise SF values are plotted as the blue points.
\label{fig17}
}
\end{figure}

\begin{figure}[t!]
\plotone{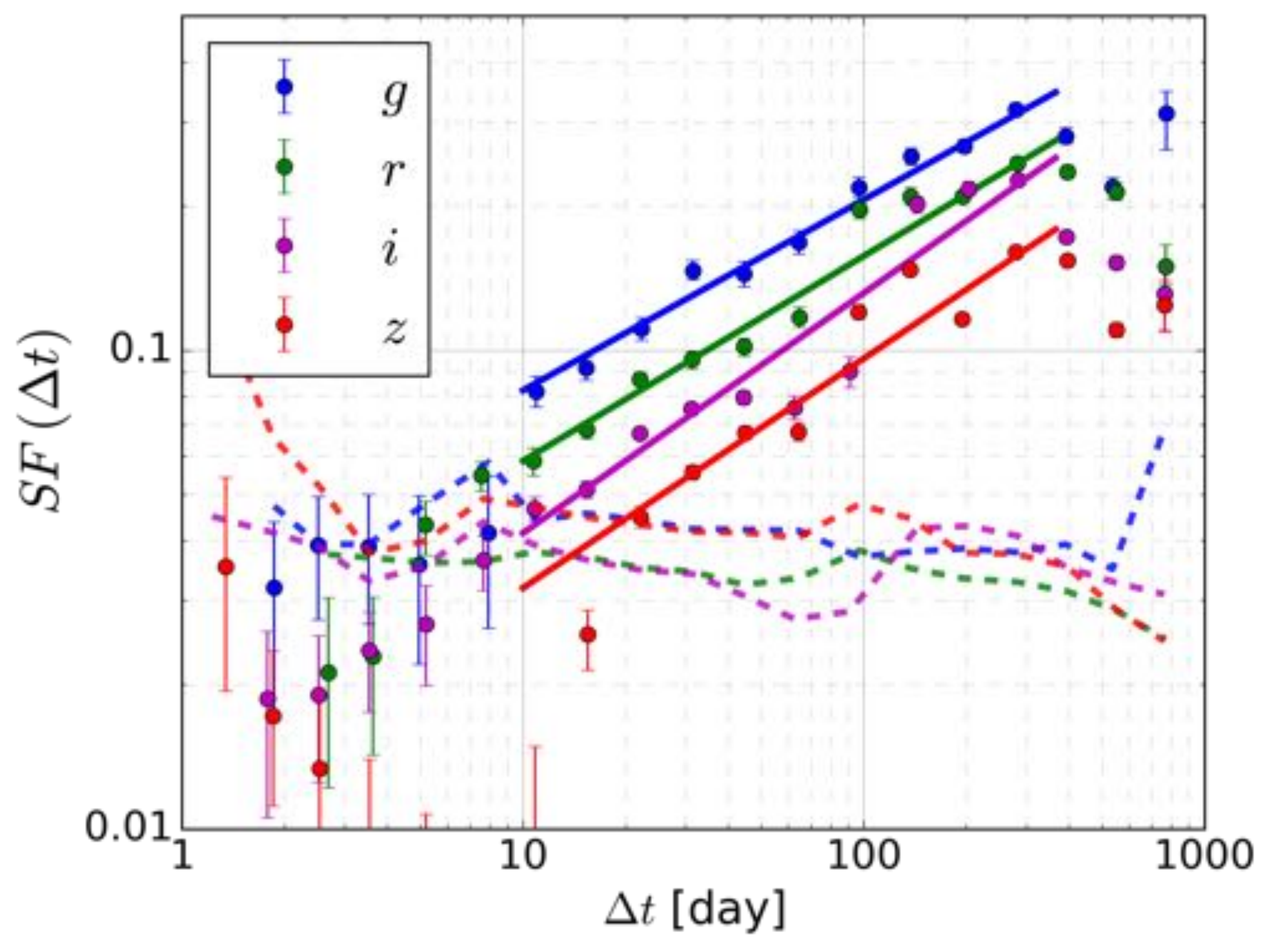}
\caption{
Net SF values (SF$_{\mathrm{net}}$) for the X-det sample in the $g$ (blue), $r$ (green), $i$ (magenta), and $z$ (red) bands.
The solid lines show the results of the best-fit models, and the dashed lines show the photometric noise SF values (SF$_{\mathrm{noise}}$) in each filter.
\label{fig18}
}
\end{figure}

The structure function (SF) is a useful tool to examine the variability properties of AGNs \citep{van04, mac12, koz16, cap17}.
The SF represents the root-mean-square (rms) of the magnitude differences $\Delta m$ of a sample in a given time interval $\Delta t$ bin, i.e., typical variability amplitude at $\Delta t$.

Practically, the SF can be calculated as
\begin{eqnarray}
{\mathrm{SF}}_{\mathrm{obs}}(\Delta t) = 0.741 \times \mathrm{IQR}
\label{eq10}
\end{eqnarray}
where IQR is the interquartile range between 25\% and 75\% of the sorted $\Delta m$ distribution in each $\Delta t$ bin, and the coefficient 0.741 is the conversion factor from the IQR to the standard deviation for a Gauss distribution.
This equation is useful since it is relatively insensitive to photometric outliers and also to the case that the distribution of $\Delta m$ is non-Gaussian \citep{mac12, koz16}; therefore, we use this equation to calculate SF in our analysis.

Since the observed SF is affected by the photometric noise, we should correct for this effect to recover the net AGN variable amplitude.
The net SF (SF$_{\mathrm{net}}$) is calculated as follows \citep{koz16}:
\begin{eqnarray}
{\mathrm{SF}}_{\mathrm{net}}(\Delta t)=\sqrt{{\mathrm{SF}}_{\mathrm{obs}}^2(\Delta t) - {\mathrm{SF}}_{\mathrm{noise}}^2(\Delta t)}.
\label{eq11}
\end{eqnarray}
To estimate the photometric noise term SF$_{\mathrm{noise}}$, we construct a control sample, which is randomly selected from the $non$-$Var$ sample, with the same distributions of magnitude and time interval as for the X-det sample, and calculate the SF from this control sample in each $\Delta t$ bin.
The error bars of SF$_{\mathrm{net}}$, SF$_{\mathrm{obs}}$, and SF$_{\mathrm{noise}}$ are estimated by a bootstrap method as follows: 
(i) For each $\Delta t$ bin, we randomly take a sample (bootstrap sample) from the original sample with the same sample size to calculate SF$_{\mathrm{obs}}$.
(ii) We also randomly select a control sample from the $non$-$Var$ sample, which has the same distributions of magnitude and time interval as for the bootstrap sample, to calculate SF$_{\mathrm{noise}}$.
(iii) Using the SF$_{\mathrm{obs}}$ and SF$_{\mathrm{noise}}$, we then calculate SF$_{\mathrm{net}}$ from Equation~(\ref{eq11}).
(iv) The processes from (i) to (iii) are conducted 1000 times and finally we calculate the scatter of the SF$_{\mathrm{obs}}$, SF$_{\mathrm{noise}}$, and SF$_{\mathrm{net}}$ in each $\Delta t$ bin.
The observed, noise, and net SF values for the $g$ band are shown in Figure~\ref{fig17}.
It is found that SF$_{\mathrm{noise}}$ is negligible at large $\Delta t$ but comparable with SF$_{\mathrm{net}}$ at $\Delta t < 10$~days.
Hereafter, we refer to the net SF as the SF unless otherwise noted.

The SF can reasonably be fitted with a power-law function,
\begin{eqnarray}
{\mathrm{SF}}(\Delta t) = {\mathrm{SF}}_{0} \left(\frac{\Delta t}{\Delta t_0}\right)^{b_{t}},
\label{eq12}
\end{eqnarray}
where SF$_{0}$ is the value of SF at a $\Delta t_0$ days time interval and $b_{t}$ is the slope of the SF$_{\mathrm{net}}$.
Here, we set $\Delta t_0$ to be $100$~days.
In this fitting, we only use the SF$_{\mathrm{net}}$ data points between $\Delta t = 10$~days and $1$~yr.
We also fit the same function to the bootstrap resampled data set for all of the iterations to estimate fitting uncertainties on the two free parameters.
Figure~\ref{fig18} shows the best-fit results for each filter.
The variability amplitudes at 100 days are SF$_{0}=0.210\pm0.003$, $0.160\pm0.002$, $0.133\pm0.001$, and $0.097\pm0.001$ for the $g$, $r$, $i$, and $z$ bands, respectively, and the power-law slopes are $b_{t}=0.411\pm0.013$, $0.440\pm0.012$, $0.511\pm0.010$, and $0.492\pm0.013$ for the $g$, $r$, $i$, and $z$ bands, respectively.

A special functional form is often assumed to explain the AGN SFs in the previous studies, following a prediction from the dumped random walk (DRW) model described as SF$(\Delta t) = {\mathrm{SF}}_{\infty}[1-\exp(-\Delta t/\tau)]^{1/2}$ \citep{kel09}.
The dumping time scale $\tau$ may be related to some physical parameters, such as black hole mass, and is typically on the order of hundreds of days \citep{mac10, mac12}.
Our SF$_{\mathrm{net}}$ in Figure~\ref{fig18} also suggests dumping around 1 yr for all four bands, but this may be due to the insufficient data sampling at $\Delta t > 1$~yr.
The SFs at such long timescales generally show unexpected breaks or wiggles due to insufficient data sampling \citep{emm10}.
Although we cannot exclude the possibility that we are detecting true dumping signatures, we do not use the DRW model fitting in this paper.

\subsection{Variability Amplitude Dependences on Physical Parameters}\label{sec4_2}

In the previous quasar studies, it is suggested that the variability amplitude of AGNs mainly depends on wavelength and AGN luminosity \citep[e.g.,][]{van04}.
It is shown that the dependences on redshift, Eddington ratio, and black hole mass are weaker than those on luminosity and wavelength dependences \citep[e.g.,][]{cap17}. 
If we make a tentative assumption that the $\Delta t$ dependence of the variability amplitude is independent of the dependences on the wavelength and luminosity, we can express the SF as follows:
\begin{eqnarray}
{\mathrm{SF}}(\Delta t, \lambda, L_{\mathrm{bol}}) &=& {\mathrm{SF}}_{0}(\lambda, L_{\mathrm{bol}}) \left(\frac{\Delta t}{\Delta t_0}\right)^{b_{t}}
\label{eq13}
\end{eqnarray}
where $\lambda$ is wavelength and $L_{\mathrm{bol}}$ is the AGN bolometric luminosity.
As the SF has units of magnitude, it is natural to express the SF$_0$ with the following form:
\begin{gather}
{\mathrm{SF}}_{0}(\lambda, L_{\mathrm{bol}}) = \ -2.5\log V_{0}(\lambda, L_{\mathrm{bol}}) \label{eq14}\\
V_{0}(\lambda, L_{\mathrm{bol}}) \propto V_{1}(\lambda)\ V_{2}(L_{\mathrm{bol}}),\label{eq15}
\end{gather}
where $V_{1}(\lambda)$ and $V_{2}(L_{\mathrm{bol}})$ denote the dependencies of the SF on wavelength and luminosity, respectively.
It is noted that the SF$_{0}$ is the SF normalized at $\Delta t_0$ and we have set the $\Delta t_0$ to be 100~days. 

\begin{figure}[t!]
\plotone{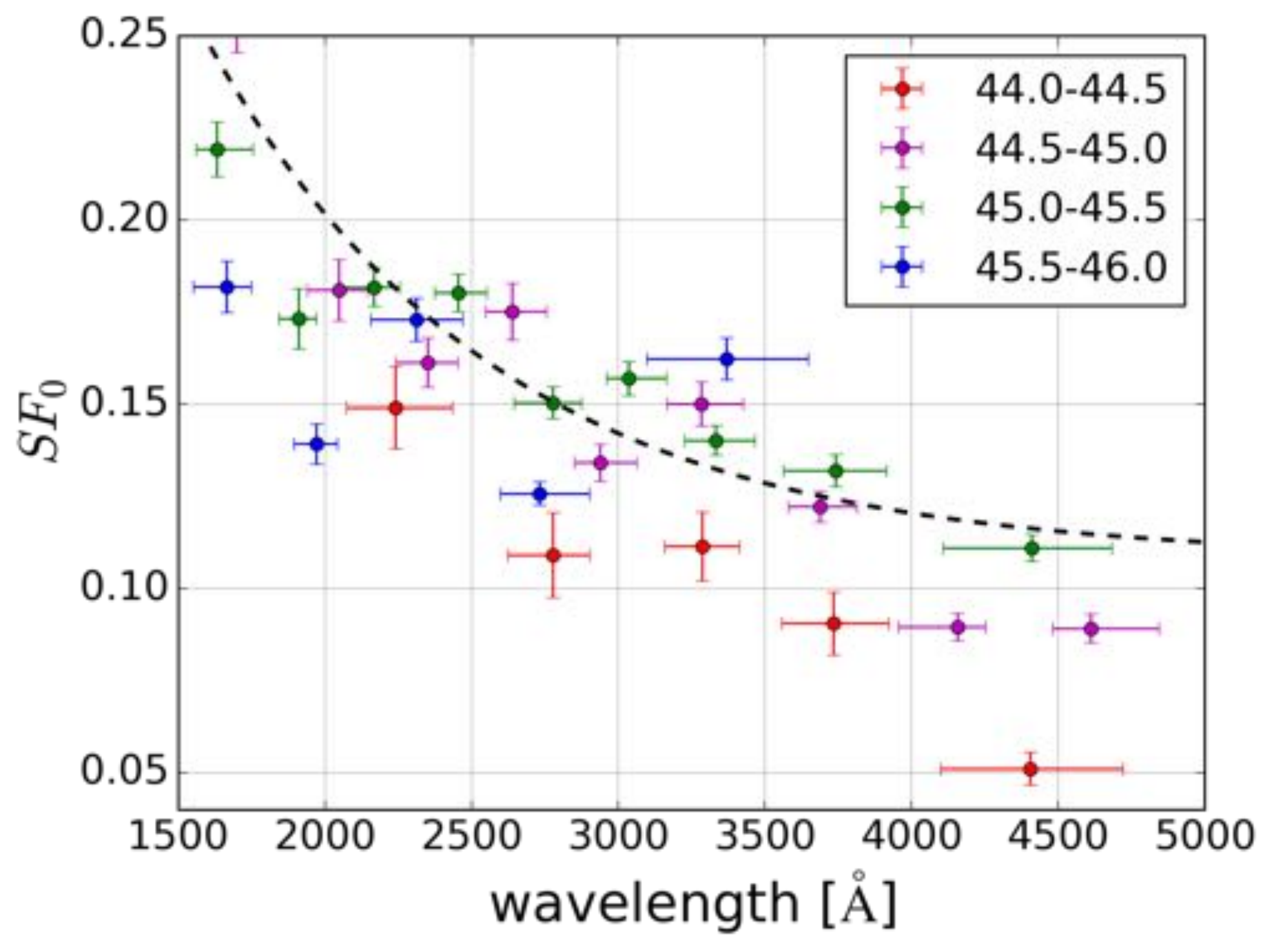}
\plotone{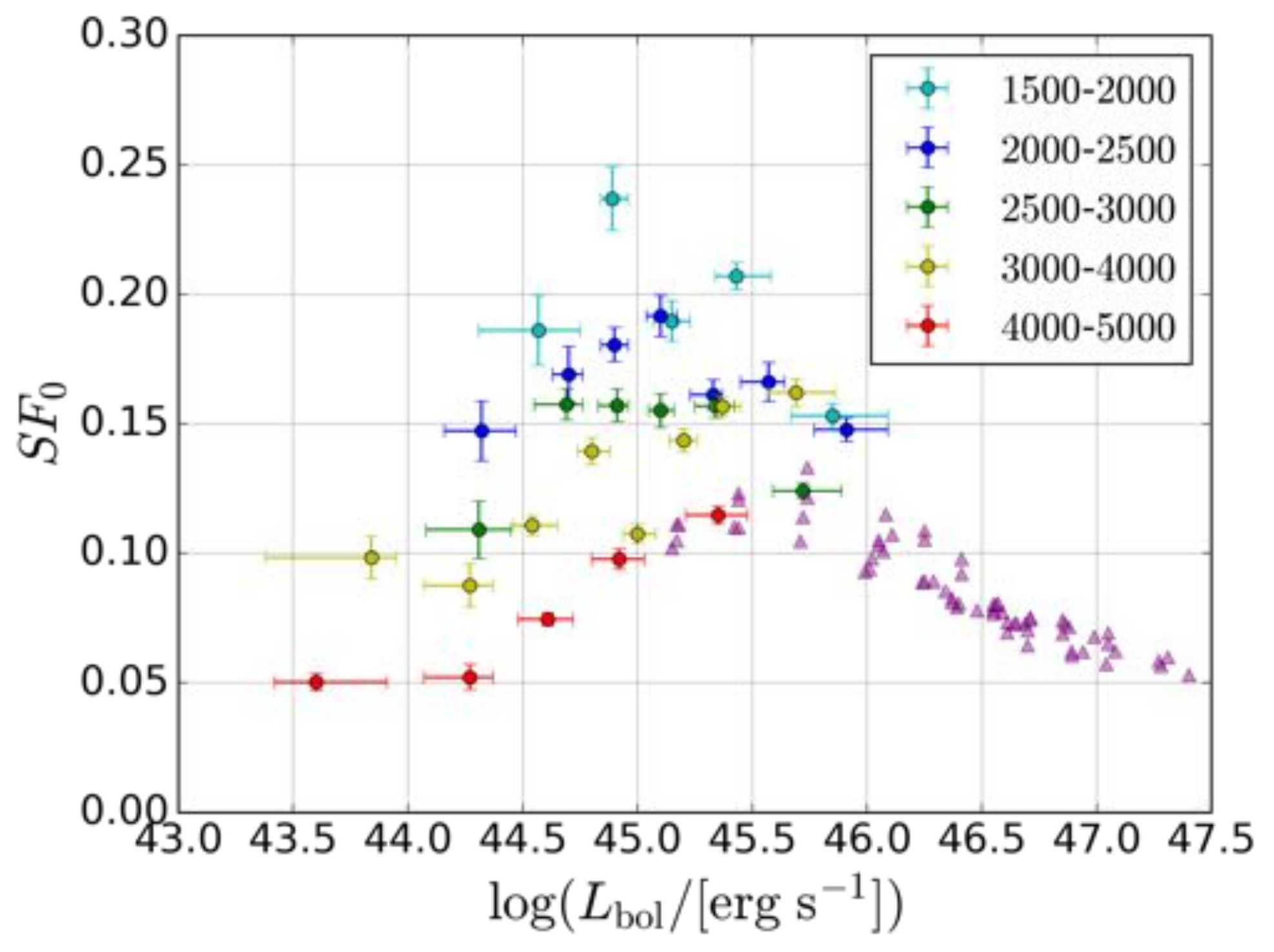}
\caption{
Variability amplitude SF$_{0}$ as a function of wavelength (top panel) and AGN bolometric luminosity (bottom panel).
(Top panel) The dashed line is the result of the previous quasar study \citep[Equation 11 in][ and scaled]{van04}.
(Bottom panel) The magenta triangles are the results of \citet{cap17} normalized at $3000~\mbox{\AA}$.
\label{fig19}
}
\end{figure}

\begin{figure*}[t!]
\gridline{
\fig{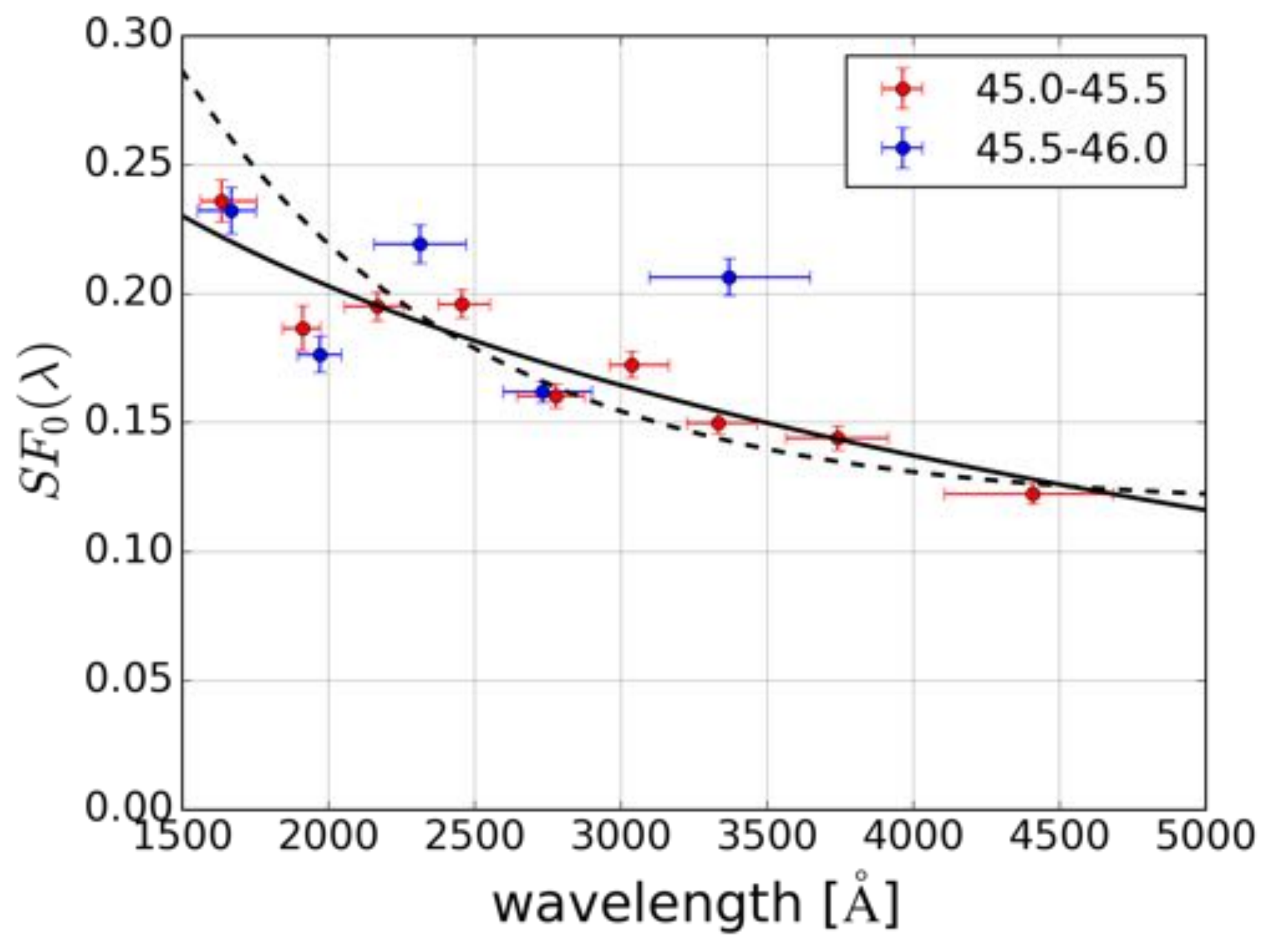}{0.33\textwidth}{(a) wavelength dependence}
\fig{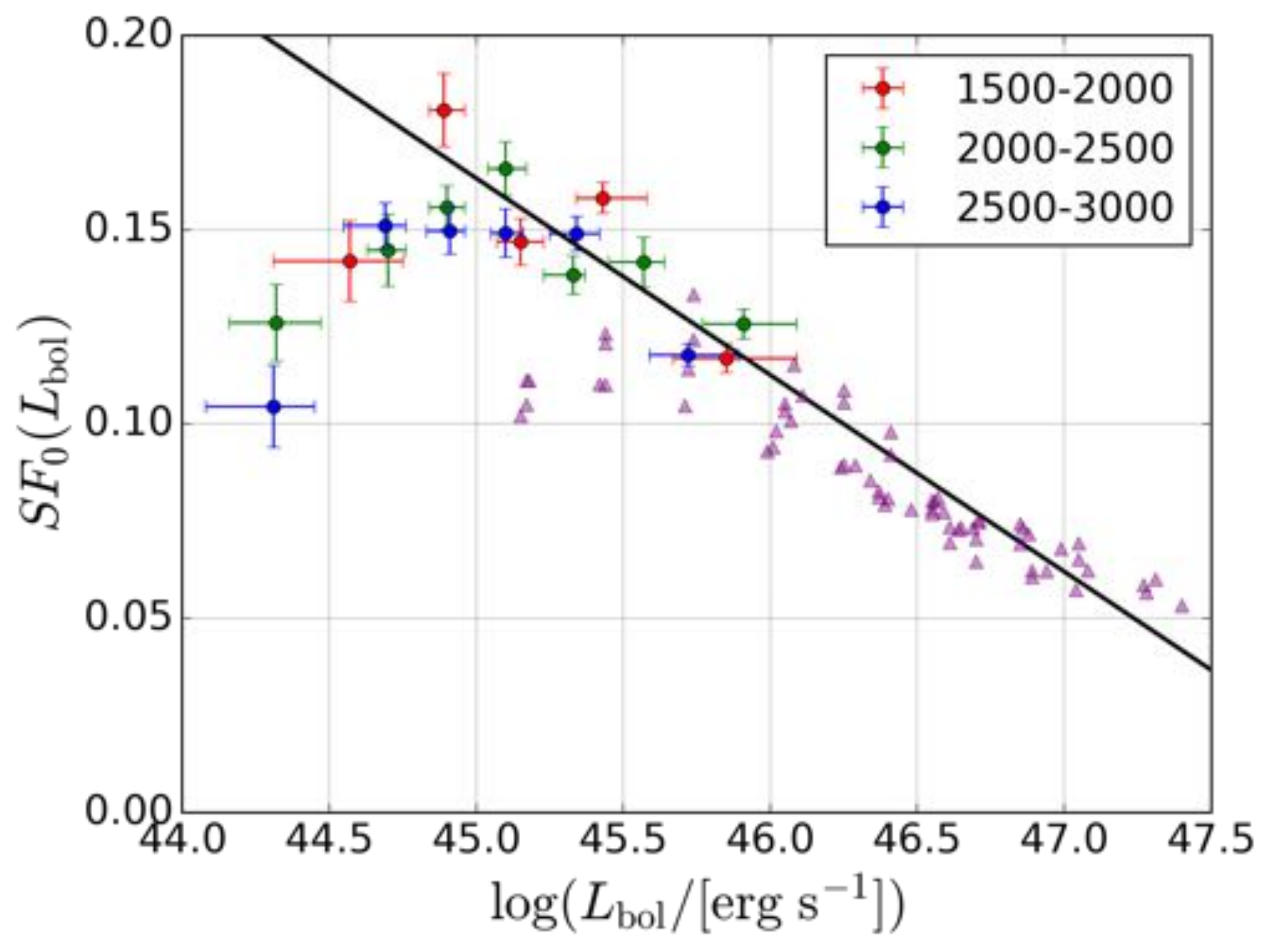}{0.33\textwidth}{(b) luminosity dependence}
\fig{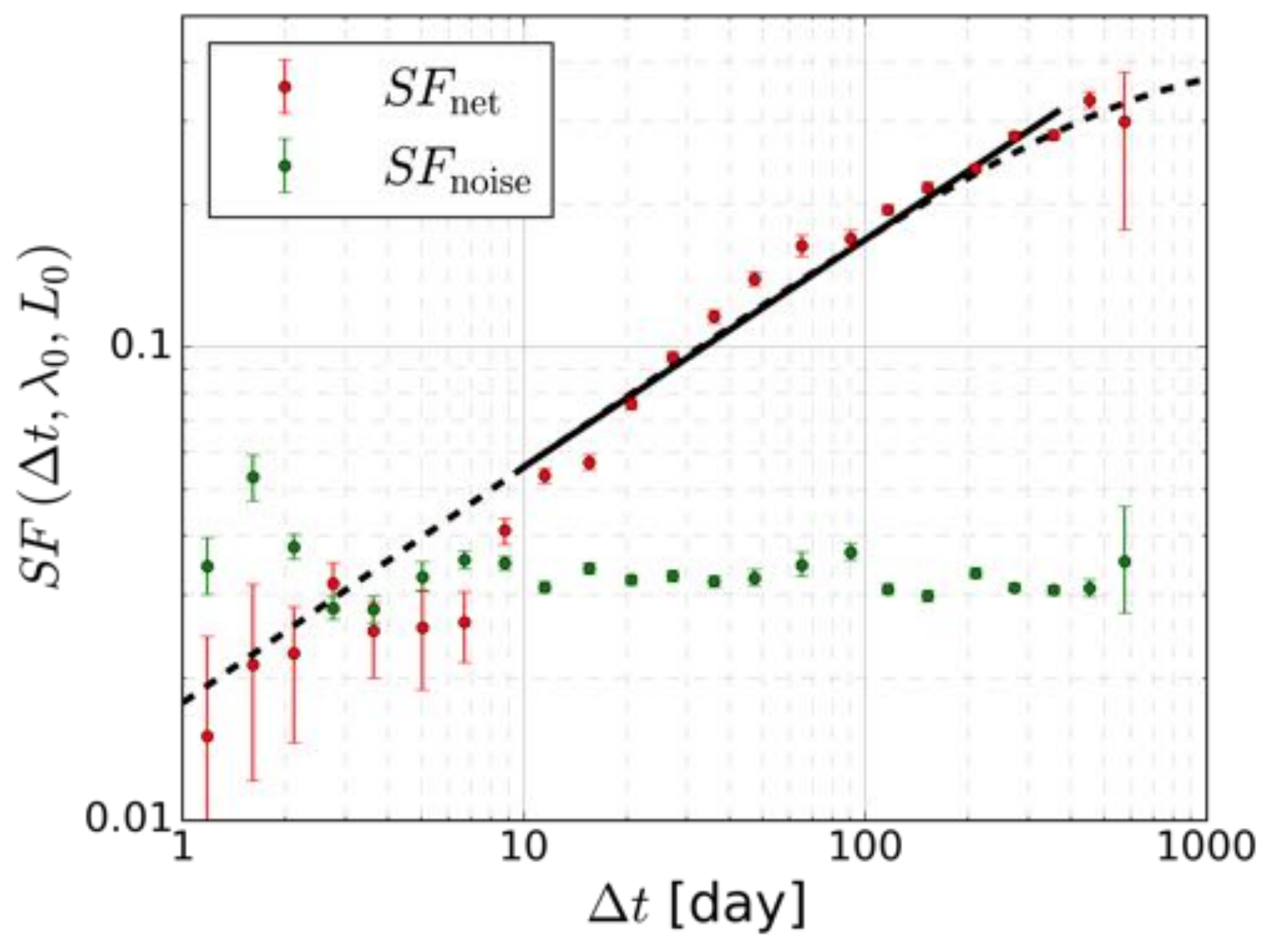}{0.33\textwidth}{(c) time interval dependence}
}
\caption{
Variability amplitude dependences of (a) wavelength, (b) AGN bolometric luminosity, and (c) time interval.
The variability amplitude is normalized at (a) $L_0=10^{45}$~erg~s$^{-1}$, (b) $\lambda_0=3000~\mbox{\AA}$, and (c) $L_0$ and $\lambda_0$, respectively.
The solid lines in the panels are the best-fit results with the models.
The dashed line in the left panel and the magenta triangles in the middle panel are the same as Figure~\ref{fig19}.
The dashed line in the right panel is the DRW model prediction, which is the case of $\tau=500$~days, scaled at 100~days.
\label{fig20}
}
\end{figure*}

We discuss the dependences of the SF on wavelength and luminosity, namely the functions represented by $V_{1}(\lambda)$ and $V_{2}(L_{\mathrm{bol}})$.
We divide the X-det sample into several luminosity and wavelength bins:
(i) luminosity bins of $\log(L_{\mathrm{bol}}$/(erg~s$^{-1}))=44.0-44.5$, $44.5-45.0$, $45.0-45.5$, and $45.5-46.0$, and (ii) wavelength bins of $\lambda=1500-2000$, $2000-2500$, $2500-3000$, $3000-4000$, and $4000-5000~{\mbox{\AA}}$.
We further divide the sample in each bin into the subsamples by the other quantity (i.e., wavelength for (i) and luminosity for (ii)).
We then calculate the SF in the same manner as described in the previous subsection and obtain the variability amplitude at $\Delta t = 100$~days (i.e., SF$_{0}$) by a power-law fitting with Equation~(\ref{eq12}).
The results are shown in Figure~\ref{fig19}; the top panel shows the wavelength dependence, and the bottom panel shows luminosity dependence.
In the top panel of Figure~\ref{fig19}, an empirical relationship between the variability amplitude and wavelength for SDSS quasars \citep[Equation (11) in][]{van04} is compared with our results.
It is clear that the variability amplitudes of the X-det sample show similar wavelength dependences as those of the SDSS quasars; the larger variability amplitude is observed at the shorter wavelength.
On the other hand, we find complex luminosity dependences of the variability amplitudes especially at $L_{\mathrm{bol}}\lesssim10^{45}$~erg~s$^{-1}$, while previous studies for quasar samples (limited to $L_{\mathrm{bol}} > 10^{45}$~erg~s$^{-1}$) show a monotonic increase of the variability amplitude with decreasing AGN luminosity \citep[$\propto L_{\mathrm{bol}}^{-0.5}$; e.g.,][]{cap17}.
The luminosity dependence of the variability amplitude is clearer in the longer wavelength samples.
The decrease of the variability amplitude for low-luminosity AGNs is barely seen in the SF$_0$ at $\lambda \sim 5000~\mbox{\AA}$ for the quasar samples presented in \citet{cap17} as shown in Figure~\ref{fig19} ($L_{\mathrm{bol}}\sim10^{45.0-45.5}$~erg~s$^{-1}$).
This luminosity dependence can naturally been explained by the larger contamination of the host galaxy light for the lower-luminosity AGNs.
In fact, \citet{she11} suggest that the contamination of host galaxy light becomes significant in the low-luminosity AGN optical spectra ($L_{\mathrm{bol}}\lesssim10^{46}$~erg~s$^{-1}$).
Since the AGN accretion disk emission generally has a blue UV$-$optical SED, the contribution of host galaxy light is relatively larger in the longer wavelengths.
The properties of the AGN host galaxy light inferred from the analysis of the luminosity dependence of the multiband variability amplitude is discussed in Section~\ref{sec5_1} in detail.

To understand the intrinsic AGN variability properties eliminating the contamination from the host galaxy light, we use the bright and short-wavelength samples where the host galaxy flux contribution can be negligible.
We calculate the intrinsic AGN dependencies of wavelength and luminosity with the following procedure.
Step (i): We use the subsamples in the two luminosity bins of $\log(L_{\mathrm{bol}}/[\mathrm{erg~s^{-1}}])=45.0-45.5$ and $45.5-46.0$ to estimate the wavelength dependence of the variability amplitude SF$_0$.
In this step, we ignore the luminosity dependence between the two luminosity bins.
Step (ii): We estimate the luminosity dependence for the subsamples in the three wavelength bins of $\lambda=1500-2000$, $2000-2500$, and $2500-3000~{\mbox{\AA}}$, after correcting the wavelength dependence.
Step (iii): We reevaluate the wavelength dependence after correcting the luminosity dependence.
Step (iv): We iterate steps (ii) and (iii) 10 times.

The dependences of wavelength and luminosity are fitted with a power-law function.
In step (i), since we ignore the luminosity dependence between the two luminosity bins, the SF$_{0}$ can be written as a function of wavelength
\begin{eqnarray}
{\mathrm{SF}}_{0}(\lambda) = -2.5\log\left[a_{\lambda}V_{1}(\lambda)\right],\ V_{1}(\lambda)= \left( \frac{\lambda}{\lambda_{0}} \right)^{b_{\lambda}},
\label{eq16}
\end{eqnarray}
where $a_{\lambda}$ is a normalization factor at the wavelength $\lambda_{0}$ and $b_{\lambda}$ is the slope of a power-law function.
We here set $\lambda_{0}$ to be $3000~\mbox{\AA}$.
The fitting with Equation~(\ref{eq16}) is conducted for all of the subsamples in the two luminosity bins.

To correct the wavelength dependence in step (ii), the following factor,
\begin{eqnarray}
C_{1}(\lambda) = \frac{{\mathrm{SF}}_{0}(\lambda_{0})}{{\mathrm{SF}}_{0}(\lambda)} = \frac{\log (a_{\lambda})}{\log \left[ a_{\lambda}V_{1}(\lambda)\right]},
\label{eq17}
\end{eqnarray}
should be applied to the SF$_{0}(\lambda, L_{\mathrm{bol}})$ in each subsample.
For simplicity, we use the median wavelength in each subsample to calculate the correction factor $C_{1}(\lambda)$.

After the SF$_0$ is normalized at $\lambda_0$, in the step (ii), the SF$_{0}$ depends on only the luminosity; thus, the luminosity dependence of SF$_0$ can be written as
\begin{eqnarray}
{\mathrm{SF}}_{0}(L_{\mathrm{bol}}) = -2.5\log\left[a_{L}V_{2}(L_{\mathrm{bol}})\right], V_{2}(L_{\mathrm{bol}})= \left( \frac{L_{\mathrm{bol}}}{L_{0}} \right)^{b_{L}}\hspace{-3mm},
\label{eq18}
\end{eqnarray}
where $a_{L}$ is a normalization factor at the $L_{0}$ (here we set $L_{0}$ to be $10^{45}$~erg~s$^{-1}$) and $b_{L}$ is the slope of a power-law function.
The fitting with Equation~(\ref{eq18}) is conducted for the subsamples with $L_{\mathrm{bol}}>10^{45}$~erg~s$^{-1}$ where the host galaxy contamination is negligible.

The correction of the luminosity dependence can be written as
\begin{eqnarray}
C_{2}(L_{\mathrm{bol}}) = \frac{{\mathrm{SF}}(L_{0})}{{\mathrm{SF}}(L_{\mathrm{bol}})} = \frac{\log (a_{L})}{\log \left[ a_{L}V_{2}(L_{\mathrm{bol}})\right]}. \label{eq19}
\end{eqnarray}
For reevaluating the wavelength dependence, we use the median bolometric luminosity in each subsample to calculate the correction factor $C_{2}(L_{\mathrm{bol}})$ in step (iii). 
After normalizing the SF$_{0}(\lambda, L_{\mathrm{bol}})$ at $L_{0}$, we fit the SF$_0$ with Equation~(\ref{eq16}).

The separated wavelength and luminosity dependences after the iterations in the step (iv) are shown in the left and the middle panels of Figure~\ref{fig20}, respectively, and the best-fitted parameters of the wavelength dependence and the luminosity dependence are summarized in Table~\ref{tbl7}.
The wavelength dependence is almost consistent with previous work for the SDSS quasars \citep{van04}, and the luminosity dependence is consistent with the result of \citet{cap17} down to $L_{\mathrm{bol}}\sim10^{45}$~erg~s$^{-1}$.
The decrement of variability amplitude at a lower-luminosity range ($L_{\mathrm{bol}}<10^{45}$~erg~s$^{-1}$) can be seen even if we consider SF in shorter wavelength bins where the AGN accretion disk emission is relatively stronger.

Finally, we use the dependences of SF$_0$ on wavelength and luminosity to calculate the intrinsic (i.e., wavelength- and luminosity-independent) dependence of SF on time interval $\Delta t$.
From Equations (\ref{eq13})$-$(\ref{eq15}), the SF can be expressed as
\begin{eqnarray}
{\mathrm{SF}}(\Delta t,\lambda,L_{\mathrm{bol}}) =\left[-2.5\log\left( k V_{1}(\lambda)V_{2}(L_{\mathrm{bol}})\right)\right] \left(\frac{\Delta t}{\Delta t_0}\right)^{b_t},
\label{eq20}
\end{eqnarray}
where $k$ is a normalization factor, which is calculated by the following equation: 
\begin{eqnarray}
{\mathrm{SF}}(\Delta t_{0},\lambda_{0},L_{0}) = {\mathrm{SF}}_0(\lambda_{0}, L_{0}) = -2.5\log(k).
\label{eq21}
\end{eqnarray}
Here, we assume the variability amplitude at $\Delta t_{0} =100$~days, $\lambda_{0}=3000~\mbox{\AA}$, and $L_{0}=10^{45}$~erg s$^{-1}$ is $-2.5\log(0.86)\sim0.164$, which is the value close to the normalizations $a_{\lambda}$ and $a_{L}$ (see Table~\ref{tbl7}).
The correction factor for the wavelength and luminosity dependences is calculated as
\begin{eqnarray}
C_0(\lambda,L_{\mathrm{bol}}) = \frac{{\mathrm{SF}}(\Delta t_0,\lambda_{0},L_{0})}{{\mathrm{SF}}(\Delta t,\lambda,L_{\mathrm{bol}})} =\frac{\log(k)}{\log \left[kV_{1}(\lambda)V_{2}(L_{\mathrm{bol}})\right]}.
\label{eq22}
\end{eqnarray}
To construct the SF normalized at $\lambda_0$ and $L_0$, we calculate the correction factor $C_0(\lambda, L_{\mathrm{bol}})$ by using $\lambda$ and $L_{\mathrm{bol}}$ for individual data points and apply this correction factor to the magnitude difference $\Delta m$ for individual data points.
The same correction factor should also be applied to the magnitude difference for the randomly selected $non$-$Var$ samples to calculate SF$_{\mathrm{noise}}$.
We here calculate the normalized SF$_{\mathrm{net}}$ for the sample with $\lambda \leq 3500~\mbox{\AA}$ and $L_{\mathrm{bol}}\geq10^{45}$~erg~s$^{-1}$, which is less affected by host galaxy contamination as shown before.
We then fit the normalized SF$_{\mathrm{net}}(\Delta t)$ with a power-law function as described in Equation~(\ref{eq12}).
The result is shown in the right panel of Figure~\ref{fig20}, and the best-fitted parameters are summarized in Table~\ref{tbl7}.
We also plot a SF of the DRW model with $\tau = 500$ days (scaled to match the observed data at $\Delta t = 100$~days).
As mentioned above, the SF of the DRW model has a functional form of $[1-\exp(-\Delta t/\tau)]^{1/2}$; thus, at the shorter time interval ($\Delta t\ll \tau$), the SF shows an asymptotic power-law with exponent 0.5.
Our result of the power-law slope of the $\Delta t$ dependence, $0.487 \pm 0.007$,  is consistent with the value expected in the DRW model, which may indicate that the AGN variability is caused by stochastic processes, such as thermal fluctuations of the accretion disk \citep{kel09,dex11}.
It is noted that the SF in the right panel of Figure~\ref{fig20} seems to show that the dumping feature is consistent with the DRW model with $\tau = 500$~days, but this could be due to the insufficient light-curve sampling at the long time interval, as mentioned before (see also Figure~\ref{fig16}).

\begin{table}
\begin{center}
\caption{Best-fit parameters for the structure function}\label{tbl7}
\begin{tabular}{ccc}
\hline
Dependence & Normalization & Slope\\
\hline
Wavelength$^{a}$  & $a_{\lambda}=0.859 \pm 0.001$ & $ b_{\lambda}=0.0872 \pm 0.0050$\\
Luminosity$^{b}$    & $a_{L}           =0.860 \pm 0.003$ & $b_{L}            =0.0202 \pm 0.0021$\\
Time interval$^{c}$ & SF$_{0} =0.170 \pm 0.001$ & $ b_{t} = 0.487 \pm 0.007$\\
\hline
\end{tabular}
\tablecomments{
$^a$ Equation~(\ref{eq16}).
$^b$ Equation~(\ref{eq18}).
$^c$ Equation~(\ref{eq12}).
}
\end{center}
\end{table}

\section{Discussion} \label{sec5}
\subsection{Host Galaxy Contribution to the Structure Function} \label{sec5_1}
\subsubsection{Host Galaxy Contamination} \label{sec5_1_1}

\begin{figure*}[t!]
\plottwo{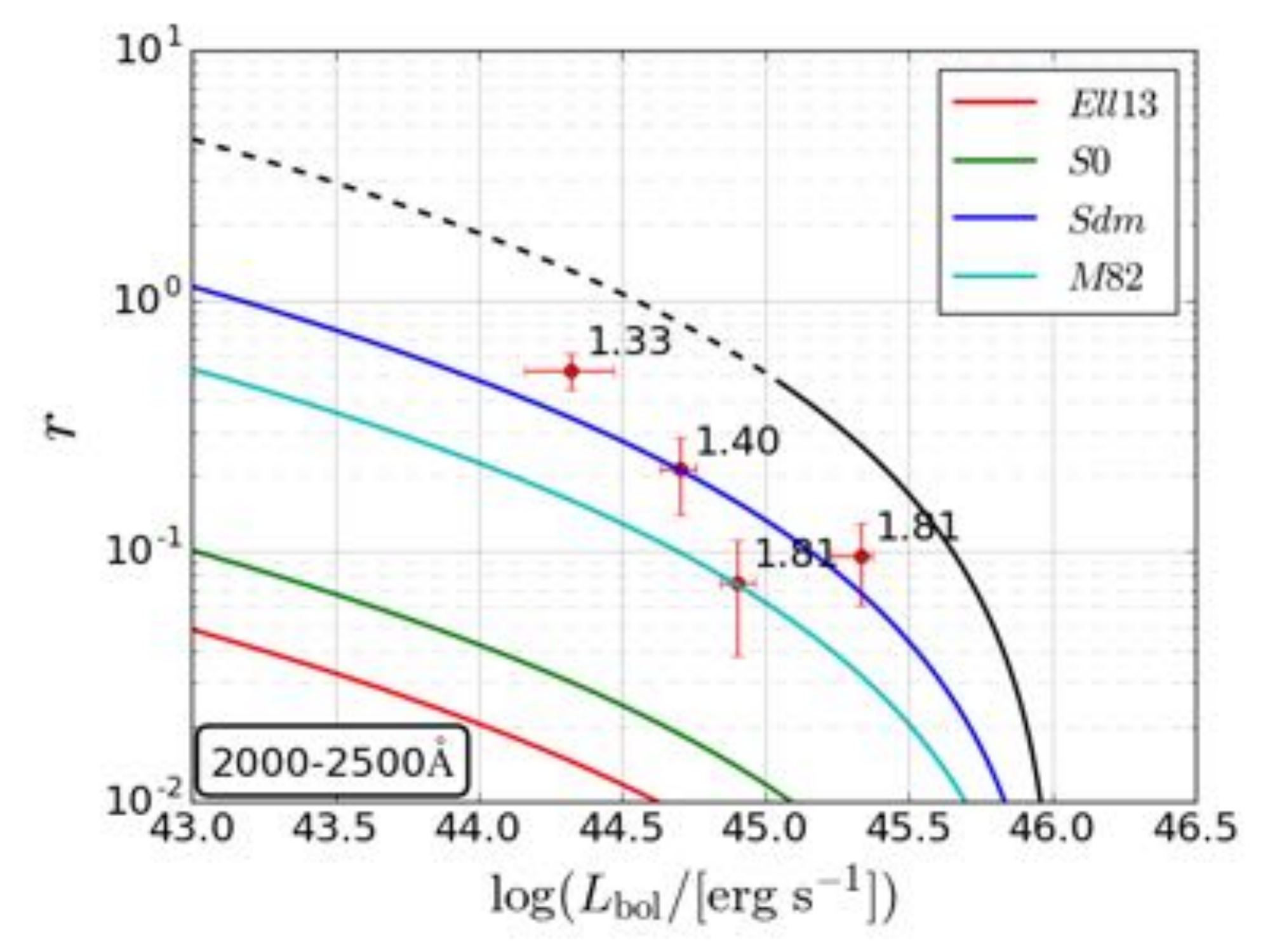}{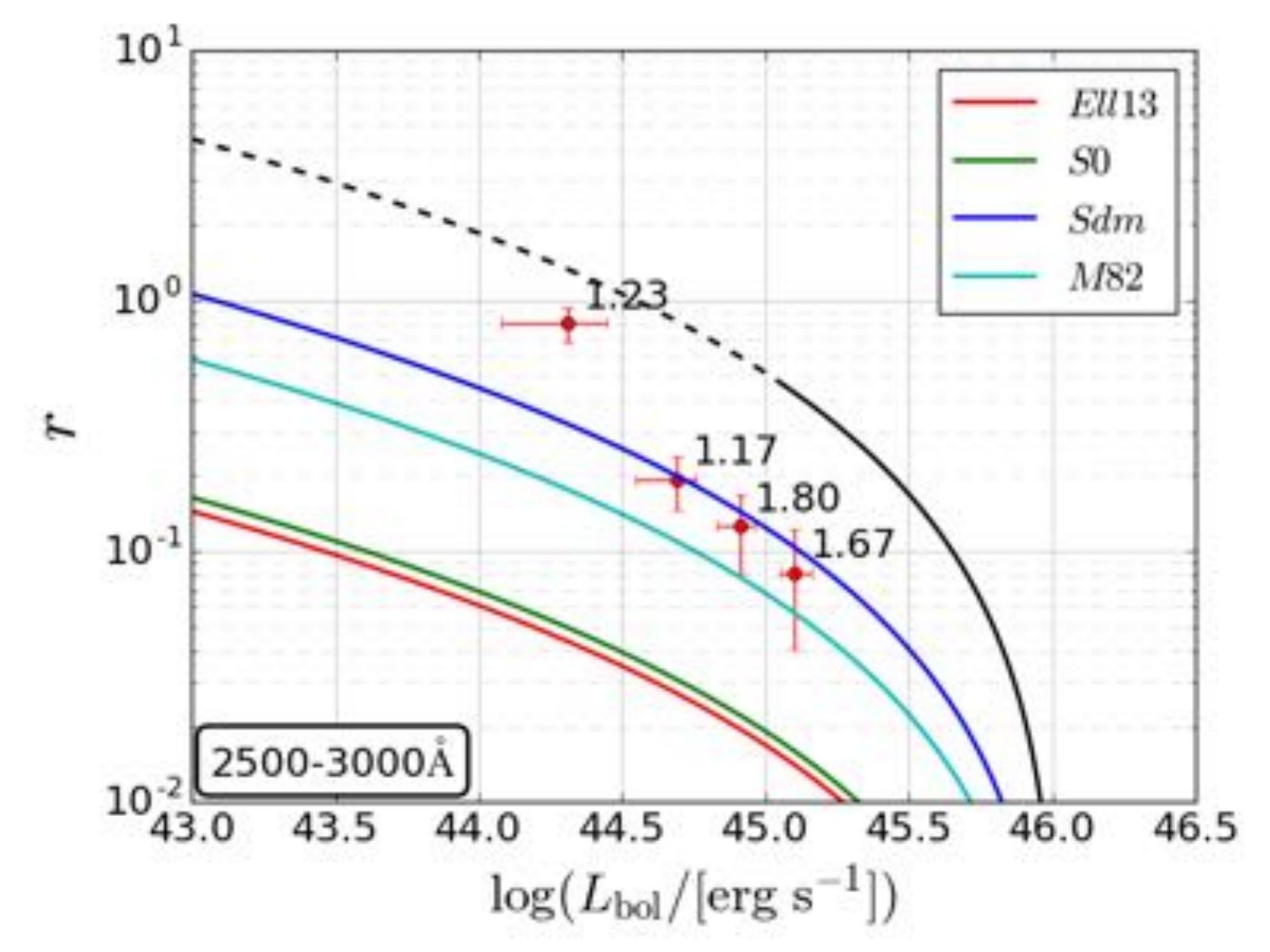}
\plottwo{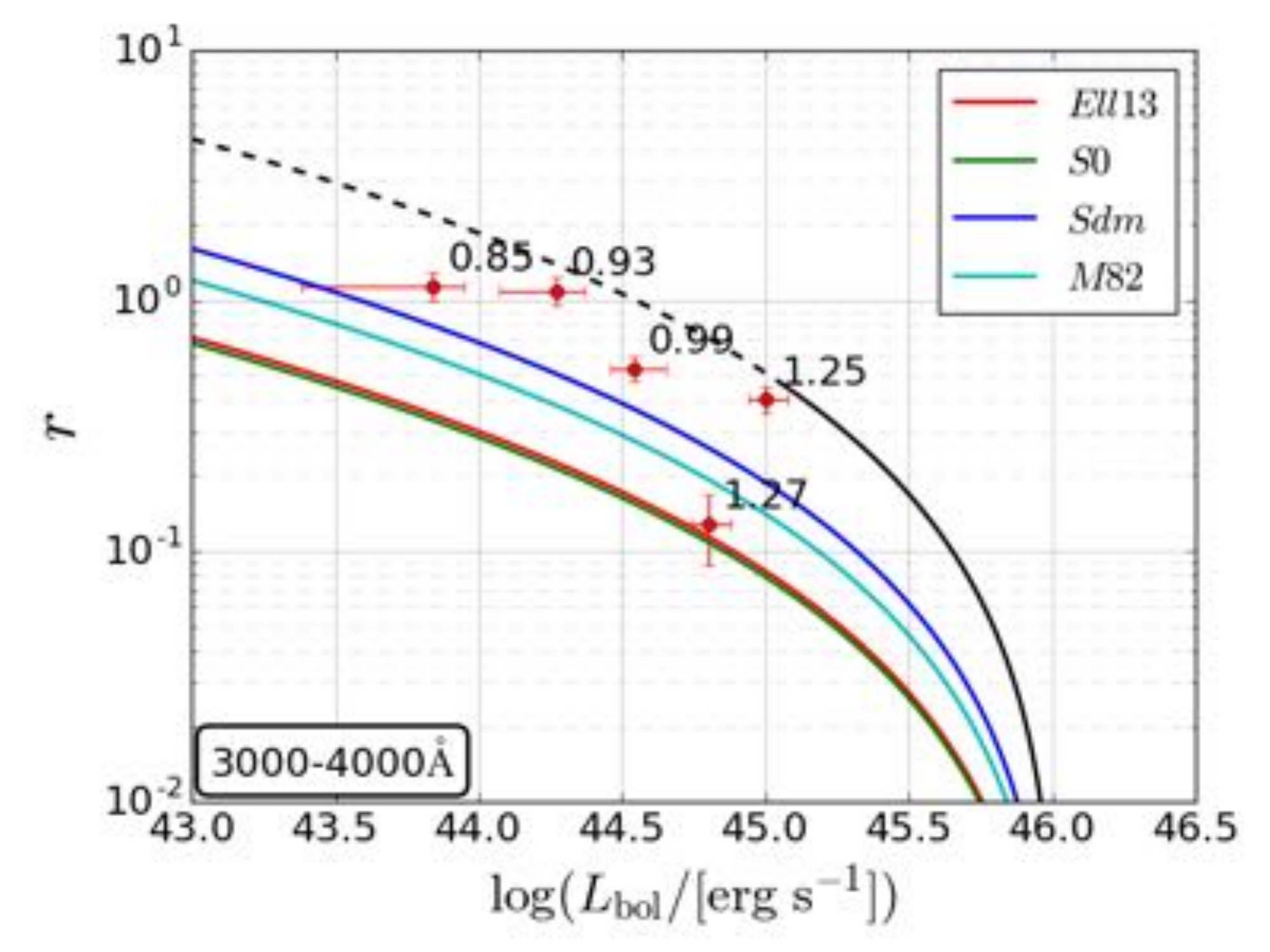}{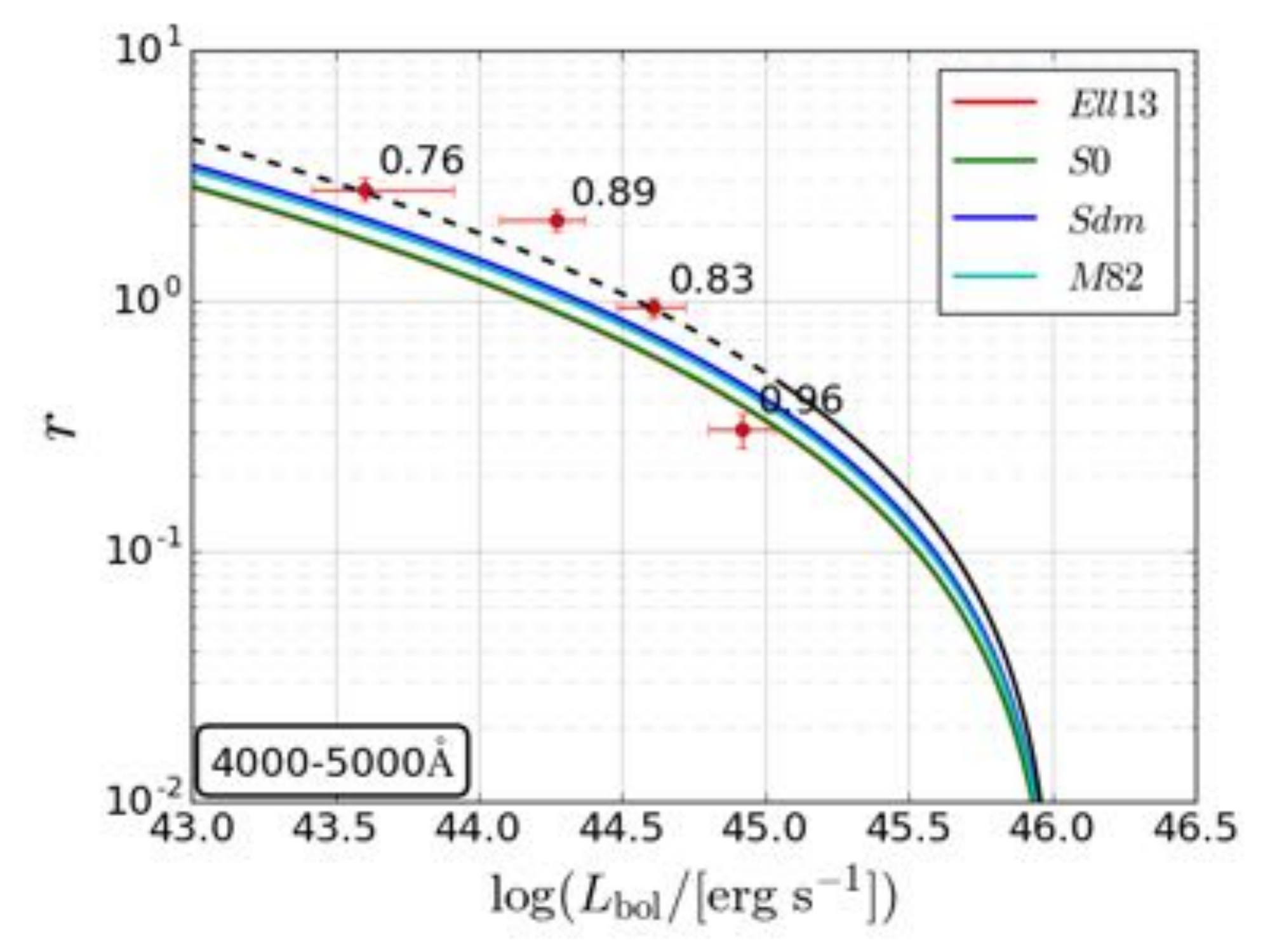}
\caption{
The average host galaxy contamination as a function of AGN bolometric luminosity.
The red points in each panel are the results of each wavelength bin (top left panel: $2000-2500~\mbox{\AA}$, top right panel: $2500-3000~\mbox{\AA}$, bottom left panel: $3000-4000~\mbox{\AA}$, bottom right panel: $4000-5000~\mbox{\AA}$).
The black solid curve is the host contamination at $5100~\mbox{\AA}$ for the SDSS quasars \citep{she11}, and the host contamination is extrapolated to a low luminosity (dashed curve; Equation~(\ref{eq30})).
The colored curves are calculated from composite spectra assuming a host contamination at $5100~\mbox{\AA}$ and a host galaxy type (red: $Ell13$, green: $S0$, blue: $Sdm$, cyan: $M82$). 
The median redshifts of the sample in each luminosity bin are shown at near data points.
\label{fig21}
}
\end{figure*}

As shown in the previous section, the host galaxy lights affect the SF at low-luminosity AGNs ($L_{\mathrm{bol}}\lesssim10^{45}$~erg~s$^{-1}$), such that the observed variability amplitude in units of magnitude decreases as the relative contribution of the host galaxy lights increases.
To understand the effect of the host galaxy lights, we introduce the host galaxy component models in the SF analysis.

As we mentioned in the previous section, the SF represents the RMS of AGN magnitude difference as a function of time interval $\Delta t$.
This means that the SF corresponds to the typical AGN magnitude difference at $\Delta t$.
For simplicity, we here treat the intrinsic AGN SF (SF$_{\mathrm{AGN}}$) in the flux form, namely,
\begin{gather}
{\mathrm{SF}}_{\mathrm{AGN}}(\Delta t) = -2.5\log\left(r_{\mathrm{AGN}}\right)\label{eq23}\\
r_{\mathrm{AGN}}(\Delta t)\equiv \frac{f_{\mathrm{AGN}}(t_2)}{f_{\mathrm{AGN}}(t_1)},\ \ \left(\Delta t\equiv t_2-t_1\right),\label{eq24}
\end{gather}
where $r_{\mathrm{AGN}}$ is the typical AGN flux ratio at a time interval $\Delta t$ and $f_{\mathrm{AGN}}(t)$ is the AGN flux at a time $t$.
On the other hand, the total (AGN+host galaxy) SF (SF$_{\mathrm{total}}$) can then be written as a function of the total flux ratio ($r_{\mathrm{total}}$) as
\begin{gather}
{\mathrm{SF}}_{\mathrm{total}}(\Delta t) = -2.5\log\left(r_{\mathrm{total}}\right)\label{eq25}\\ 
r_{\mathrm{total}}(\Delta t) \equiv \frac{f_{\mathrm{AGN}}(t_2)+f_{\mathrm{host}}}{f_{\mathrm{AGN}}(t_1)+f_{\mathrm{host}}}, \label{eq26}
\end{gather}
where $f_{\mathrm{host}}$ is the host galaxy flux.
Now, we introduce the average host contamination as
\begin{eqnarray}
r\equiv\frac{f_{\mathrm{host}}}{\langle f_{\mathrm{AGN}}(t)\rangle}, \label{eq27}
\end{eqnarray}
where $\langle f_{\mathrm{AGN}}(t)\rangle$ is the time-averaged AGN flux.
We assume that $\langle f_{\mathrm{AGN}}(t)\rangle$ is the arithmetic mean; $\langle f_{\mathrm{AGN}}(t) \rangle = (f_{\mathrm{AGN}}(t_1)+f_{\mathrm{AGN}}(t_2))/2$.
Using Equations (\ref{eq24}) and (\ref{eq27}), Equation~(\ref{eq26}) can be approximated as
\begin{eqnarray}
r_{\mathrm{total}}(\Delta t) \sim \frac{r_{\mathrm{AGN}}(\Delta t)+r}{1+r}. \label{eq28}
\end{eqnarray}
Thus, the host contamination $r$ can be estimated from the AGN intrinsic SF and the total SF values as
\begin{eqnarray}
r \sim \frac{10^{-0.4{\mathrm{SF}}_{\mathrm{total}}(\Delta t)}-10^{-0.4{\mathrm{SF}}_{\mathrm{AGN}}(\Delta t)}}{1-10^{-0.4{\mathrm{SF}}_{\mathrm{total}}(\Delta t)}}. \label{eq29}
\end{eqnarray}
In this paper, we consider $\Delta t=100$~days.
Here, we assume that the intrinsic AGN luminosity dependence (i.e., Equation~(\ref{eq18})) on the SF$_0$ (i.e., SF$_{\mathrm{AGN}}$) obtained for the luminous objects (the solid line in the middle panel of Figure~\ref{fig20}) can be extrapolated to the less-luminous objects, which is suggested from the previous studies.
\citet{aiz14} conduct AGN/host spectral decomposition by using eigenspectra for $\sim5500$ SDSS QSOs at $z\lesssim0.84$ to investigate the luminosity dependence of SF in the $g$, $r$, $i$ bands.
Their result shows that the luminosity dependence continues down to the rest-frame $i$ band absolute magnitude of $M_{i}\sim-18$, which roughly corresponds to an AGN bolometric luminosity of $\sim10^{43.5}$~erg~s$^{-1}$); although, they do not correct for the wavelength dependence.
\citet{hei16} also conduct AGN/host decomposition through SED fitting for $\sim1000$ variability-selected AGNs at $z < 1$ from the PS1 survey and show that the fractional maximum differential-flux of the AGN light curves are anticorrelated with the AGN bolometric luminosity (indicating the variability amplitude $\propto L_{\mathrm{bol}}^{-0.5}$) and this anticorrelation continues to hold down to an AGN bolometric luminosity of $\sim10^{43.5}$~erg~s$^{-1}$.

We use the SF$_0$ for the subsamples in the four wavelength bins of $2000-2500$, $2500-3000$, $3000-4000$, and $4000-5000~\mbox{\AA}$ (Section~\ref{sec4_2}), as the SF$_{\mathrm{total}}$. 
We then calculate the host contamination $r$ from Equation~(\ref{eq29}).
The host contaminations $r$ as a function of AGN bolometric luminosity for each wavelength bin is shown in Figure~\ref{fig21}.
Figure~\ref{fig21} shows that the host contamination increases as the AGN luminosity decreases, and the host contamination is higher at the longer wavelengths.
The derived luminosity dependence of the host contamination is consistent with that of \citet{she11}, who provide an empirical relationship between the host contamination at $5100~\mbox{\AA}$ and the total (AGN+host galaxy) monochromatic luminosity at $5100~\mbox{\AA}$ ($L_{5100\text{\AA}}^{\mathrm{total}}$) for the SDSS quasars with luminosities of $L_{\mathrm{bol}}>10^{45}$~erg~s$^{-1}$ (the black line in Figure~\ref{fig21}), expressed as
\begin{eqnarray}
r_{5100\text{\AA}} = 0.8052 -1.5502x + 0.9121x^2 - 0.1577x^3,
\label{eq30}
\end{eqnarray}
where $x + 44 \equiv \log(L_{5100\text{\AA}}^{\mathrm{total}}/{\mathrm {[erg~s^{-1}]}}) < 45.053$, and the contribution of the host galaxy light can be ignored at at $ (L_{5100\text{\AA}}^{\mathrm{total}} > 10^{45.053}$ erg~s$^{-1}$.

\begin{figure}[t!]
\plotone{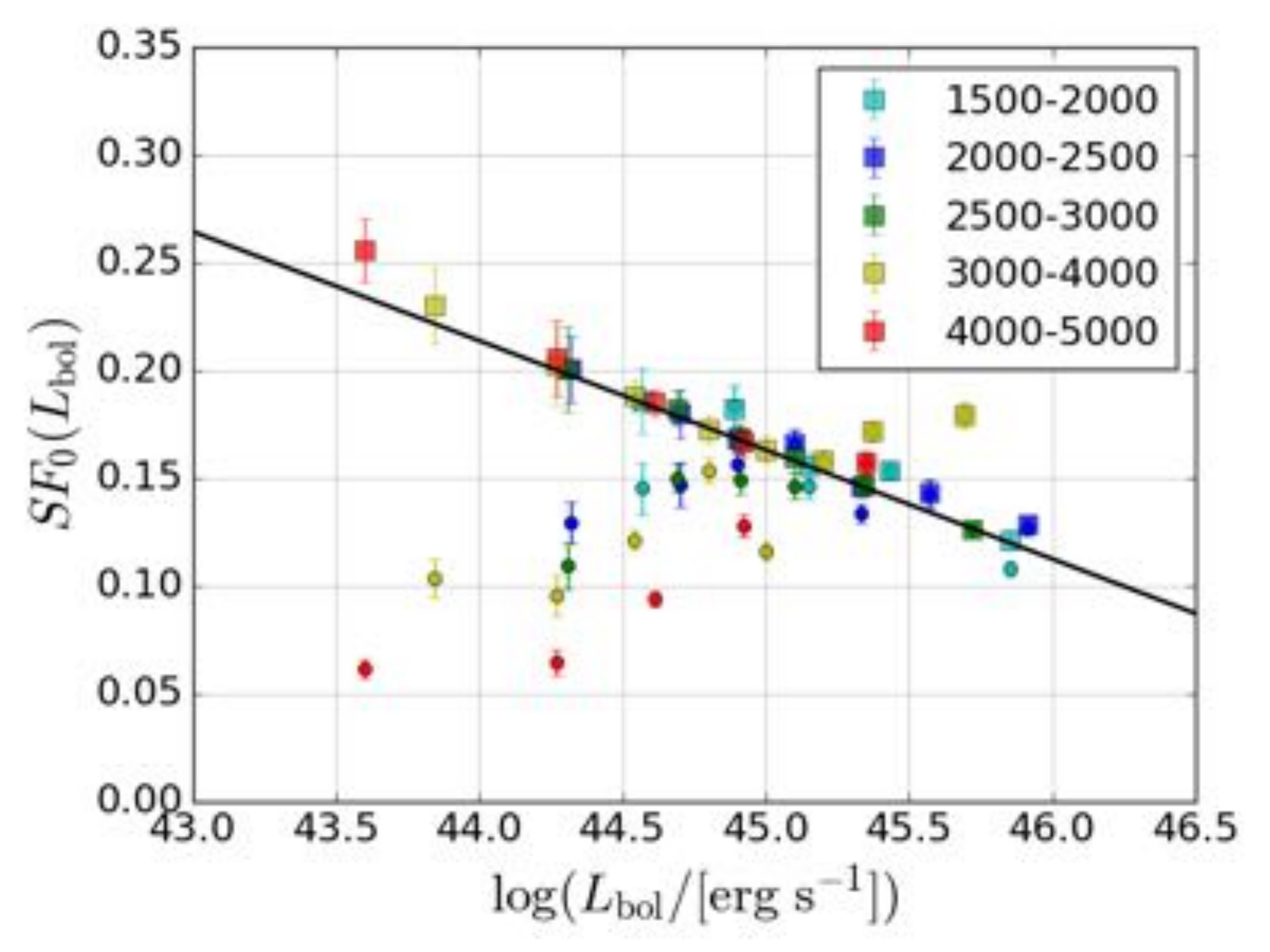}
\caption{
Variability amplitudes, which are  normalized at $3000~\mbox{\AA}$, as a function of AGN bolometric luminosity for samples of (cyan) $1500-2000$, (blue) $2000-2500$, (green) $2500-3000$, (yellow) $3000-4000$, and (red) $4000-5000~\mbox{\AA}$ bins.
The circles are the total SF (SF$_{\mathrm{total}}$) and the squares are the SF$_0$ after subtraction of the host galaxy flux.
The black solid line is the same as the line in the middle panel of Figure~\ref{fig20}.
It is clearly shown that the variability amplitudes after subtraction of the host galaxy flux can reproduce the intrinsic AGN SF (SF$_{\mathrm{AGN}}$).
\label{fig22}
}
\end{figure}

Finally, we need to check whether the approximation described in Equation~(\ref{eq28}) is invalid.
We first use the host contamination $r$ for the subsamples in each wavelength bin to calculate the host galaxy flux as
\begin{eqnarray}
f_{\mathrm{host}} = \left\langle{f_{\mathrm{total}}(t)}\right\rangle \times \frac{r}{1+r},
\label{eq31}
\end{eqnarray}
where $\langle f_{\mathrm{total}}(t)\rangle$ is the time-averaged total (i.e., observed) flux of individual objects in each subsample.
Then, we subtract the host galaxy flux from the total flux to calculate the AGN flux in each epoch.
Using the AGN flux, we reconstruct the luminosity dependence of the SF$_0$ in the same manner as described in Section~\ref{sec4_2} to check whether we can reproduce the intrinsic AGN dependence, i.e., Equation~(\ref{eq18}).
Figure~\ref{fig22} shows the luminosity dependence of the SF$_0$ after subtraction of the host galaxy flux.
It is clear that the SF$_{\mathrm{AGN}}$ (the black solid line in Figure~\ref{fig22}) can be recovered by the SF$_0$ after subtraction of the host galaxy flux, suggesting that the approximation of Equation~(\ref{eq28}) is reasonable. 
In other words, we can easily estimate the typical host contamination from the SF analysis.

\subsubsection{Constraints on the AGN Host Galaxy Type at High Redshift} \label{sec5_1_2}

To identify the typical spectral type of the host galaxies of our variability-selected AGNs, we use the composite spectra with type-I quasars and different types of galaxy SEDs to calculate host contamination in each wavelength bin as shown in Figure~\ref{fig21}.
Here, we use $QSO1$ (type-I QSO) for the quasar SED template and $Ell13$ (13~Gyr old elliptical), $S0$ (spiral 0), $Sdm$ (spiral dm), and $M82$ (starburst) for the host galaxy SED templates, which are presented in \citet{pol07}.
We assume the host contamination at $5100~\mbox{\AA}$ as a function of the luminosity presented in \citet{she11} (Equation~(\ref{eq30})).
The bolometric correction factor for the $5100~\mbox{\AA}$ monochromatic luminosity is assumed to be a constant value, 9.26 \citep{she11}.
Under these assumptions, we construct luminosity-dependent AGN+host galaxy composite SEDs and calculate the contamination in each wavelength bin ($2000-2500$, $2500-3000$, $3000-4000$, and $4000-5000~\mbox{\AA}$), which is shown in Figure~\ref{fig21}.
The median redshifts in each subclass are also shown near the data points in Figure~\ref{fig21}.
It is clearly seen in Figure~\ref{fig21} that in shorter wavelength bins ($\leq3000~\mbox{\AA}$), the host contaminations of young stellar systems, like $Sdm$ and $M82$, are larger than those of old stellar systems, such as $Ell13$ and $S0$, due to the dominance of strong UV$-$optical radiation from the massive stars in young stellar systems.
Figure~\ref{fig21} suggests that the low-luminosity variability-selected AGNs ($L_{\mathrm{bol}}<10^{45.5}$~erg~s$^{-1}$) at high redshift ($0.8\lesssim z\lesssim1.8$) are hosted in the young stellar population systems such as $Sdm$ and $M82$.

\subsection{Interpretation of the X-Ray Absorbed Variable AGNs}\label{sec5_2}

The X-undet  samples in our variability-selected AGNs show the harder stacked X-ray spectrum (i.e., larger gas column density) compared to the X-det sample as described in Section~\ref{sec3_2}.
Among the X-undet samples, the low-mass ($M_{\star} < 10^{10}~M_{\odot}$) subsamples show larger column density ($N_{\mathrm{H}} \gtrsim 10^{23}$~cm${}^{-2}$; see Figure~\ref{fig12}).
Previous studies find that at least $10\%$ of optical spectroscopically identified type-I AGNs are X-ray absorbed \citep{per04,toz06,taj07,mer14,shi18}.
Our X-undet objects constitute $10\%$ of the entire variability-selected AGN sample, which is consistent with these previous studies for the optical spectroscopically identified type-I AGNs.

What are these optically unobscured type-I AGNs with significant X-ray absorption?
One explanation for these objects is to consider a putative `neutral gas torus', which is a geometrically thick, dust-free absorption material colocated with or inside of the BLR.
The neutral gas torus is assumed to have larger opening angles than those of the dusty torus.
If we see the objects from intermediate viewing angles, we can observe them as X-ray absorbed optically unobscured type-I AGNs \citep{dav15, liu18}.

Another possibility is the presence of `shielding gas' in the inner dusty torus, which is related to disk outflows.
A fraction of AGNs show outflow signature, which are observed as broad absorption line (BAL) quasars.
These objects are considered to have high Eddington ratios $\lambda_{\mathrm{Edd}}\gtrsim0.1$; see, e.g., \citealt{gan07}.
To check whether the X-undet samples contain a large number of high Eddington ratio objects that are probably associated with strong disk outflows, we estimate the Eddington ratios of the X-undet samples.
Although there is no direct information about the Eddington ratio for each individual X-undet object, we can estimate the Eddington ratios by using AGN bolometric luminosity ($L_{\mathrm{bol}}$), stellar mass ($M_{\star}$), bulge-to-total stellar mass ratio (B/T), and black hole mass-to-bulge stellar mass ratio ($M_{\mathrm{BH}}/M_{\mathrm{bulge}}$), as follows:
\begin{eqnarray}
\lambda_{\mathrm{Edd}} \equiv \frac{L_{\mathrm{bol}}}{L_{\mathrm{Edd}}}=\frac{L_{\mathrm{bol}}}{1.26\times10^{38} \left(M_{\mathrm{BH}}/M_{\odot}\right)} \hspace{7mm}\nonumber\\
= 0.011\left(\frac{L_{\mathrm{bol}}}{10^{43}\ \mathrm{erg\  s^{-1}}}\right)\left(\frac{M_{\star}}{10^{10}\ M_{\odot}}\right)^{-1}\nonumber\\
\times\left(\frac{B/T}{0.5}\right)^{-1}\left(\frac{M_{\mathrm{BH}}/M_{\mathrm{bulge}}}{0.0014}\right)^{-1}.
\label{eq32}
\end{eqnarray}
Here, we only consider the low-mass ($M_{\star}<10^{10}~M_{\odot}$) samples for low-z ($z\leq0.7$) and high-z ($0.7<z\leq2.0$), which are crucially affected by strong X-ray absorption (Figure~\ref{fig12}).
We use the median values of the stellar mass of the samples, $\log(M_{\star}/M_{\odot})=9.10$ and $9.14$, for low-z and high-z bins, respectively.
We assume that the bulge-to-total stellar mass ratio ($B/T$) is 0.5, which is the intermediate value between early-type and late-type galaxies, and the black hole mass-to-bulge stellar mass ratio ($M_{\mathrm{BH}}/M_{\mathrm{bulge}}$) is $0.14\%$ \citep{har04}.
The AGN bolometric luminosities listed in Table \ref{tbl5} are used for the calculation.
The calculated Eddington ratios are $0.036$ and $0.46$ for the low-z and high-z bins, respectively.
In the low-z bin, the Eddington ratio is slightly lower than that expected for BAL quasars.
On the other hand, in the high-z bin, the Eddington ratio is comparable to that expected for a BAL quasar; thus, it is possible to launch powerful gas outflows.
Additionally, in such a high accretion state, the inner region of the accretion disk can be significantly puffed up due to enhanced radiation pressure in the disk \citep{abr88}, which is predicted as a narrow line Seyfert 1 (NLS1). 
This thick disk can also absorb the X-ray emission, resulting in a hard X-ray spectrum \citep{luo15}.

The origin of the X-undet objects in our variability-selected AGNs is still unclear.
To put more stringent constraints on the nature of these objects (specifically, to examine black hole masses and Eddington ratios), future deep optical spectroscopic follow-up observations are needed.

\section{Summary} \label{sec6}

In this paper, we have investigated the AGN optical variability properties especially for the less-luminous objects detected by the Subaru HSC SSP survey data set in the COSMOS field.
Our variability analysis has been conducted for the $\sim3$~yr data with the four optical filters ($g$, $r$, $i$, and $z$ bands), where the single epoch limiting magnitude is $\sim 25$ mag.
Combining multiple variability selection criteria using single-band variability amplitudes, cross-correlation of multiband light curves, and visual inspection, we have found 491variability-selected AGNs, out of which 441 ($\sim90\%$) objects are detected in the Chandra X-ray imaging.
These variability-selected objects cover a wide range of bolometric luminosity of $L_{\mathrm{bol}}=10^{43.0-46.5}$~erg~s$^{-1}$ and redshift up to $4.26$.

We have conducted an X-ray stacking analysis for the X-undet sample in our variable AGNs and have detected the X-ray signals, which are lower than the Chandra X-ray detection limits for individual sources.
The X-undet sample has harder stacked X-ray spectra compared to the X-det sample,  possibly due to absorption in the soft band X-ray flux.
We have suggested that the X-ray emissions of the X-undet sample are absorbed in the neutral torus, outflowing gas, or the puffed-up accretion disk.

We have shown that the dust covering factor of our variability sample has a similar luminosity dependence to the X-ray absorbed fraction, suggesting that both absorbers of optical and X-ray spectra are regulated by the same physical mechanism, and the geometry of both absorbers gradually changes with increasing AGN luminosity.
We have also found that the dust covering factor of our variability sample is slightly lower than that of optical spectroscopically or photometrically identified AGNs, which is possibly due to the detection of BLAGNs with a line width of $\mathrm{FWHM}<2000$~km~s$^{-1}$ in our variability-selected AGN sample.

We have also shown that a certain fraction of our variability sample is not selected as AGNs in the MIR color-color diagnostics due to the large flux contamination from the host galaxy light.

Based on structure function analysis, we have found that the variability amplitude (at $\Delta t = 100$~days) of the X-det sample in our variable AGNs is anticorrelated with wavelength and AGN bolometric luminosity.
The variability amplitude is correlated with the time interval $\Delta t$ with a power-law slope of $0.487$, which is consistent with the expectation from the DRW model, indicating that the AGN variability is caused by the stochastic processes in the accretion disk. 

At the very low-luminosity range ($L_{\mathrm{bol}} < 10^{45}$~erg~s$^{-1}$), we have found that the observed variability amplitude (at $\Delta t = 100$~days) decreases as the AGN luminosity decreases.
This can naturally be interpreted as the host galaxy flux contamination being more significant for the low-luminosity AGNs, which results in the decrease of the variability amplitude.
Since this decrement is related to the ratio of the host galaxy light to AGN light, we have tried to calculate the host galaxy fraction from the observed variability amplitude and found that the host galaxy fraction increases as the AGN luminosity decreases.
This trend is consistent with previous quasar studies, suggesting that the decrement of the variability amplitude is a good estimator of the typical host galaxy fraction at a given AGN luminosity.
The host galaxy fraction depends not only on the AGN luminosity but also on the wavelength, i.e., it depends on the type of host galaxy.
Compared with the host galaxy fraction calculated from the AGN+host galaxy composite spectra, we have shown that the typical host galaxies of the variability-selected AGNs at $z\gtrsim0.8$ have young stellar populations.
These results suggest that less-luminous AGNs ($L_{\mathrm{bol}}\lesssim10^{45}$~erg~s$^{-1}$) at high redshift ($0.8\lesssim z\lesssim 1.8$) are preferentially hosted in star-forming galaxies.

\acknowledgments
We thank Masaomi Tanaka, Kohei Ichikawa, Masayuki Akiyama, Toshihiro Kawaguchi, Takamitsu Miyaji, Neven Caplar, and Matthew Graham for valuable discussions.
This work was supported by JSPS KAKENHI Grant Numbers JP16H03958, JP17J01884.
 
The Hyper Suprime-Cam (HSC) collaboration includes the astronomical communities of Japan and Taiwan, and Princeton University. The HSC instrumentation and software were developed by the National Astronomical Observatory of Japan (NAOJ), the Kavli Institute for the Physics and Mathematics of the Universe (Kavli IPMU), the University of Tokyo, the High Energy Accelerator Research Organization (KEK), the Academia Sinica Institute for Astronomy and Astrophysics in Taiwan (ASIAA), and Princeton University. Funding was contributed by the FIRST program from Japanese Cabinet Office, the Ministry of Education, Culture, Sports, Science and Technology (MEXT), the Japan Society for the Promotion of Science (JSPS), Japan Science and Technology Agency (JST), the Toray Science Foundation, NAOJ, Kavli IPMU, KEK, ASIAA, and Princeton University. 

This paper makes use of software developed for the Large Synoptic Survey Telescope. We thank the LSST Project for making their code available as free software at \url{http://dm.lsst.org}.

The Pan-STARRS1 Surveys (PS1) have been made possible through contributions of the Institute for Astronomy, the University of Hawaii, the Pan-STARRS Project Office, the Max-Planck Society and its participating institutes, the Max Planck Institute for Astronomy, Heidelberg and the Max Planck Institute for Extraterrestrial Physics, Garching, The Johns Hopkins University, Durham University, the University of Edinburgh, Queen’s University Belfast, the Harvard-Smithsonian Center for Astrophysics, the Las Cumbres Observatory Global Telescope Network Incorporated, the National Central University of Taiwan, the Space Telescope Science Institute, the National Aeronautics and Space Administration under grant No. NNX08AR22G issued through the Planetary Science Division of the NASA Science Mission Directorate, the National Science Foundation under grant No. AST-1238877, the University of Maryland, and Eotvos Lorand University (ELTE) and the Los Alamos National Laboratory.

Based on data collected at the Subaru Telescope and retrieved from the HSC data archive system, which is operated by Subaru Telescope and Astronomy Data Center at National Astronomical Observatory of Japan.
\vspace{5mm}
\facility{Subaru (HSC)}
\software{
Astropy \citep{astropy_13, astropy_18},  
NumPy \citep{numpy_06},
SciPy \citep{scipy_19},
Matplotlib \citep{matplotlib_07},
PyAstronomy \citep{pyastronomy_19},
APLpy \citep{aplpy_12},
SExtractor \citep{sextractor_96},
IRAF \citep{iraf_86, iraf_93},
CSTACK \citep{miy08},
BEHR \citep{par06}
}
\bibliography{main.bbl}

\end{document}